\documentclass[a4paper,fleqn,usenatbib]{mnras}
\usepackage[utf8]{inputenc}
\usepackage[T1]{fontenc}
\usepackage{ae,aecompl}
\usepackage{epsf}
\usepackage{graphics,graphicx}
\usepackage{color}
\usepackage{amssymb}
\usepackage{xcolor}
\usepackage{amsmath,amssymb}
\usepackage{longtable}
\usepackage{deluxetable}
\usepackage{pdflscape}
\usepackage{hyperref}
\hypersetup{colorlinks,citecolor=blue}
\def\arcsec{$^{\prime\prime}$}
\def\~{$\sim$}

\def\Lq{\textquotedblleft}

\def\kms{km~s$^{-1}$}
\newcommand{\myemail}{yogesh.chandola@pmo.ac.cn, yogesh.chandola@gmail.com}
\title[H{\sc i} absorption towards radio AGNs]{ H{\sc i} absorption towards radio Active Galactic Nuclei of different accretion modes}
\author[Chandola Y., Saikia D.J. \& Li Di]{Yogesh Chandola$^{1,2,3}$
\thanks{\myemail}, D.J. Saikia$^{2}$ and Di Li$^{3,4,5}$\\
$^{1}$Purple Mountain Observatory, Chinese Academy of Sciences, 10, Yuan Hua  Road, Qixia District, Nanjing, 210023, China \\
$^{2}$Inter-University Centre for Astronomy and Astrophysics (IUCAA), Post Bag 4, Ganeshkhind, Pune, 411007, India\\ 
$^{3}$National Astronomical Observatories, Chinese Academy of Sciences, 20A, Datun Road, Chaoyang District, Beijing, 100101, China \\
$^{4}$NAOC-UKZN Computational Astrophysics Centre, University of KwaZulu-Natal, Durban 4000, South Africa \\
$^{5}$University of Chinese Academy of Sciences, Beijing, 100049, China}
\date{\today}
\pubyear{2019}
 
\begin{document}
\label{firstpage}
\pagerange{\pageref{firstpage}--\pageref{lastpage}}
\maketitle
\begin{abstract}
	We present results of H{\sc i} absorption experiment done using the Giant Metrewave Radio Telescope (GMRT) towards 27 low- and intermediate-luminosity ($P_{\rm 1.4 GHz}$ $\sim$10$^{23}$-10$^{26}$ W Hz$^{-1}$) radio active galactic nuclei (AGN), classified as either  low excitation radio galaxies (LERGs) or high excitation radio galaxies (HERGs) and with \emph{WISE} colour W2[4.6 $\mu$m]$-$W3[12 $\mu$m]$>$ 2. We report H{\sc i} absorption detection towards seven radio AGNs, six of which are new. Combined with other sources from literature classified as LERGs or HERGs, we confirm our earlier result that compact radio AGNs with \emph{WISE} colour W2$-$W3 $>$2 have higher detection rates compared to those with W2$-$W3 $<$2. We find that H{\sc i} absorption detection rate is higher for HERGs (37.0$^{+15.8}_{-11.5}$ per cent) compared to LERGs (22.0$^{+3.9}_{-3.4}$ per cent), mainly due to a larger fraction of HERGs being gas and dust rich with a younger stellar population compared to LERGs. However, for similar compact radio structures and host galaxies with \emph{WISE} colours W2$-$W3$>$2, we don't find any significant difference in detection rates of two types of AGNs implying detection of H{\sc i} gas may not necessarily mean high excitation mode AGN. We further analysed the kinematics towards these sources.  We find that while LERGs show a wide range in the shift of centroid velocities ($\sim$ $-$479 to $+$356 km s$^{-1}$) relative to the optical systemic velocity,  most of the HERGs  have centroid velocity shift  less than 200 km s$^{-1}$, possibly due to differences in jet-interstellar medium interaction. 
\end{abstract}
\begin{keywords}
galaxies: active -- galaxies: general -- galaxies: nuclei -- infrared: galaxies -- radio lines: galaxies -- radio continuum: galaxies 
\end{keywords}
\section{Introduction} 
     
      Almost all massive galaxies with a bulge have a supermassive black hole (SMBH) present at their centres \citep{2013ARA&A..51..511K}, which is a key component of Active Galactic Nuclei (AGN) models. However, triggering of AGN activity is not limited only to the presence of a SMBH. It also depends on the fueling mechanisms, which are found to be different according to the different types of accretion modes. In terms of Eddington accretion rate, it can be broadly classified into hyper-accretion or super Eddington accretion \citep{1989PASJ...41.1215A}; thin disk or radiatively efficient mode characterised by accretion of cold gas and a high accretion rate \citep{1973A&A....24..337S,1973blho.conf..343N}; and hot accretion or radiatively inefficient mode, characterised by the accretion of hot gas and a low accretion rate  \citep{1994ApJ...428L..13N, 1995ApJ...452..710N, 2014ARA&A..52..529Y}. AGNs with cold accretion modes are known to be efficiently fueled by cold interstellar medium (ISM) gas and have geometrically thin and optically thick accretion disks, surrounded by a torus of dust and cold molecular/atomic gas. Due to the presence of strong optical emission lines, these AGNs are also known as high-excitation mode AGNs \citep{2010A&A...509A...6B}.  These AGNs are also known as radiative mode AGNs as they emit efficiently radiative power across the whole electromagnetic spectrum.  Based upon the orientation of torus with respect to the observer's line of sight to the central AGN, these systems are classified into Type 2 (obscured) and Type 1 (unobscured) AGNs. Obscured AGNs with the presence of dusty torus can be probed with the help of mid-IR, X-ray and radio data \citep{2013MNRAS.434..941M, 2014IAUS..304..209M, 2016MNRAS.462.2631M, 2018ARA&A..56..625H}. Different selection criteria based on all-sky mid-IR \emph{Wide-Field Infrared Survey Explorer} (WISE; \citealt{2010AJ....140.1868W}) colour-colour diagram W1[3.4 $\mu$m]$-$W2[4.6 $\mu$m] vs W2[4.6 $\mu$m]$-$W3[12 $\mu$m] have been discussed in the literature to identify high accretion mode obscured AGNs \citep{2009MNRAS.392..617D, 2012ApJ...753...30S, 2012ApJ...748...80D, 2013AJ....145...55Y, 2014MNRAS.438.1149G, 2015MNRAS.452.3776G,2018MNRAS.473.5210K, 2018MNRAS.480.3201K}. However, it is worth noting that some amount of obscuration can be due to dust in host galaxy ISM and edge-on orientation of host galaxies. Triggering of nuclear activity in high-excitation mode AGNs is generally considered to be through mergers and interactions; the evidence for which reduces with radio power \citep{2019MNRAS.487.5490P}. On the other hand, hot accretion mode AGNs  are fueled by the cooling of hot halo gas \citep{1994ApJ...428L..13N, 1995ApJ...452..710N, 2014ARA&A..52..529Y}. In these systems, the accretion disk is geometrically thick and optically thin with no torus surrounding it \citep{1994ApJ...428L..13N, 1995ApJ...452..710N, 2014ARA&A..52..529Y}. These sources are also referred to as jet-mode AGNs \citep{2014ARA&A..52..589H}.
  \begin{figure*}
  	\centering
  	\includegraphics[scale=0.45]{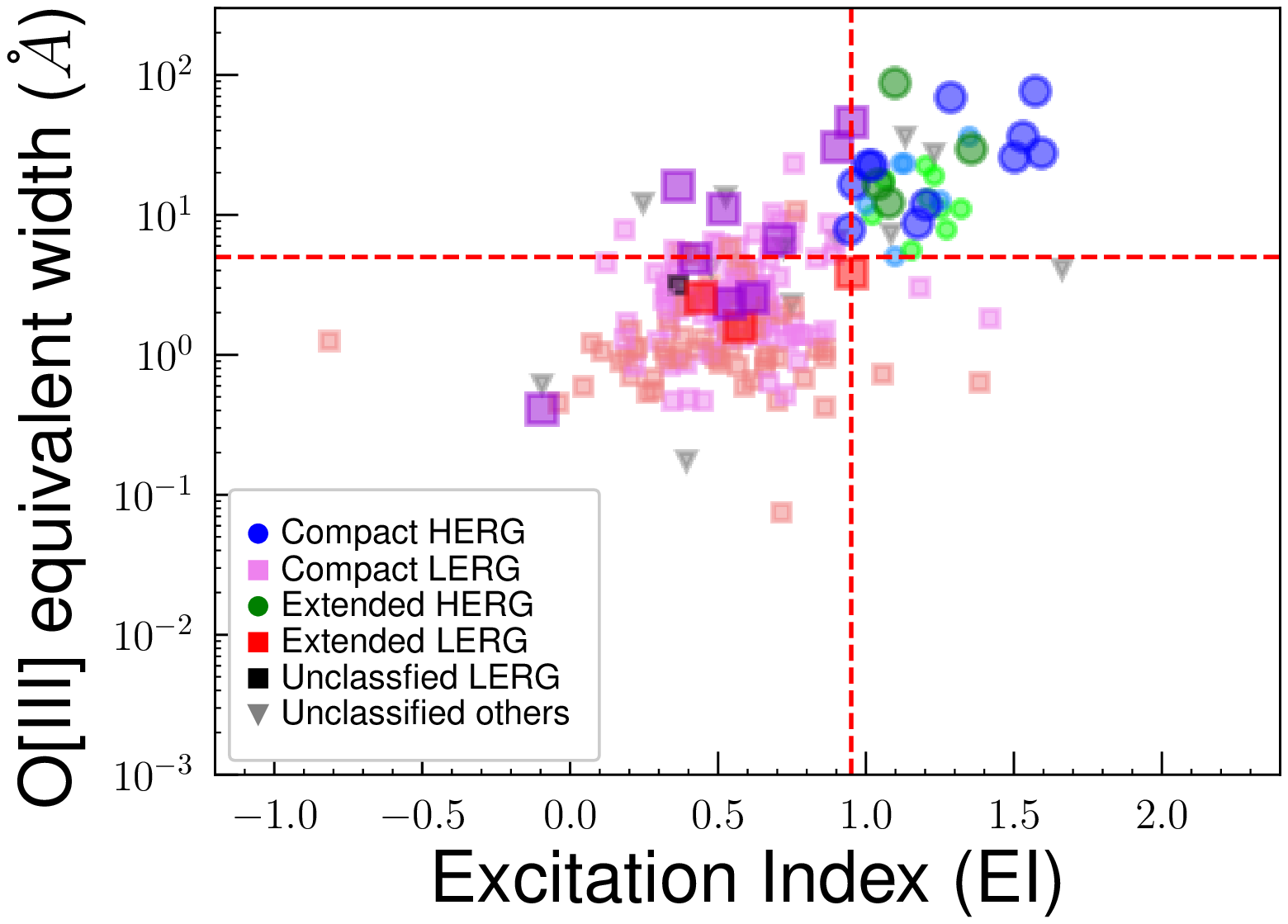}
  	\includegraphics[scale=0.45]{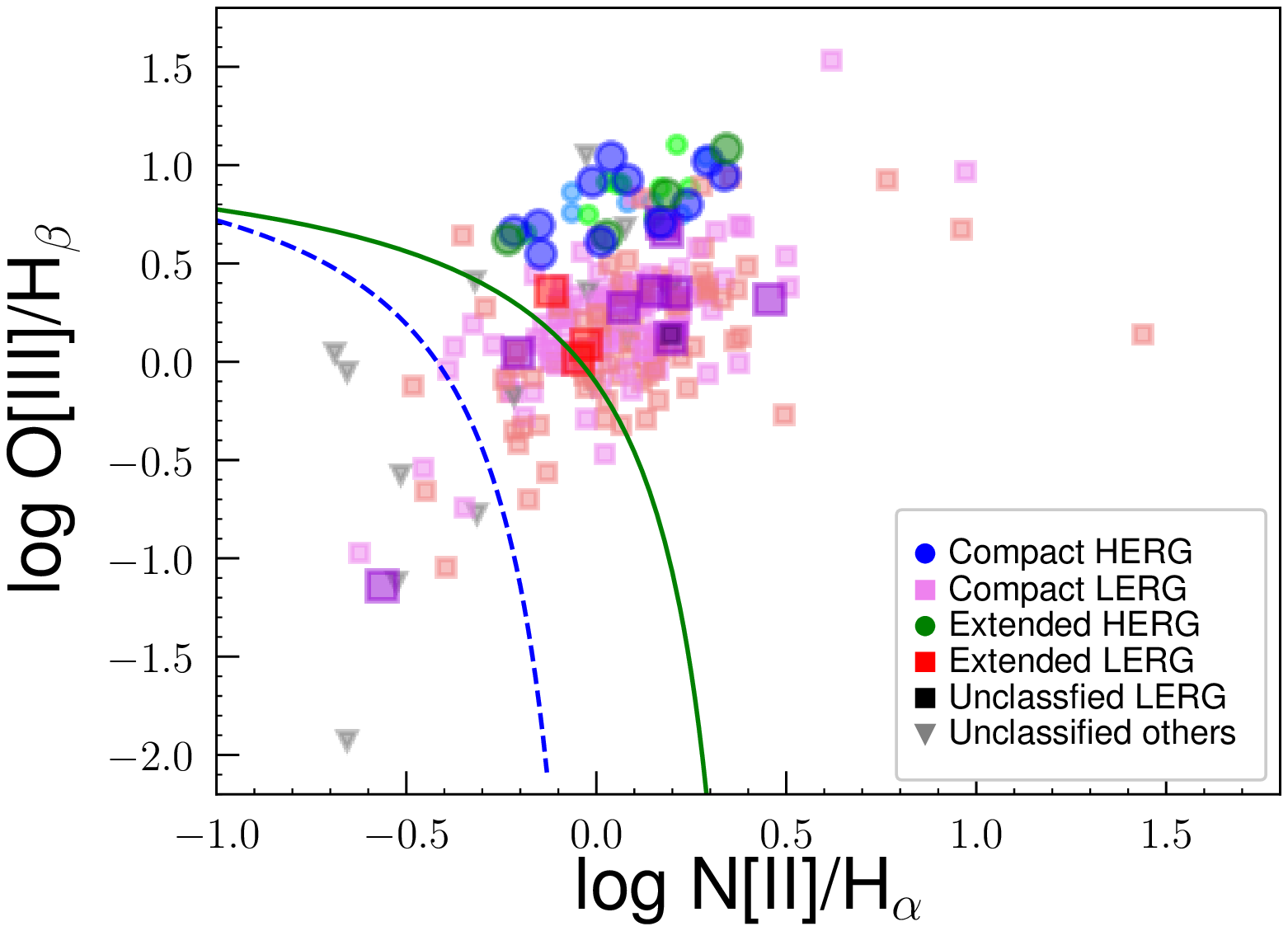}
  	\caption{\textbf{Left:} O[III] equivalent width versus excitation index (for 211 sources with all six emission lines used to calculate excitation index). The \Lq excitation index" has been defined by \protect\cite{2010A&A...509A...6B} as EI $= \rm{log}_{10}(\rm{[OIII]}/H\beta) -\frac{1}{3}\left[\rm{log}_{10}({\rm [NII]} / H\alpha) + \rm{log}_{10}(\rm{[SII]} / H\alpha) + \rm{log}_{10}(\rm{[OI]} /H\alpha)\right]$. Compact and extended HERGs are shown with blue and green circles respectively, while compact, extended and structurally unclassified LERGs are shown with violet, red and black colored squares respectively. Unclassified sources are shown as grey downward triangles. Sources observed with the GMRT are shown with larger symbols with darker shades.  Smaller symbols with lighter shades represent sources from \protect\cite{2017A&A...604A..43M}. While the horizontal line is for O [III] equivalent width = 5 Å, the vertical line depicts EI = 0.95. \textbf{Right:} log O [III]/H$_{\beta}$ versus log N [II]/H$_{\alpha}$ for 238 sources which have the required spectral information. While the dashed blue curve is the \protect \cite{2003MNRAS.346.1055K} dividing line between star-forming (SF) and composite (SF+AGN) galaxies, the solid green curve shows the \protect \cite{2001ApJS..132...37K, 2006MNRAS.372..961K} dividing line between AGNs and composite galaxies. Symbols have the same meaning as in the left plot.}
  	\label{fig1}
  \end{figure*} 

      Nearly 10-20 per cent of all types of AGNs are known to show radio activity in the form of jets extending from parsec to megaparsec scales. Triggering of radio AGN activity may depend on their black hole mass and spin along with the accretion modes and nearby environment of hosts \citep{2014ARA&A..52..589H}. Radio AGNs with high-excitation emission lines and cold accretion mode are termed as High Excitation Radio Galaxies (HERGs) while those with weak emission lines are known as Low Excitation Radio Galaxies (LERGs). In the nearby Universe (z $<$ 0.3), LERGs dominate at low (P$_{\rm 1.4 GHz} \sim$ 10$^{23}$ -10$^{24.3}$ WHz$^{-1}$) and intermediate radio powers (P$_{\rm 1.4 GHz} \sim$ 10$^{24.3}$ -10$^{26}$ WHz$^{-1}$) while HERGs are dominant at high radio powers (P$_{\rm 1.4 GHz} >$ 10$^{26}$ WHz$^{-1}$) \citep{2012MNRAS.421.1569B}. LERGs are hosted by massive early-type galaxies with higher SMBH mass as compared to HERGs. LERG hosts have an older stellar population and lower accretion rates as compared to HERGs \citep{2018MNRAS.480..358W}.  \cite{2018MNRAS.475.3429W} studied properties of intermediate to high redshift LERG and HERG hosts, to find that LERG luminosity function shows negative evolution with increasing redshift (z$\sim$0.5-2) implying decrease in LERG population and HERG as a dominant population at higher redshifts for all values of radio power. This is consistent with the scenario of AGN-host galaxy co-evolution with redshift. Mechanical feedback from radio AGNs is one of the important factors, which can significantly impact this co-evolution. \cite{2018MNRAS.480..358W} find $\sim$ 10\% of accretion power is released as mechanical feedback via radio jets for almost all sources in their sample irrespective of accretion modes. 

		   In radio AGNs, evidence of feedback has been found as outflows of ionized as well as neutral atomic and molecular gas (e.g. PKS1549-79, \citealt{2001MNRAS.327..227T}; 3C305 , \citealt{2005A&A...439..521M};  IC5063, \citealt{2007A&A...476..735M}).  In the ultra-luminous infra-red galaxy (ULIRG) 4C12.50, \cite{2013Sci...341.1082M} reported ongoing parsec scale atomic hydrogen outflow due to jet-cloud interaction coinciding with molecular CO outflow (\citealt{2012A&A...541L...7D}). In some cases, evidence of jet-cloud interactions has been found in the form of largely blueshifted H{\sc i} absorption profiles with widths up to several 1000 \kms \citep{2013MNRAS.435L..58M, 2018MNRAS.473...59A}. Outflowing gas in different phases, especially the cold gas, may have a significant effect on star formation. However, how this may affect the star-formation of host galaxies is still a subject of debate \citep{2017NatAs...1E.165H}. Some studies suggest that AGN outflows deplete the cold gas reservoir in host galaxies, leading to suppressed star-formation  \citep{2011A&A...532L...3L, 2014A&A...562A..21C}. Massive outflow of molecular gas $\dot{M}$ $\thickapprox$ 110 M$_{\odot}$yr$^{-1}$ was detected in NGC1266 but only 2 M$_{\odot}$yr$^{-1}$ could escape the galaxy (\citealt{2015ApJ...798...31A}). In this galaxy, star formation rate was found to be  suppressed by a factor of $\thickapprox$ 50 (Kennicutt-Schmidt law) (\citealt{1998ApJ...498..541K}) compared with a  normal star-forming galaxy, possibly due to mechanical interaction. However, in the galaxy Mrk 231, despite a massive outflow of molecular gas, star-formation efficiency is consistent with the Kennicutt-Schmidt law (\citealt{2015ApJ...801L..17A}). Also, the jet-ISM interaction may have an important role in determining the evolutionary path (\citealt{2010MNRAS.408.2261K}; \citealt{2012ApJ...760...77A}) and structures of radio sources (\citealt{2000A&A...363..507G}). 
            
  H{\sc i} 21-cm absorption towards radio AGNs is a useful technique which has been used to probe the presence of H{\sc i} gas in the circumnuclear environment of these sources \citep{2003A&A...404..861V,2004MNRAS.352...49B, 2006MNRAS.373..972G,2010MNRAS.403..269C, 2011MNRAS.418.1787C, 2012BASI...40..139C, 2013MNRAS.429.2380C, 2010ApJ...715L.117S,  2015MNRAS.453.1249A, 2017A&A...604A..43M, 2017MNRAS.465.5011A,2018MNRAS.473...59A, 2019MNRAS.489.4926G,2019ApJS..245....3G}. Studying cold H{\sc i} gas and its kinematics in the host galaxies and nearby environment of radio AGNs is useful for understanding the fueling and triggering of radio AGN activity in these sources. It is also useful for understanding the feedback effect from central AGN that may affect the star-formation activity in the host galaxy. Several intrinsic physical properties, as well as geometrical factors, which affect the probability of detecting H{\sc i} absorption, have been discussed in the literature. These factors include the size of radio structure, radio power, central AGN characteristics and properties of host galaxy interstellar medium (please see recent review by \citealt{2018A&ARv..26....4M}).

In an earlier paper, we explored the dependence of H{\sc i} absorption properties of the sources on HERGs/LERGs and hence on the accretion mode, source size, and nature of the host galaxy as reflected by the \emph{WISE} infrared colours \citep{2017MNRAS.465..997C}. We considered the sample of 100 sources observed uniformly by  \cite{2015A&A...575A..44G} and the classification into high-excitation and low-excitation radio galaxies done by \cite{2012MNRAS.421.1569B}. We found H{\sc i} detection rates have a significant dependence on \emph{WISE} W2$-$W3 colours with those with W2$-$W3 > 2 and compact radio structures having detection rates $\sim$70 per cent. We also found a trend, which shows HERGs have higher H{\sc i} absorption detection rates as compared to LERGs as most of them are with W2$-$W3 >2. However, for all HERGs and LERGs with W2$-$W3 >2, there was no significant difference in their detection rates. We also reported a trend that H{\sc i} absorption lines associated with HERGs have a narrower range in shift with respect to the velocity corresponding to the red-shift derived from optical emission lines as compared to LERGs. Our study was limited by a small number of 11 HERGs because this low and intermediate radio-power (P$_{\rm 1.4 GHz}$ $\sim$ 10$^{23}$-10$^{26}$ WHz$^{-1}$) sample was dominated by LERGs. In this paper, we extend our study to a larger number of low- and intermediate-luminosity HERGs from our observations with the Giant Metrewave Radio Telescope (GMRT). In addition, we also consider sources from the literature to compare with LERGs, to explore differences in the kinematics of H{\sc i} gas in the hosts of these two types of AGNs.

     This paper is organized as follows. In Section 2, we describe the sample. In Section 3, we explain the observational details and data reduction method. In Section 4, we report the results from our study and  discuss these in the Section 5. We summarise our findings in section 6. In this paper, we adopt a cosmological model with $H_{\rm o}$=70 km s$^{-1}$ Mpc$^{-1}$, $\Omega_{\rm m}$=0.3 and $\Omega_\Lambda$=0.7. \emph{WISE} colours in this paper are in Vega magnitude.

\section{Sample} 
From the sample of radio AGNs compiled by \cite{2012yCat..74211569B, 2012MNRAS.421.1569B} using the SDSS optical spectroscopic data (DR7; \citealt{2009ApJS..182..543A}) and the \emph{Faint Images of the Radio Sky at Twenty Centimeters survey} (FIRST; \citealt{1995ApJ...450..559B}) and \emph{NRAO VLA Sky Survey} (NVSS; \citealt{1998AJ....115.1693C}), we compiled a sample for H{\sc i} absorption observations, which satisfy the following criteria: S(1400 MHz) $>$ 80 mJy, redshift z$<$0.2, \emph{WISE} colour W2$-$W3 $>$ 2 and classified as either a LERG or HERG. \emph{WISE} colour W2$-$W3 $>$ 2 selection was done in order to increase the chances of H{\sc i} absorption detection. This resulted in a sample of 71 sources, of which 29 were HERGs and 42 LERGs. With the allocated time we observed 30 sources with the Giant Metrewave Radio Telescope (GMRT). Of these 30, 3 were found later to have significant redshift uncertainties and have not been used in this analysis. Of the remaining 27 sources listed in Table~\ref{sample}, 12 have been classified as LERGs and 15 as HERGs by \cite{2012MNRAS.421.1569B}. Of these 27, 7 sources, 4 LERGs and 3 HERGs, have baseline ripples in the spectra, and are not used in our final analysis (also see Section \ref{sec:results}). 

In addition to the sources observed with the GMRT, we also considered 219 radio AGNs from a larger sample studied in H{\sc i} absorption by \cite{2017A&A...604A..43M} using Westerbork Synthesis Radio Telescope (WSRT) and common with the sample of \cite{2012MNRAS.421.1569B} for this study. Of these 219 radio AGNs, 189 are classified as LERGs and 15 as HERGs by \cite{2012MNRAS.421.1569B} while the remaining 15 are unclassified. In this sample, 5 sources, all LERGs, (J0816+3804, J0906+4636, J1400+5216, J1435+5051 and J1447+4047) are common with our GMRT observed sample in Table~\ref{sample}. By combining the two samples, we have now a sample of 241 low- and intermediate- radio power (P$_{\rm 1.4 GHz} \sim$ 10$^{22.5}$-10$^{26.2}$ WHz$^{-1}$) AGNs, which include 196 LERGs and 30 HERGs, to study the H{\sc i} absorption properties. Of these 241 sources, we have  shown in Fig.~\ref{fig1} the classification of 211 radio AGNs by \cite{2012MNRAS.421.1569B} in O[{\sc iii}] equivalent width vs. excitation index plot and 238 radio AGNs in BPT  \citep{1981PASP...93....5B} diagram  using the SDSS spectral line catalog data \protect\footnote{http://wwwmpa.mpa-garching.mpg.de/SDSS/DR7/} from Max Planck Institute for  Astrophysics and John Hopkins University group (hereafter MPA-JHU;\citealt{2004astro.ph..6220B}). The remaining three sources, J1030+4113, J1325+4920 and J1516+2919, have no detection of H$_{\alpha}$ emission line in SDSS DR7 and hence not shown in these plots. Of these 3, J1030+4113 and J1325+4920 are LERGs while J1516+2919 is unclassified.

 Of the five sources which are in common, all of which are LERGs, J0816+3804, J1400+5216 and J1447+4047 are undetected in both our GMRT observations as well as in the sample of \cite{2017A&A...604A..43M}. The present observations of J0906+4636 were affected by a bandpass ripple while earlier \cite{2011MNRAS.418.1787C} reported a detection and \cite{2015A&A...575A..44G} a non-detection. For the fifth source J1435+5051 both \cite{2011MNRAS.418.1787C} and the present 
 
\begin{table*}
	\caption{Radio sources observed with the GMRT.}
	\begin{center}
		\scriptsize{ 
			\begin{tabular}{  l l l l lllll }
				\hline
				(1) & (2) & (3) & (4)& (5) & (6)& (7) & (8) \\
				Source name &  R.A. (J2000) & Dec (J2000)  & Redshift & LERG/HERG & C/E &Radio spectral class$^{\dagger}$ & log P$_{\rm 1.4 GHz}$$^{*}$  \\
				& & & & & & &  W Hz$^{-1}$ \\
				
				\hline
				J0028+0055 & 00:28:33.454 & +00:55:10.95 & 0.10429 & LERG & C & CSS & 	24.63\\
				J0813+0734 & 08:13:23.752 & +07:34:05.69 & 0.11239 & LERG & C & CSS & 	25.26\\
				J0816+3804 & 08:16:01.822 & +38:04:14.30 & 0.17275 & LERG & E & -- &   	25.26\\
				J0832+1832 & 08:32:16.042 & +18:32:12.08 & 0.15411 & HERG & C & GPS/CFS? & 	25.64\\
				J0853+0927 & 08:53:23.429 & +09:27:44.33 & 0.11569 & HERG & C & CFS & 	24.52\\
				J0906+4636 & 09:06:15.544 & +46:36:19.04 & 0.0847 & LERG & C & CFS &  	24.70\\
				J0912+5320 & 09:12:01.404 & +53:20:35.36 & 0.10173 & HERG & C & CFS & 	24.45\\
				J1056+1419 & 10:56:38.839 & +14:19:30.39 & 0.08127 & LERG & C & CSS & 	24.43\\
				J1058+5628 & 10:58:37.724 & +56:28:11.26 & 0.14324 & LERG & C & CFS & 	25.12\\
				J1107+1825 & 11:07:01.191 & +18:25:48.80 & 0.17856 & HERG & C & CSS & 	25.04\\
				J1110+2131 & 11:10:20.462 & +21:31:45.82 & 0.13461 & HERG & C & CSS & 	25.06\\
				J1156+2632 & 11:56:54.672 & +26:32:32.38 & 0.15625 & HERG & C & CFS & 	24.69\\
				J1217$-$0337 & 12:17:55.297 & $-$03:37:23.16 & 0.18229 & HERG & E & -- &   	25.23\\
				J1328+1738 & 13:28:59.269 & +17:38:42.31 & 0.18035 & HERG & C & CFS & 	25.19\\
				J1341+5344 & 13:41:34.847 & +53:44:43.82 & 0.14094 & HERG & E & -- &   	25.70\\
				J1350+0940 & 13:50:22.142 & +09:40:10.64 & 0.13255 & LERG & C & GPS/CFS? & 	24.98\\
				J1352-0156 & 13:52:23.470 & $-$01:56:48.32 & 0.16694 & HERG & E & -- & 		25.56\\
				J1400+5216 & 14:00:51.581 & +52:16:06.55 & 0.11789 & LERG & C & CSS & 	24.74\\
				J1410+1438 & 14:10:28.055 & +14:38:40.22 & 0.14419 & LERG & E & -- & 		25.34\\
				J1435+5051 & 14:35:21.664 & +50:51:22.16 & 0.09969 & LERG & C & CFS & 	24.51\\
				J1447+4047 & 14:47:12.766 & +40:47:45.02 & 0.19515 & LERG & E & -- & 		25.58\\
				J1449+4221 & 14:49:20.710 & +42:21:01.39 & 0.17862 & HERG & C & CFS & 	24.93\\
				J1534+2330 & 15:34:57.247 & +23:30:11.48 & 0.0184 & LERG & C & CFS & 		23.40\\
				J1538+5525 & 15:38:36.083 & +55:25:41.39 & 0.19117 & HERG & C & CSS & 	25.32\\
				J1543+0235 & 15:43:17.401 & +02:35:52.04 & 0.18793 & HERG & E & -- & 		25.58\\
				J1559+5330 & 15:59:27.661 & +53:30:53.52 & 0.17918 & HERG & C & CSS & 	25.09\\
				J2133$-$0712 & 21:33:33.322 & $-$07:12:49.30 & 0.08654 & HERG & C & CSS & 	24.54\\
				
				\hline
			\end{tabular}
		}
	\end{center}
	\vspace{-0.4 cm}
	\begin{flushleft}
		Column 1: source name; column 2: right ascension (J2000);
		column 3: declination (J2000);
		column 4; redshift;
		column 5: LERG and HERG classification based on \cite{2012MNRAS.421.1569B};
		column 6: radio structural classification as compact (C) and extended (E), sources with $\lesssim$ 20 kpc in linear projected sizes are classified as compact and  larger ones as extended;
		column 7: radio spectral class;
		column 8: logarithm of luminosity at 1.4 GHz in units of W Hz$^{-1}$.\\
		
		$^{\dagger}$:  Using the radio flux densities obtained from ViZieR photometry viewer (\url{http://vizier.u-strasbg.fr/vizier/sed/}). CFS: Compact Flat Spectrum; CSS: Compact Steep Spectrum; GPS: Gigahertz Peaked Spectrum. Sources with spectral index ($\alpha$) $\leq$ 0.5 are classified as flat spectrum sources while sources with $\alpha$ $>$ 0.5 are classified as steep spectrum sources, where S$_{\nu}$ $\propto$ $\nu^{-\alpha}$. \\
		$^{*}$: Using integrated flux densities listed in Table~\ref{results}. 
	\end{flushleft}
	\label{sample}
\end{table*}
\begin{table*}
	\caption{Observational details of the search for associated H{\sc i} absorption.}
	\begin{center}
		\scriptsize{
			\begin{tabular}{ l l c c c r l l c }
				
				\hline
				(1)& (2)& (3)& (4)& (5)& (6) \\
				Source name & Ob. frequency (MHz)       & Flux Cal.     & Phase Cal.                   & Ob. Date          & Hours       \\
				\hline
				J0028+0055 & 1286.26      & 3C48        & 0022+002                   & November 10, 2016 &  4   \\
				J0813+0734 & 1276.90      & 3C147       & 0745+101                   & November 11, 2016 & 3     \\
				J0816+3804 & 1211.18      & 3C48,3C147       & 0741+312                   & November 12, 2016 & 4     \\
				J0832+1832  & 1230.74    & 3C147       & 0842+185                   & May 15, 2016       & 3           \\
				J0853+0927 & 1273.12      & 3C48, 3C286  & 0842+185                   & November 01, 2016 & 6          \\
				J0906+4636 & 1309.49      & 3C147       & 0713+438                   & December 09, 2016 & 4     \\
				J0912+5320 & 1289.25      & 3C147       & 0614+607                   & November 02, 2016 & 5            \\
				J1056+1419 & 1313.65      & 3C147, 3C286       & 1120+143        & December 02, 2016 & 4       \\
				J1058+5628 & 1242.44      & 3C147       & 1035+564 & December 02, 2016 & 4       \\
				J1107+1825  & 1205.20       & 3C48        & 1120+143                   & May 29, 2016       & 4             \\
				J1110+2131 & 1251.89      & 3C286  & 1120+143                   & July 31, 2016      & 3             \\
				J1156+2632 & 1228.46      & 3C286       & 1221+282                   & November 13, 2016 & 6      \\
				J1217$-$0337  & 1201.40       & 3C147, 3C286       & 1150$-$003                   & July 09, 2016      & 4           \\
				J1328+1738 & 1203.38      & 3C286       & 1347+122                   & December 18, 2016 & 5        \\
				J1341+5344 & 1244.94      & 3C286       & 1400+621                   & November 13, 2016 & 3       \\
				J1350+0940 & 1254.17      & 3C286       & 1445+099                   & November 15, 2016 & 3        \\
				J1352$-$0156  & 1217.21       & 3C286       & 1445+099          & July 09, 2016      & 3            \\
				J1400+5216 & 1270.61      & 3C147, 3C286 & 1400+621                   & November 12, 2016 & 4        \\
				J1410+1438 & 1241.41      & 3C286       & 1445+099                   & November 14, 2016 & 3       \\
				J1435+5051 & 1291.64      & 3C286       & 1438+621                   & December 17, 2016 & 4        \\
				J1447+4047 & 1188.47      & 3C286       & 1602+334 & December 18, 2016 & 3         \\
				J1449+4221 & 1205.14      & 3C286       & 1602+334 & September 01, 2016       & 4               \\
				J1534+2330 & 1394.74      & 3C286       & 1609+266                   & December 20,2016 & 3         \\
				J1538+5525  & 1192.46      & 3C286       & 1438+621         & July 09, 2016      & 4               \\
				J1543+0235 & 1195.70       & 3C286       & 1445+099          & June 06, 2016       & 3          \\
				J1559+5330  & 1204.57      & 3C286       & 1438+621                   & June 16-17, 2016   & 4             \\
				J2133$-$0712 & 1307.27      & 3C48        & 2136+006                   &  May 13, 2016       & 4      \\

				\hline
			\end{tabular}
		}
	\end{center}
	\begin{flushleft}
		Column 1: source name; column 2: central line frequency used for observation in MHz;
		column 3: flux density/bandpass calibrator;
		column 4: Phase calibrator; column 5: date of observation;
		column 6: total observation time in hours.
	\end{flushleft}
	\label{GMRTobs}
\end{table*}
\begin{landscape}
	\begin{deluxetable}{lrlrrrrrrrrrr}
		\tablewidth{0pt}
		\tabletypesize{\scriptsize}
		\tablecaption{Results of the search for associated H{\sc i} absorption.}

		\startdata
		\hline 
		(1) & (2)  & (3) & (4) & (5)  & (6) & (7) & (8)           & (9)         & (10) & (11)      & (12) &  (13)   \\ 
		Source & Vel. res.  & Beam size & S$_{\rm int}$ & S$_{\rm peak}$  & $\Delta$S$_{\rm rms}$ & $\tau_{\rm rms}$ & $\int \tau$ dv           & $N$(H{\sc i})         & No. & Velo.      & FWHM       &  $\tau_{\rm peak}$ \\
		name &   &  &  &  & &  &            &          & Gauss & shift      &       &   \\
		
		& km s$^{-1}$  & \arcsec $\times$ \arcsec, P.A. ($^{\circ}$) & mJy & mJy/b  & mJy/b &  & km s$^{-1}$          & 10$^{20}$ cm$^{-2}$         &  & km s$^{-1}$      &km s$^{-1}$       &     \\   
		
		\hline
		J0028+0055 & 7.67  & 2.69$\times$2.23, 80.56  	& 183.9 & 124.1  & 1.09 & 0.009      & $<$1.2           & $<$2.1        & 0 &        &       &          \\
		J0813+0734 & 7.74  & 2.63$\times$1.68, 57.89  	& 676.6 & 547.3 & -- & --  &-- &-- & -- & &   &     \\
		
		J0816+3804 & 8.18  & 2.60$\times$1.55, 25.40  	& 277.8  & 21.6  & 3.51 & 0.166     & $<$22.8         & $<$ 41.5      & 0 &        &       &         \\
		J0832+1832 &  8.02     & 2.35$\times$1.64, 67.88  	& 823.0  & 758.4 & -- & --   & --   &    --    &  -- &        &       &          \\
		J0853+0927 & 7.73  & 2.08$\times$1.55, 51.98  	& 111.7  & 101.4  & 1.34 & 0.01	     & 10.5$\pm$0.8    & 19.0$\pm$1.4    & 1 & $-$351.1$\pm$40.1 & 144.2$\pm$68.5 & 0.022$\pm$0.005   \\
		&       & 			       	         	&       &        &        &      &           		         &                  & 2 & $-$206.2$\pm$15.9 & 132.9$\pm$25.9 & 0.050$\pm$0.007 \\
		J0906+4636 & 8.19  & 2.91$\times$1.82, 38.67  	& 312.8 & 308.6 &  --    & --     & --   & -- &  --      &       &     &    \\
		J0912+5320 & 7.65  & 2.82$\times$1.67, 67.77  	& 121.2 & 121.7 & 1.15 & 0.009     & 1.8$\pm$0.3    & 3.4$\pm$0.5   & 1 & 25.6$\pm$3.2   & 41.2$\pm$7.4  & 0.042$\pm$0.007  \\
		
		J1056+1419 & 7.5   & 2.40$\times$1.72, 59.33 	& 185.8 & 166.7 & 2.04 & 0.012     & 7.0$\pm$0.8     & 12.8$\pm$1.4   & 1 & $-$2.7$\pm$6.0 & 161.9$\pm$15.0 & 0.036$\pm$0.003 \\
		&       & 			       	        	&       &        &      &       &                   		 &                 & 2 & 158.2$\pm$4.0  & 32.3$\pm$9.6  & 0.024$\pm$0.006  \\
		J1058+5628 & 7.9  & 3.22$\times$1.57, 66.09 	& 270.7 & 260.5 & -- & --     & --           & --        & -- &        &       &        \\
		J1107+1825 &  8.19   & 2.79$\times$1.84, 89.97 	& 156.4 & 128.6 &  -- &  --  &  -- & -- & --   &        &       &        \\
		J1110+2131 & 7.91 & 2.22$\times$2.05, 27.99 	& 285.5 & 270.6 & 1.18 & 0.004     & $<$0.6           & $<$1.1         & 0 &        &       &        \\
		J1156+2632 & 8.0  & 2.15$\times$1.72, $-$87.73 	& 86.35  & 84.8   & 1.27 & 0.015     & $<$2.0          & $<$3.7          & 0 &        &       &        \\
		J1217$-$0337 & 8.23 & 2.48$\times$2.15, $-$5.39 	& 235.8 & 159.8  & 1.52 & 0.01      & $<$1.3           & $<$2.4         & 0 &        &       &        \\
		J1328+1738 & 8.19 & 2.21$\times$1.71, 77.22 	& 205.4 & 193.1 & 1.76 & 0.009  & $<$1.3    &$<$2.4    & 0 &  &   &   \\
		J1341+5344 & 7.92 & 3.34$\times$1.69, 63.5 	& 1260.2 & 14.1 & 1.69 & 0.124     & $<$16.7          & $<$30.4        & 0 &        &       &        \\
		J1350+0940 & 7.85 & 2.85$\times$1.88, 86.04 	& 239.2  & 235.9  & -- & -- & --&  -- & -- & &  &   \\
		J1352$-$0156 & 8.15 & 3.44$\times$1.96, 54.12 	& 616.0 & 361.2 & 1.63 & 0.005     & 5.8$\pm$0.4     & 10.6$\pm$0.7   & 1 & 44.0$\pm$13.9 & 299.6$\pm$27.8 & 0.013$\pm$0.001  \\
		&      &     			       	        	&       &        &      &       &                   &        & 2 & 146.4$\pm$1.3  & 38.5$\pm$1.5  & 0.041$\pm$0.003 \\
		J1400+5216 & 7.76 & 3.08$\times$1.36, 54.79 	& 179.3 & 166.6 & 2.82 & 0.017     & $<$2.3           & $<$4.1        & 0 &        &       &        \\
		J1410+1438 & 7.96 & 3.74$\times$1.93, 87.42 	& 480.4 & 125.6 & 1.31 & 0.01      & $<$1.4           & $<$2.6         & 0 &        &       &        \\
		J1435+5051 & 7.65 & 2.77$\times$1.61, 55.47 	& 147.3 & 147.0 & 1.15 & 0.008     & $<$1.0          & $<$1.9        & 0 &        &       &        \\
		J1447+4047 & 8.3  & 2.48$\times$1.84, 19.73 	& 456.7  & 44.5  & 1.35 & 0.031     & $<$4.2           & $<$7.7          & 0 &        &       &        \\
		J1449+4221 & 8.2  & 2.57$\times$2.27, $-$63.0 	& 111.0 & 111.2  & 1.33 & 0.012     & $<$1.7           & $<$3.0         & 0 &        &       &        \\
		J1534+2330 & 7.06 & 2.55$\times$1.81, 69.46 	& 349.3  & 277.1  & 1.21 & 0.004     & 89.4$\pm$0.5    & 163.0$\pm$0.9      & 1 &$-$325.9$\pm$14.3 & 139.3$\pm$25.3 & 0.024$\pm$0.003  \\
		&  & 	& 	 & 	 &	 &	 &  &  & 2 &$-$219.5$\pm$3.9 & 85.8$\pm$11.5  & 0.046$\pm$0.008  \\
		&	 &	&	 &   &	 & 	 &  &  & 3 &$-$97.4$\pm$0.7  & 24.8$\pm$2.0  & 0.057$\pm$0.004  \\
		&	 &	&  	 &   & 	 & 	 &  &  & 4 &$-$65.4$\pm$1.6  & 255.6$\pm$6.5 & 0.285$\pm$0.002  \\
		&	 &	&	 &   & 	 & 	 &  &  & 5 &   77.0$\pm$0.8  & 17.8$\pm$2.0  & 0.045$\pm$0.004  \\
		&   &	& 	 &    &	 &	 &  &   & 6 & 156.0$\pm$11.8  & 121.2$\pm$27.6 & 0.014$\pm$0.003  \\
		J1538+5525 & 8.23 & 3.06$\times$1.82, 44.2 	& 259.6  & 182.4 & 0.84 & 0.005 & 2.0$\pm$0.3      & 3.6$\pm$0.6  	   & 1 &$-$105.3$\pm$2.3 & 18.3$\pm$5.9 & 0.021$\pm$0.006  \\
		&      & 			       	       	&        &      &        &       &                   &               			   & 2 &$-$48.8$\pm$23.0  & 186.0$\pm$48.4 & 0.008$\pm$0.002  \\
		J1543+0235 & 8.28 &     3.12$\times$2.35, $-$82.92 & 495.2 & 297.2 & -- 	&  -- &  --     &   --   & -- &         &       &        \\
		J1559+5330 & 8.21 & 3.22$\times$1.64, 11.53 	& 177.7 & 168.2 & 3.06 & 0.01  & $<$1.4           & $<$2.6            & 0 &         &       &       \\
		J2133$-$0712 & 14.85& 2.59$\times$2.11, 37.1 	& 213.7  & 211.3  & 0.64 & 0.003 & 5.5$\pm$0.1     & 10.1$\pm$0.3      & 1 & 50.6$\pm$0.5    & 51.8$\pm$1.2  & 0.119$\pm$0.002 \\
		
		\hline 
		\enddata
		\begin{flushleft}
			Column 1: source name; column 2: velocity resolution; column 3:  spatial resolution in arcseconds and position angle (P.A.) in degrees; column 4: integrated flux density in mJy; 
			column 5:  Peak flux density in mJy/beam for the continuum image made using line free channels;
			column 6:  r.m.s.in line free channels; column 7: r.m.s in optical depth for line free channels ; column 8: Integrated optical depth, and column 9:  H{\sc i} column density in units of 
			10$^{20}$ cm$^{-2}$, assuming ${T}_{\rm s}$=100 K and a covering factor, $f_c$ of unity; upper limits are 3$\sigma$ values,
			assuming FWHM $\Delta v$=100 \kms ; column 10: No. of Gaussian component; columns 11 to 13: fit parameters of Gaussian profiles, velocity shift w.r.t. systemic velocity, Full Width Half Maximum  and peak optical depth.\\
		\end{flushleft}
		\label{results}
	\end{deluxetable}
\end{landscape}
\begin{figure*}
	\centering
	\hbox{ 
		\includegraphics[scale=0.37]{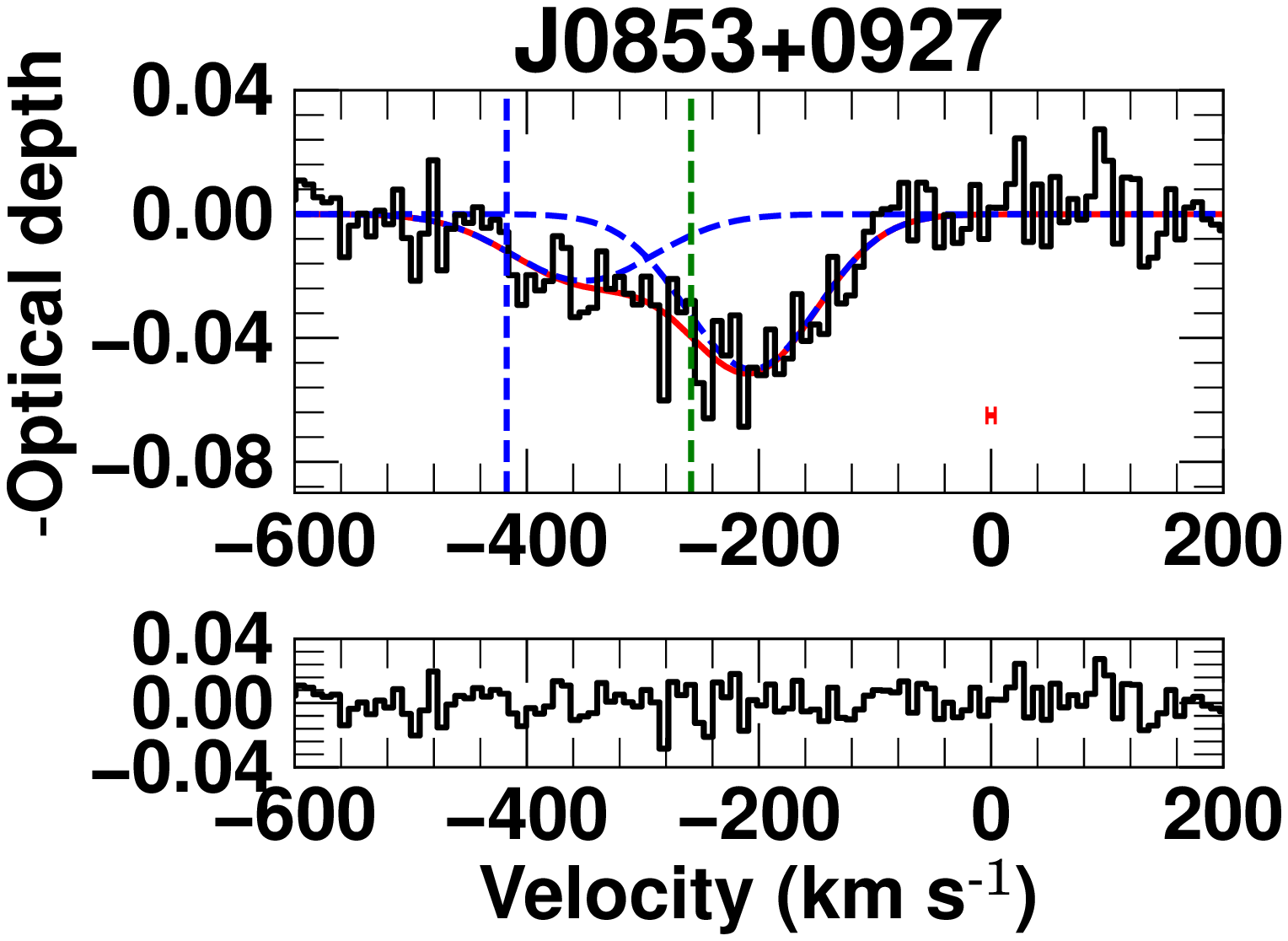}
		\includegraphics[scale=0.37]{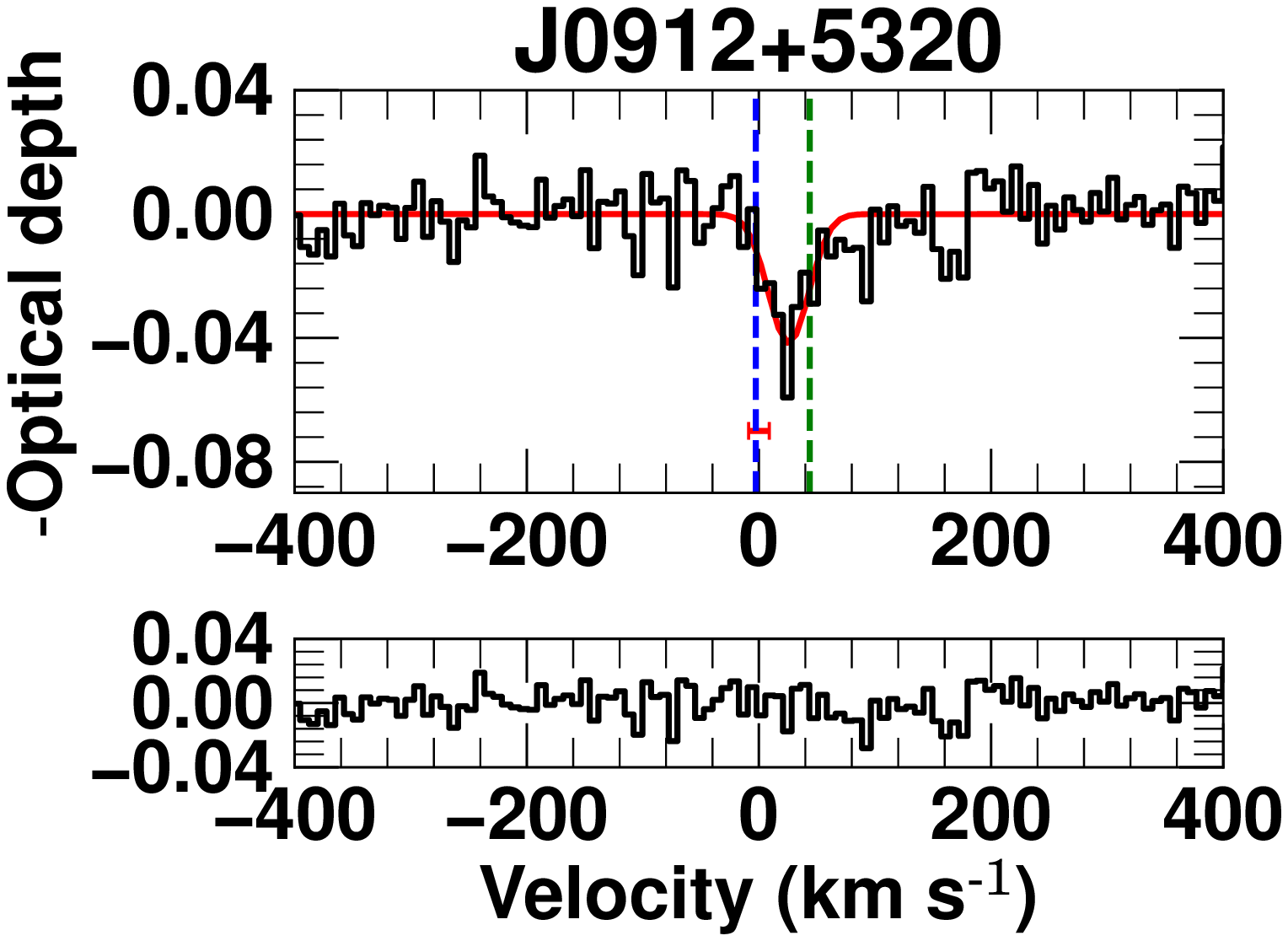}
		\includegraphics[scale=0.37]{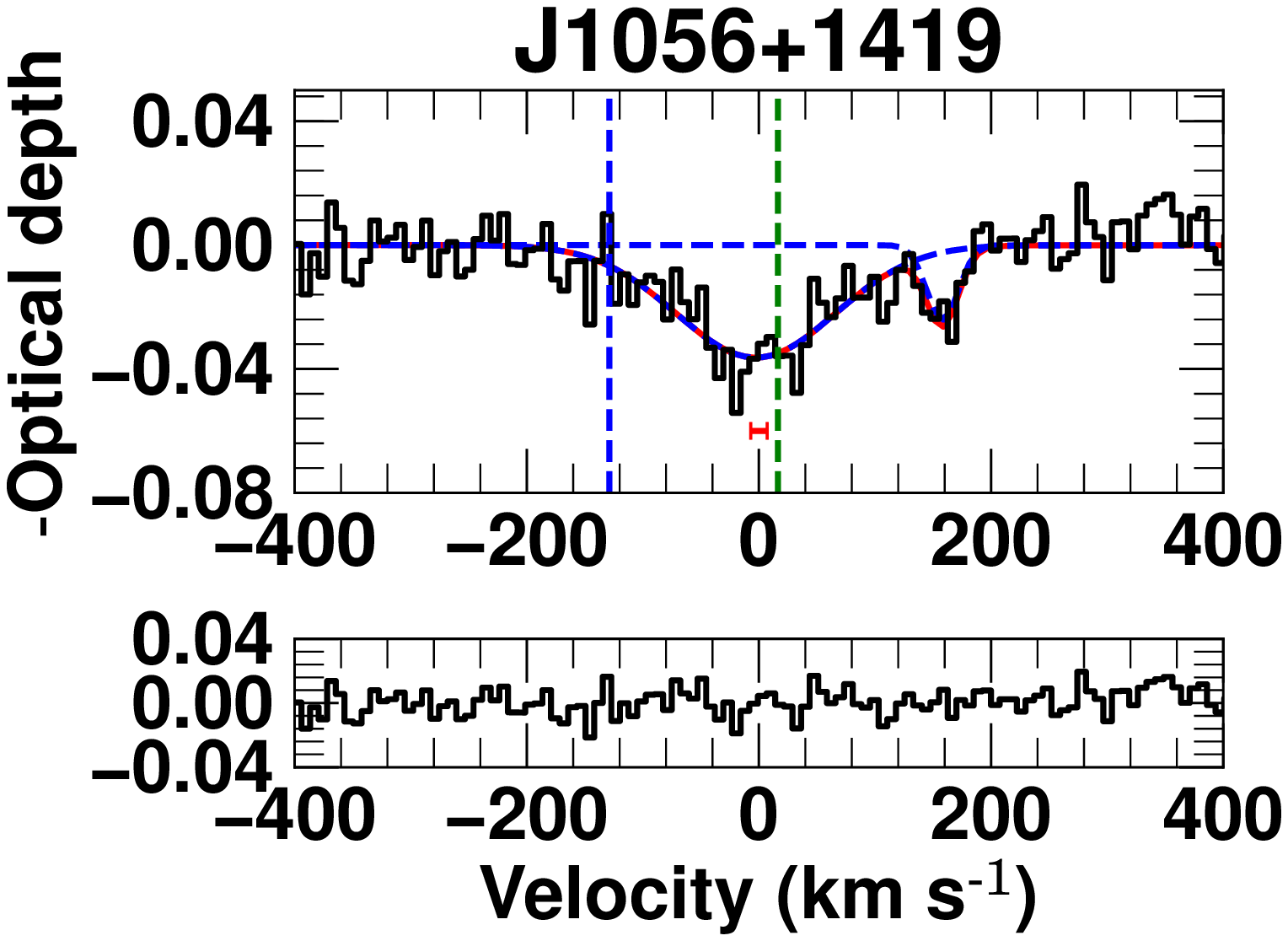} 
	}
	\hbox{
		\includegraphics[scale=0.37]{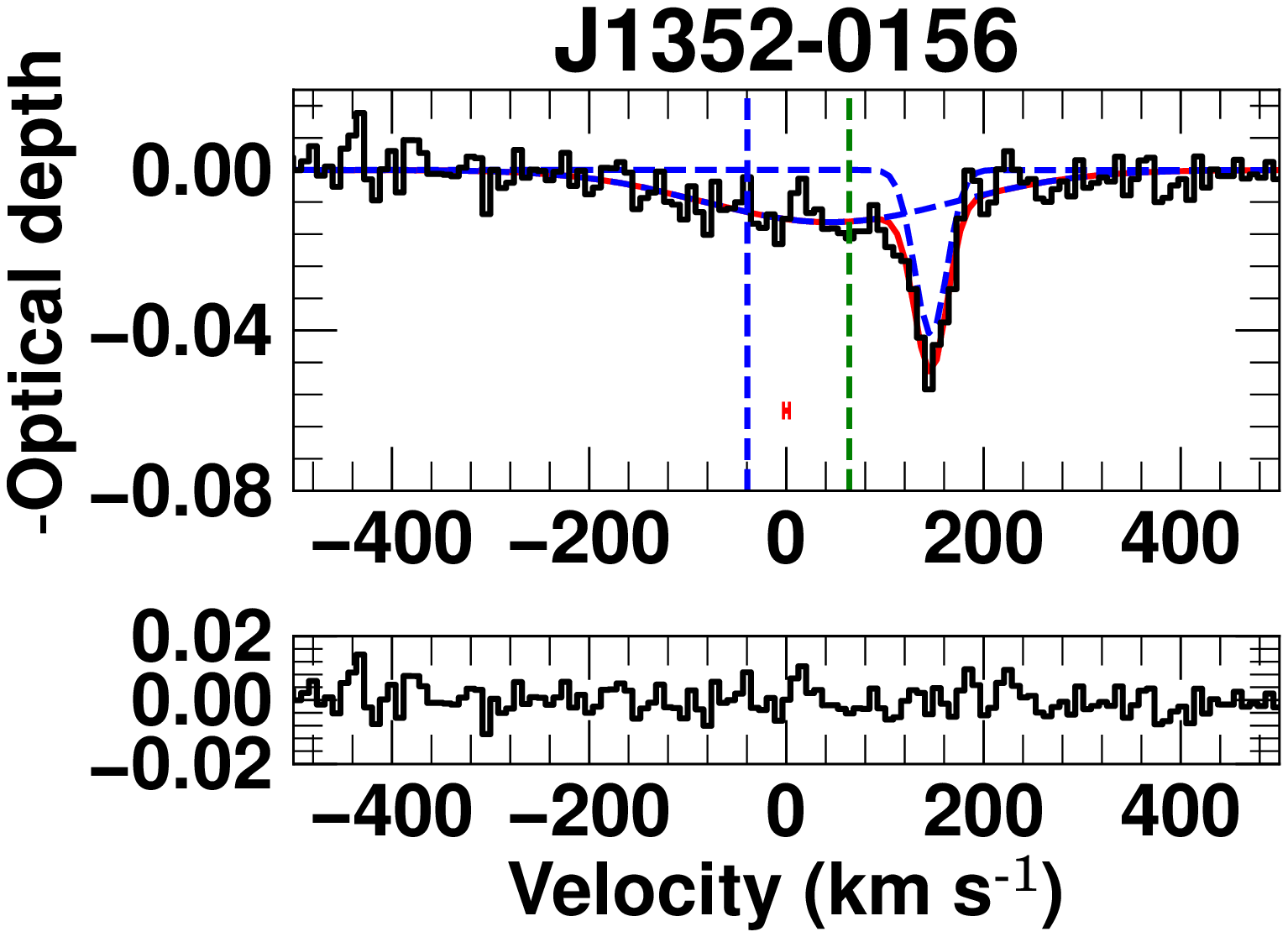}
		\includegraphics[scale=0.37]{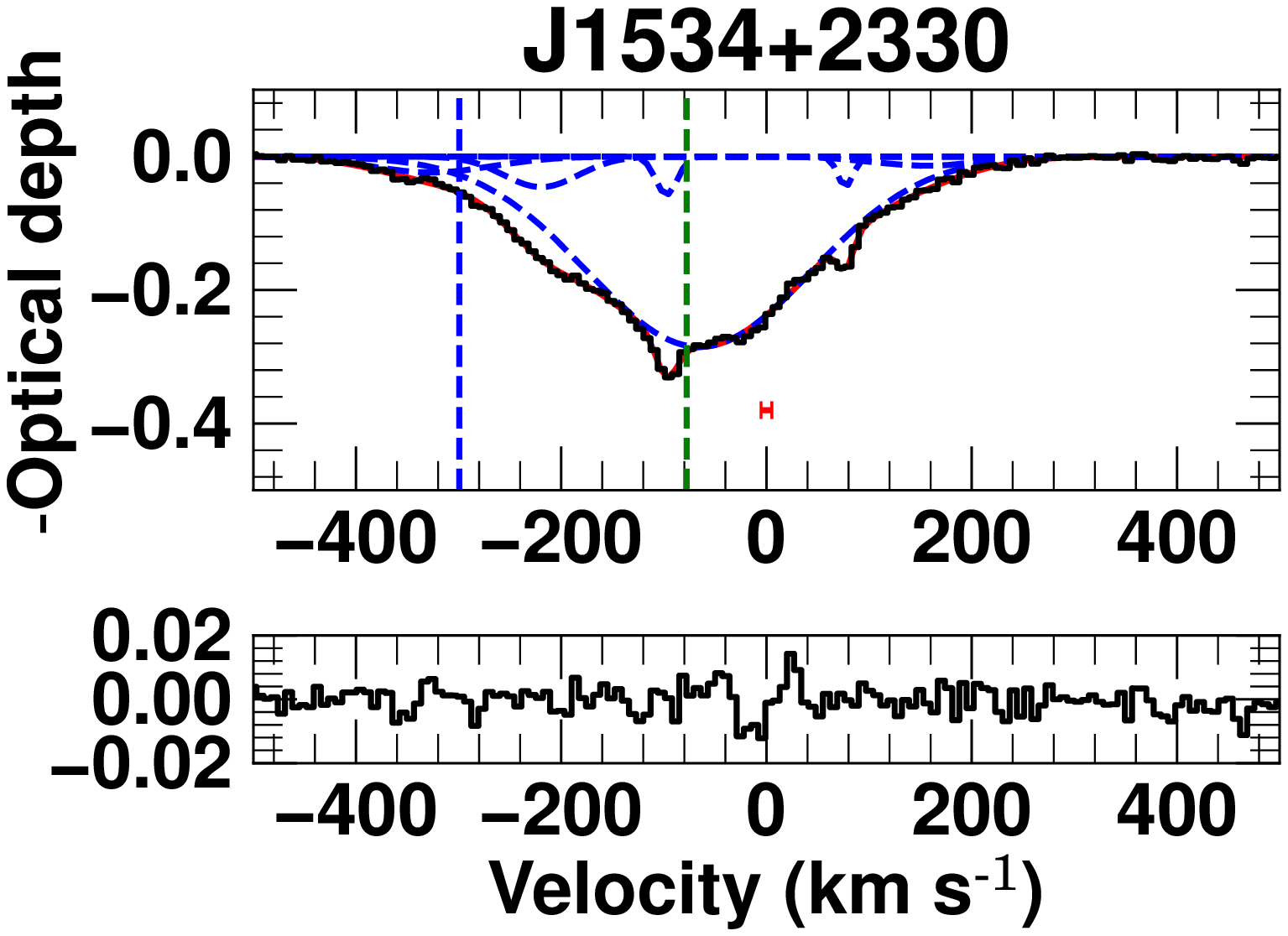}
		\includegraphics[scale=0.37]{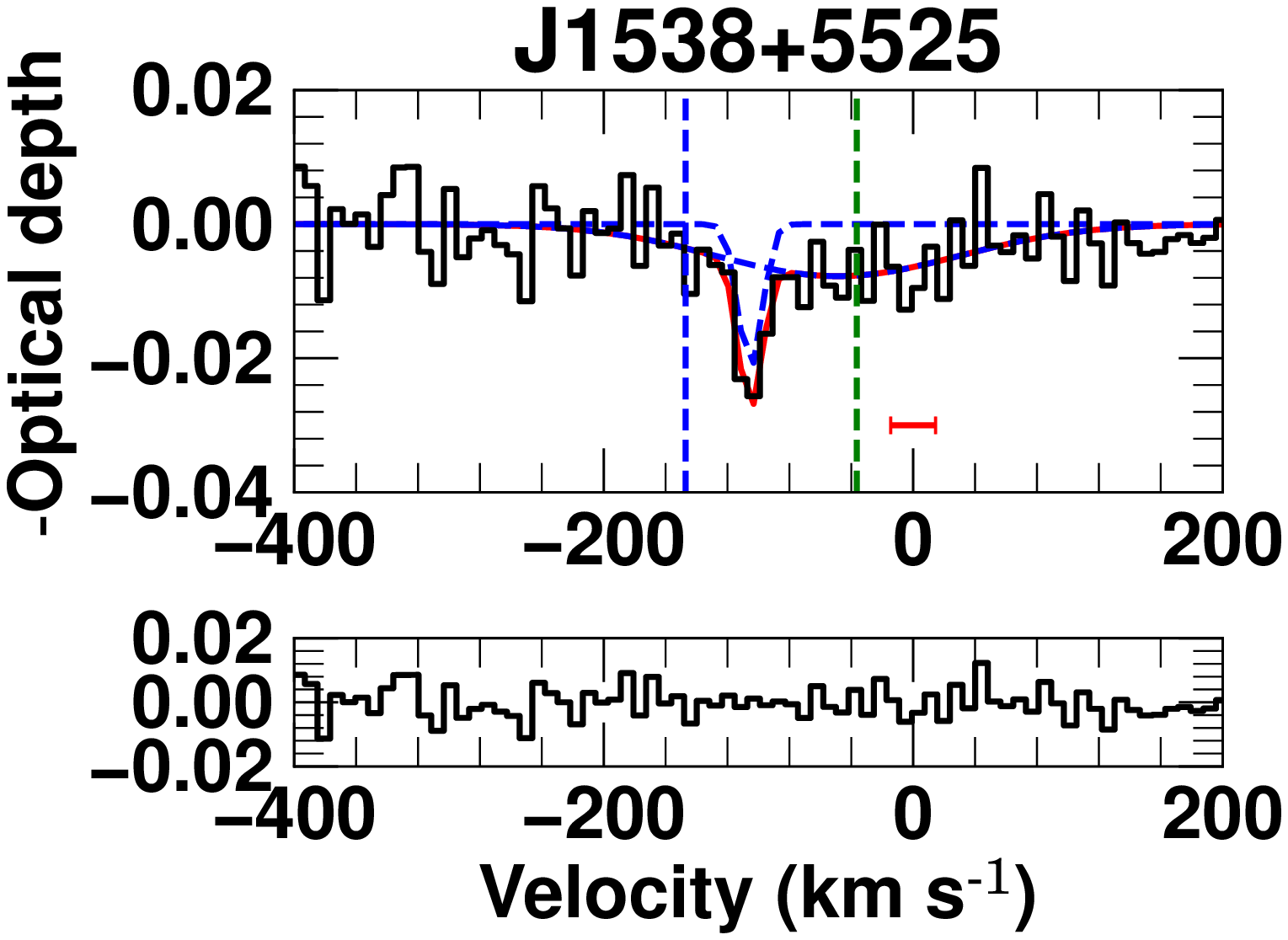}
	}
	
	\hbox{
		\includegraphics[scale=0.37]{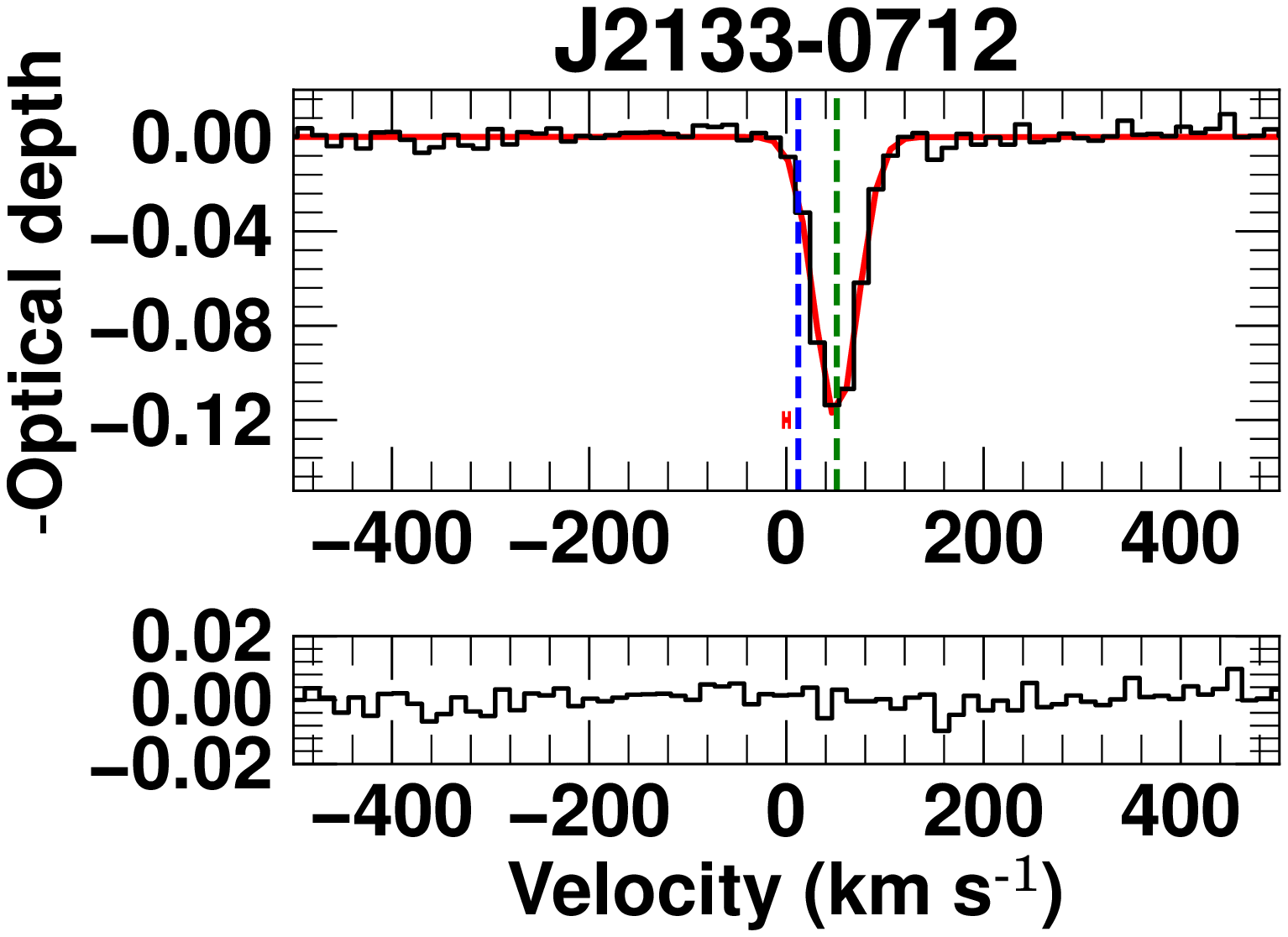} 
	}
	
	\caption{H{\sc i} absorption profiles towards brightest pixel in sources detected with H{\sc i} absorption. X-axes show velocity shift w.r.t. optical systemic velocity and Y-axes show optical depth. Zero represent the optical systemic velocity corresponding to optical redshift. Gaussian profile fit to the profiles are shown with red colour lines, where individual components are shown with blue colour dashed lines. Residuals from fit are shown at the bottom panels of each plot. Error bars in red colour at the bottom of each plot show the error in optical redshift. Green and blue  dashed vertical lines show position of $V_{\rm centroid}$ and $-V_{\rm FW20}$ respectively for Busy function fit.}
	\label{fig2}
\end{figure*} 

\begin{figure*}
	\begin{center}
		
		\hbox{ 
			\includegraphics[scale=0.3]{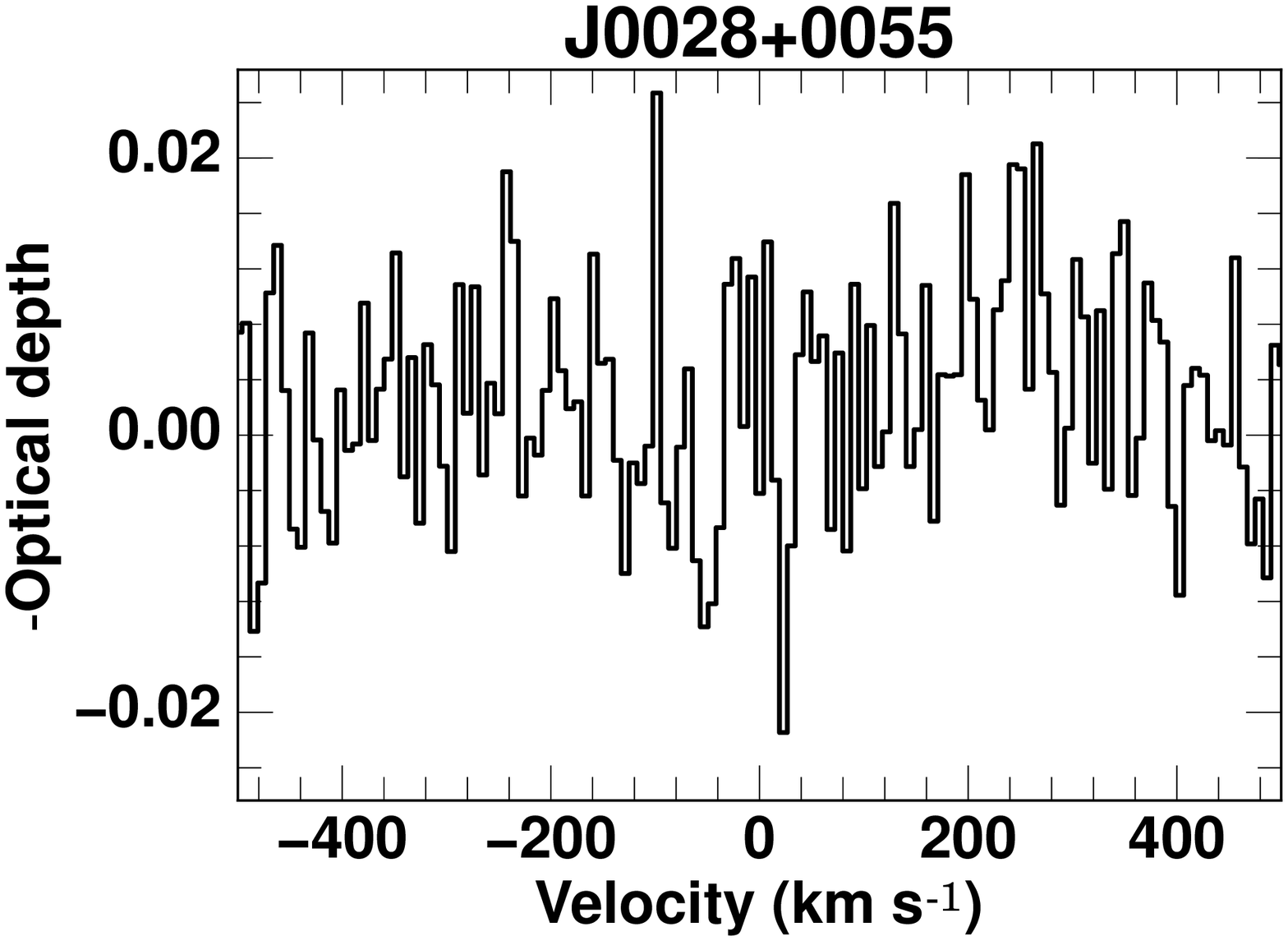}
			\includegraphics[scale=0.3]{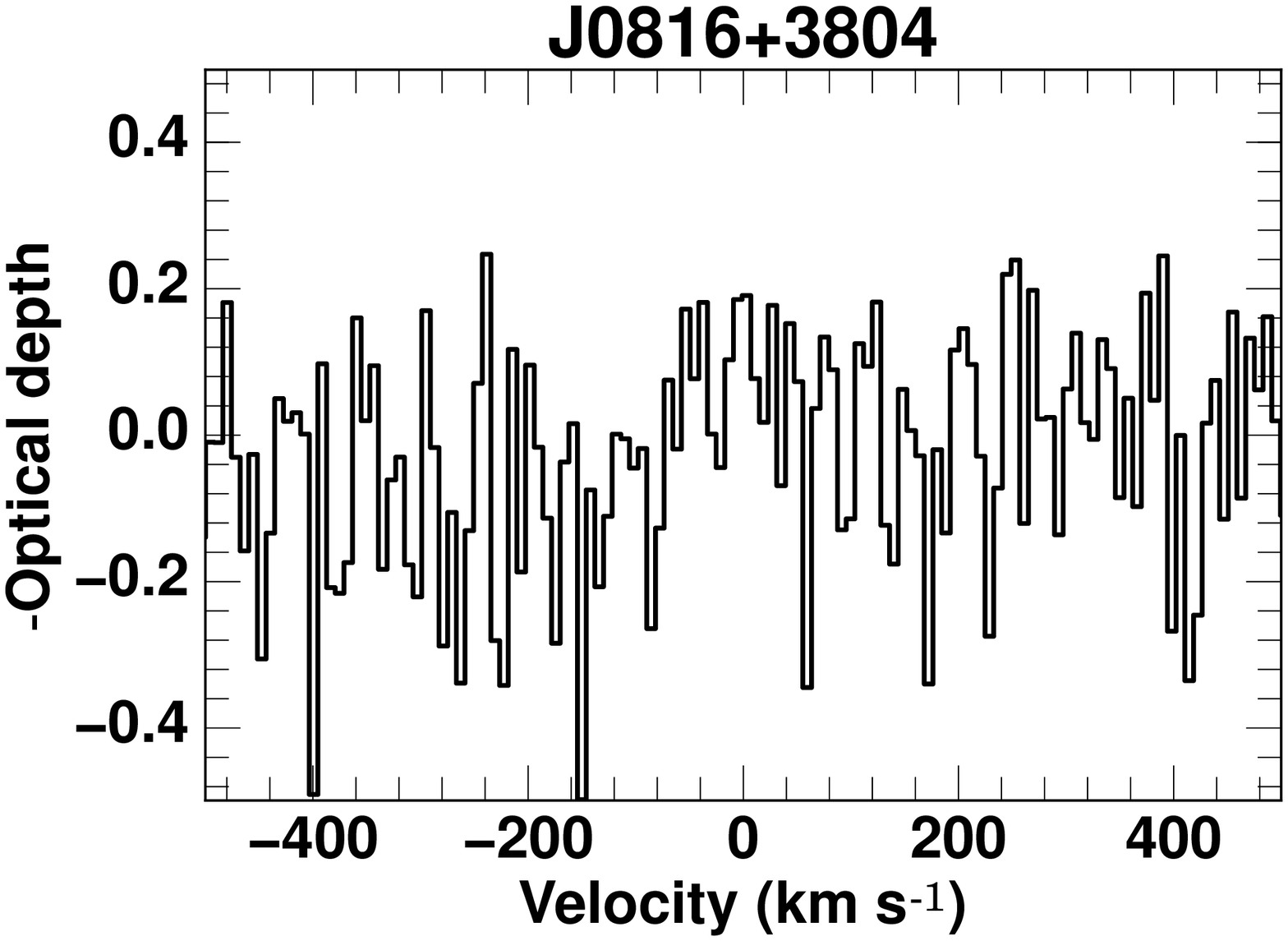}
			\includegraphics[scale=0.3]{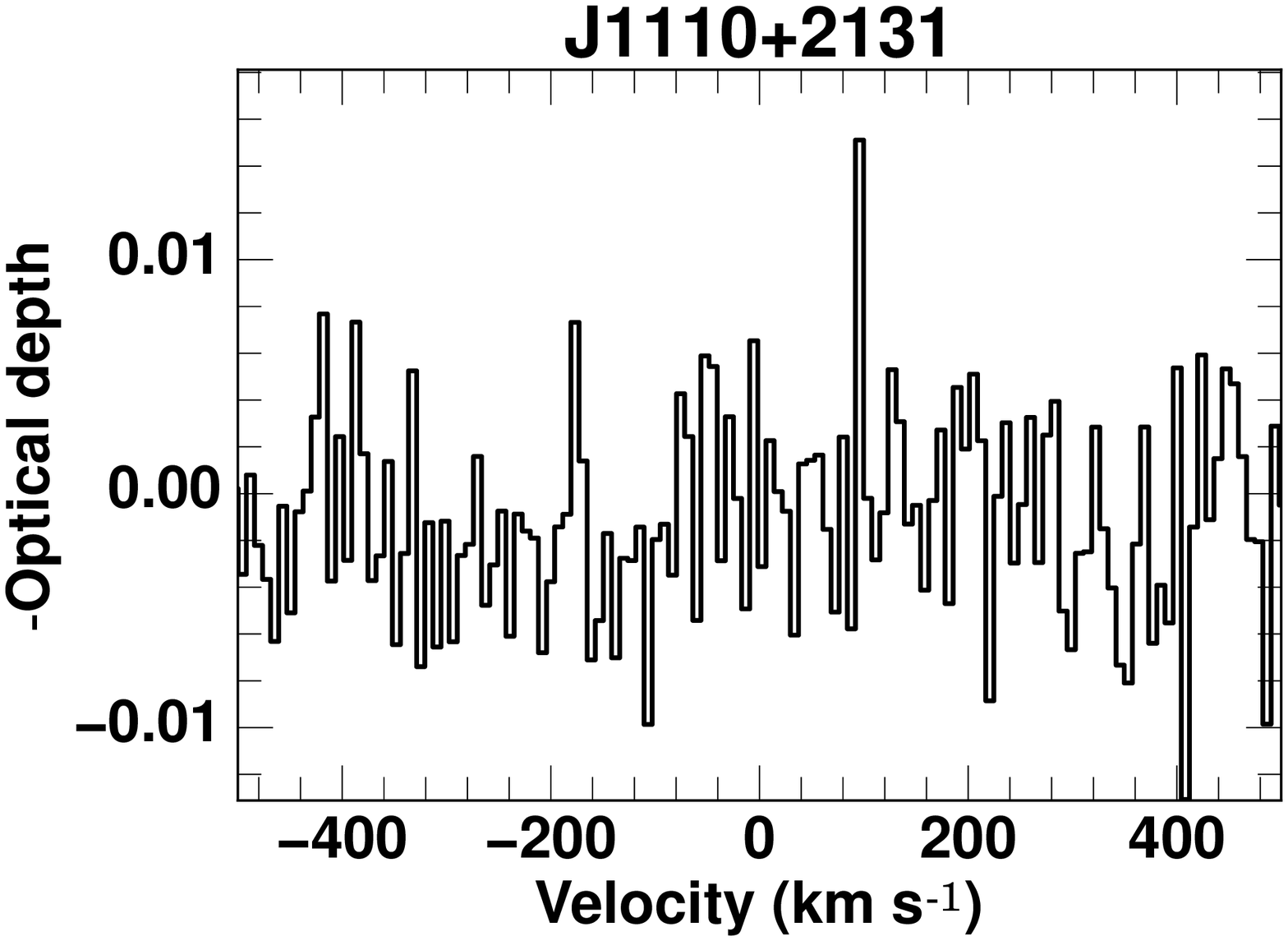}
		}
		
		\hbox{
			\includegraphics[scale=0.3]{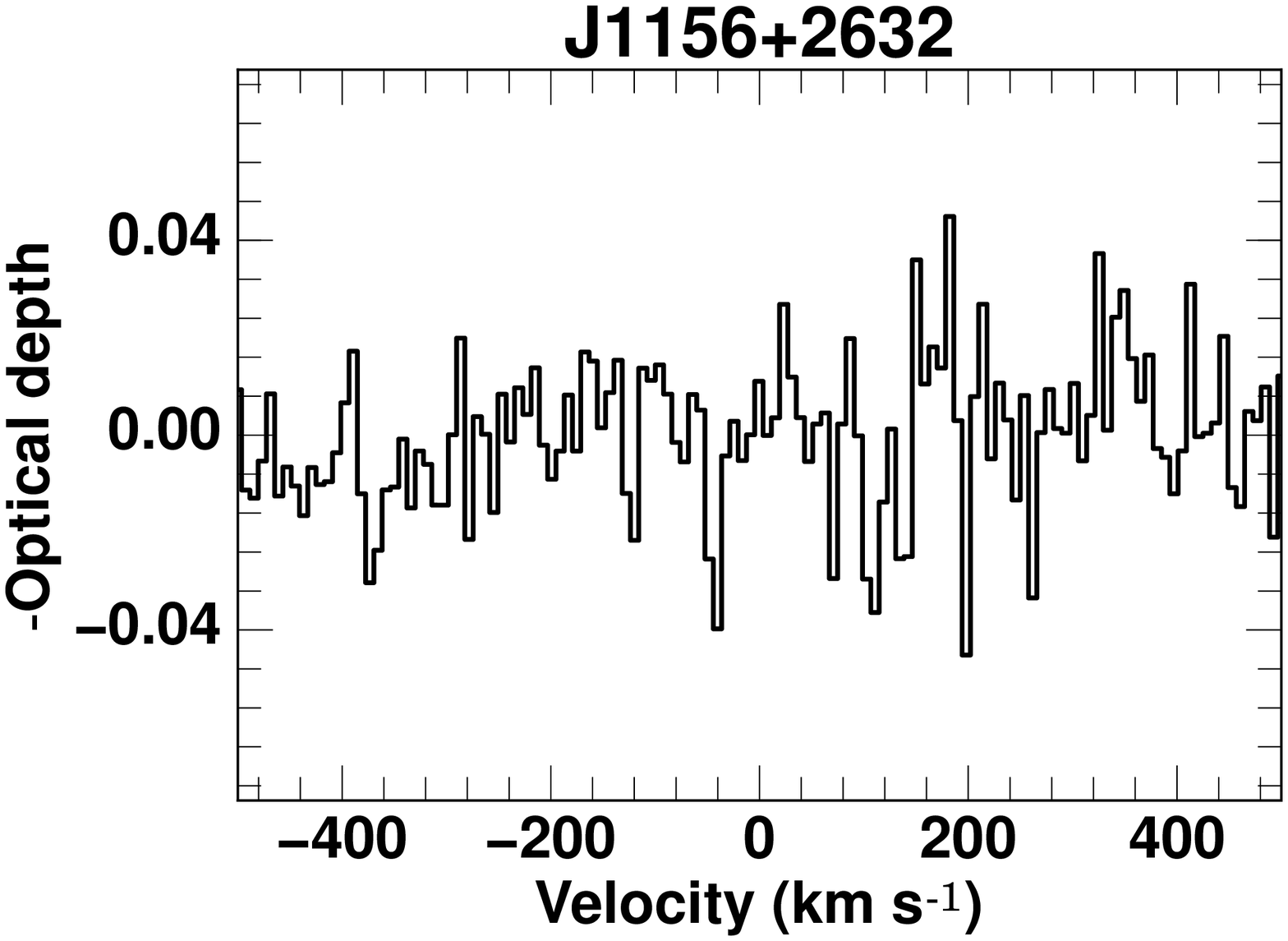}
			\includegraphics[scale=0.3]{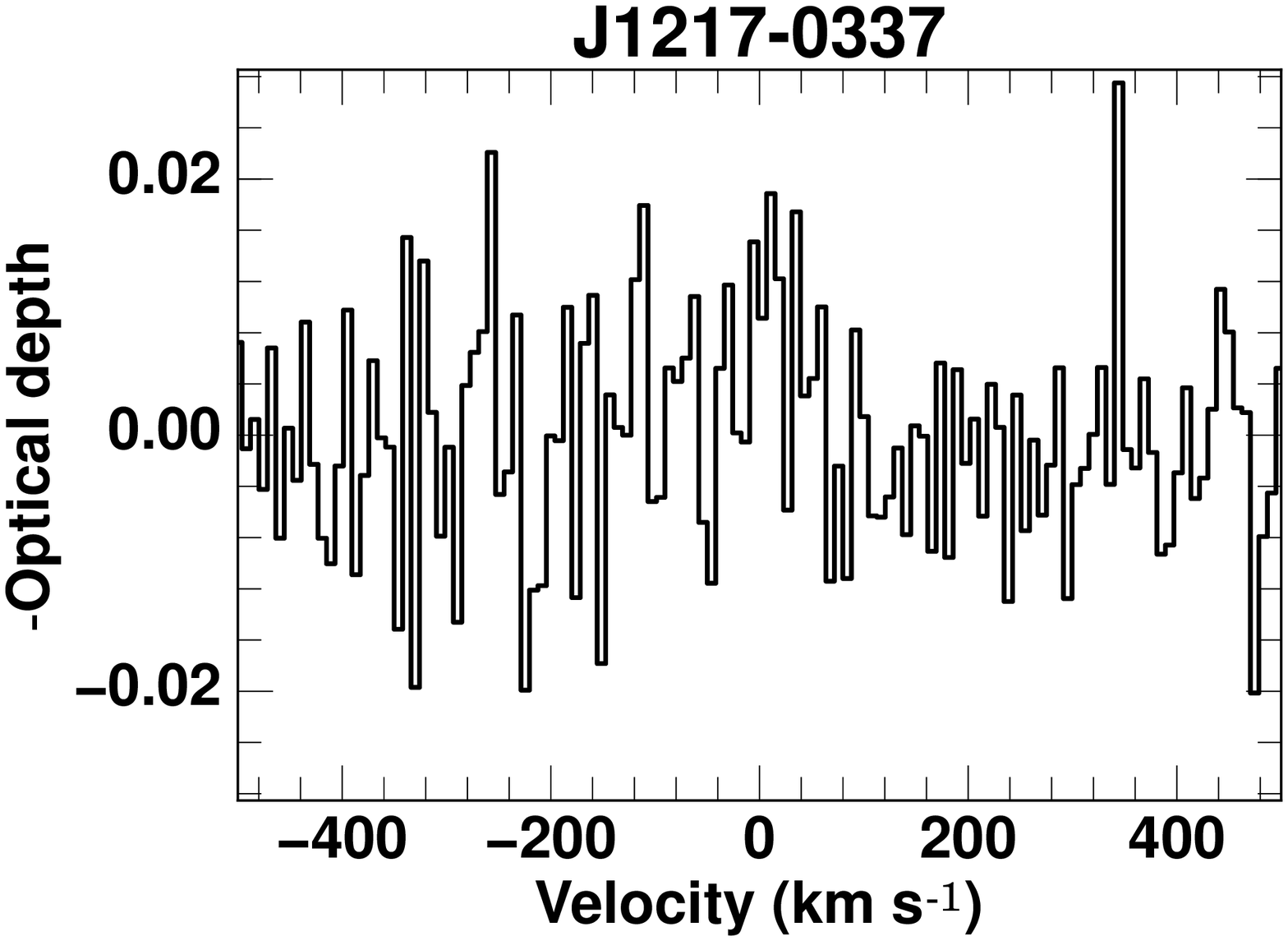}
			\includegraphics[scale=0.3]{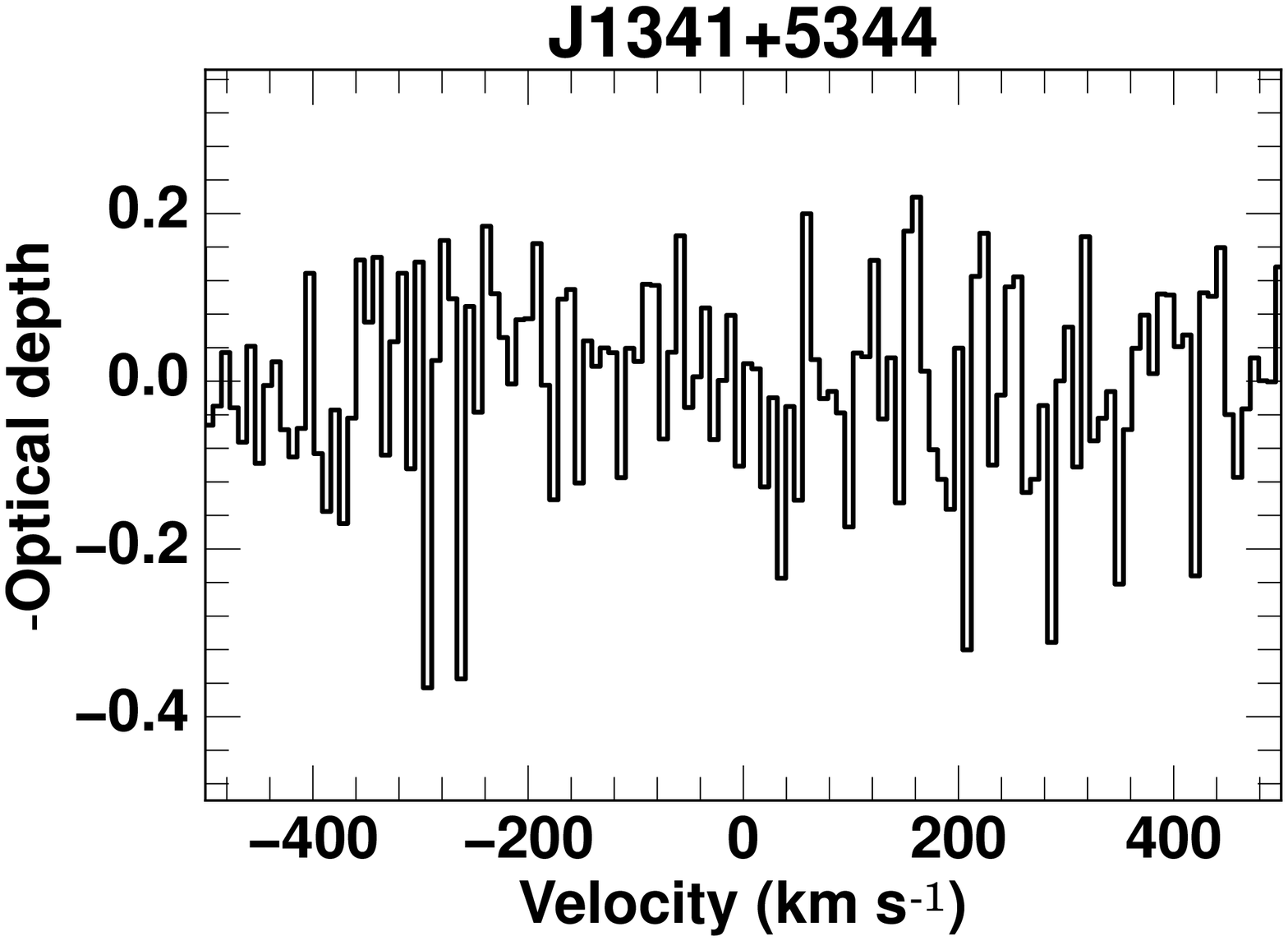}
		}
		\hbox{	
			\includegraphics[scale=0.3]{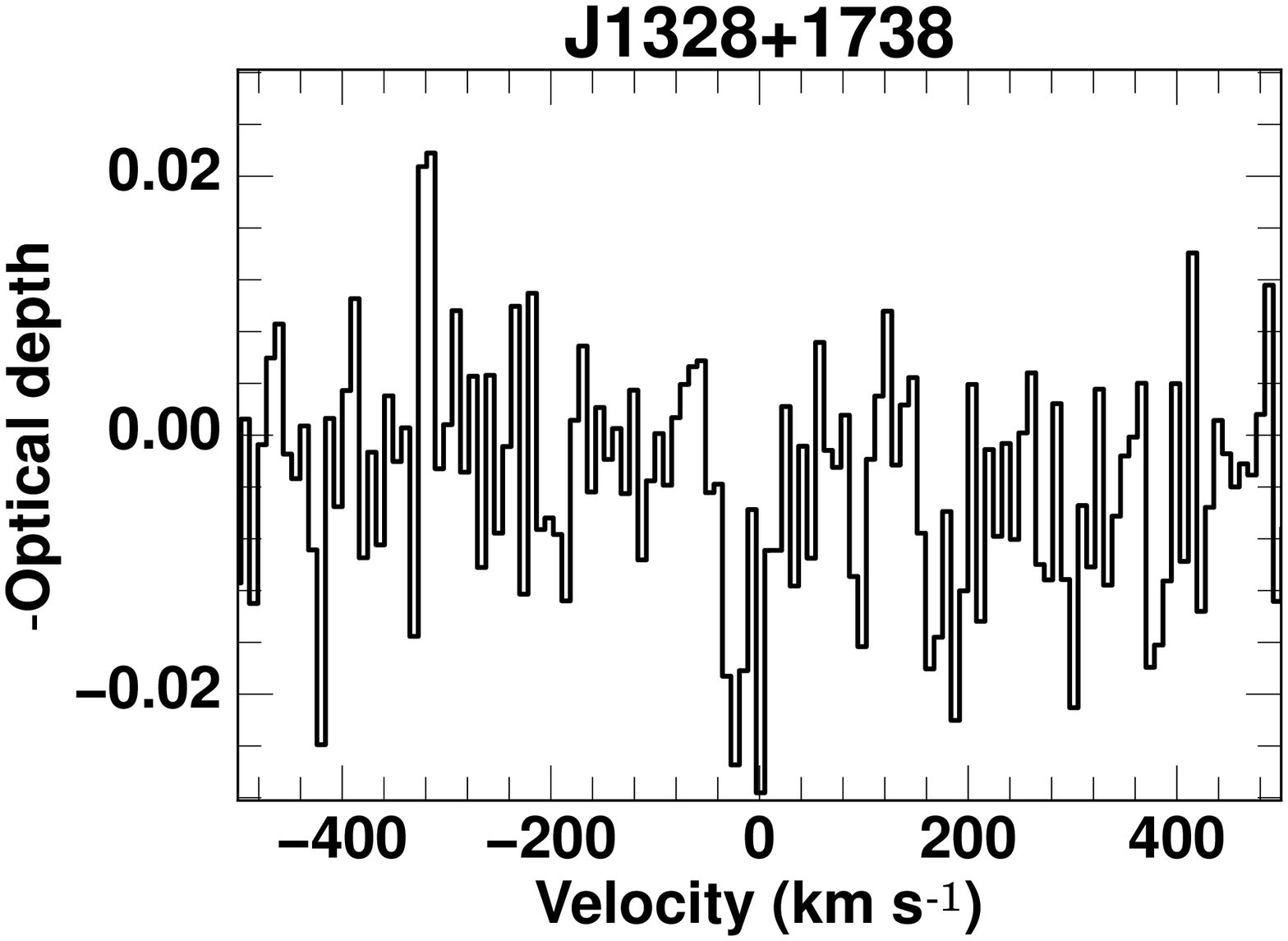}
			\includegraphics[scale=0.3]{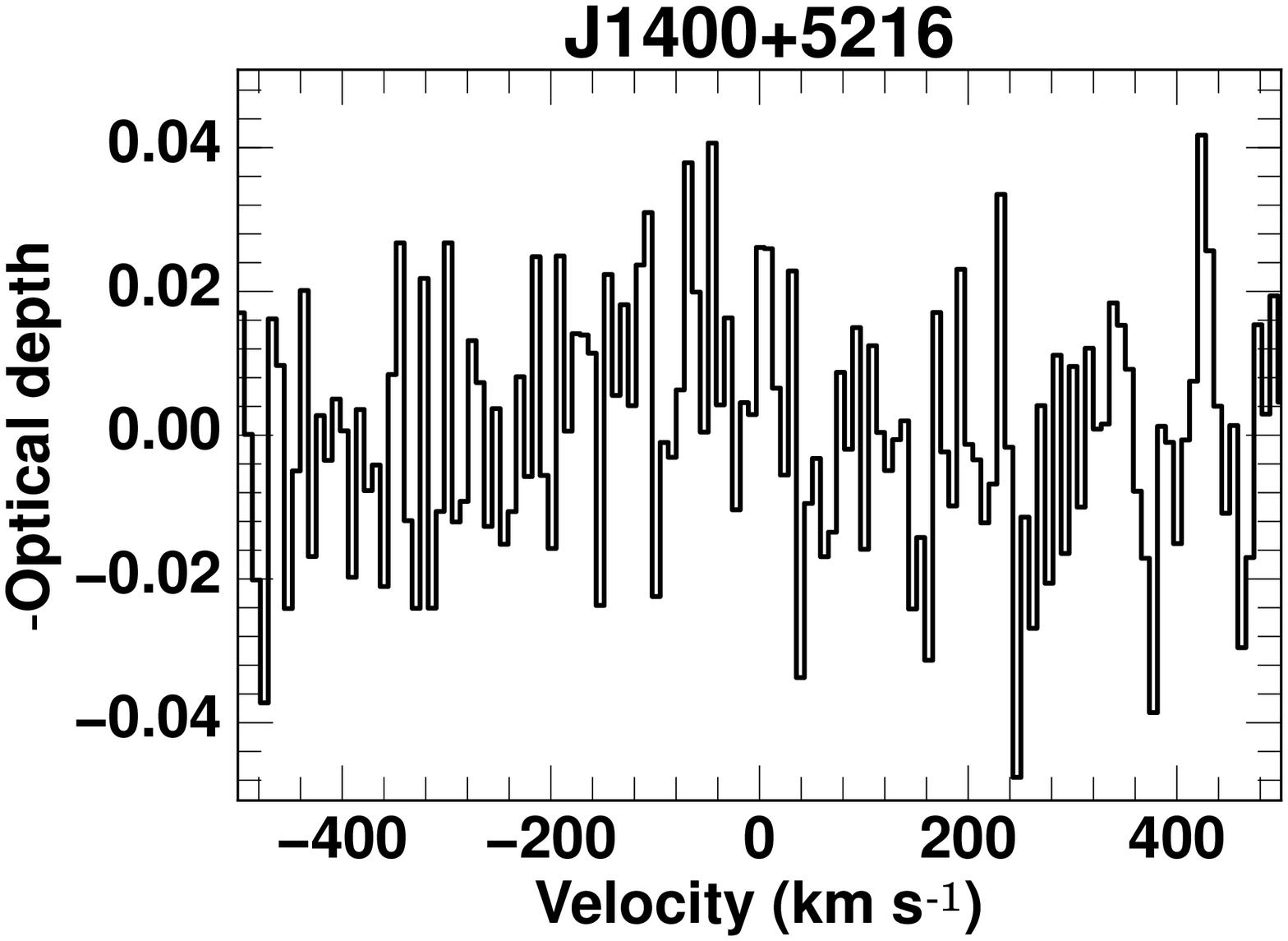}
			\includegraphics[scale=0.3]{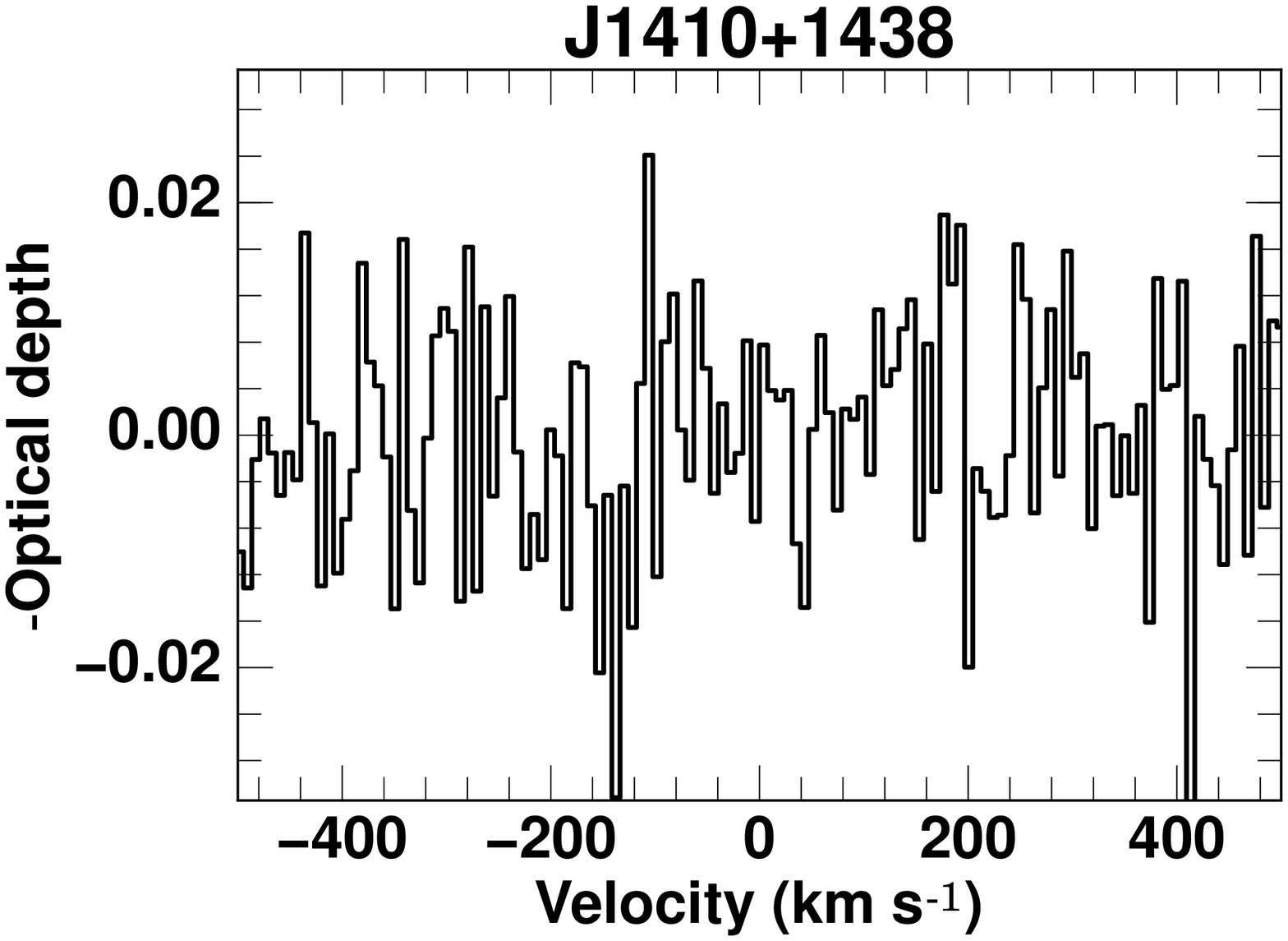}
		}
		
		\hbox{  
			\includegraphics[scale=0.3]{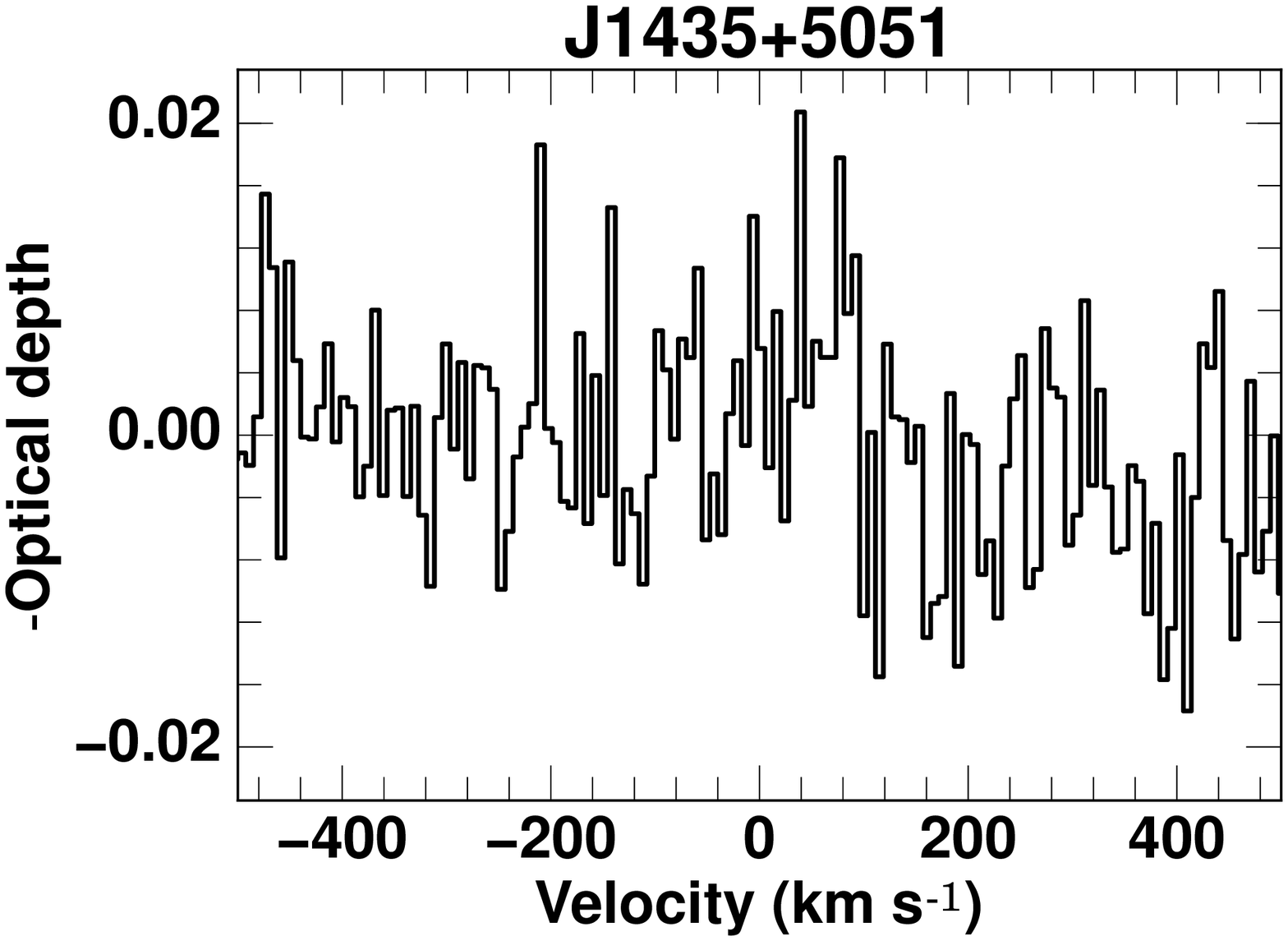}
			\includegraphics[scale=0.3]{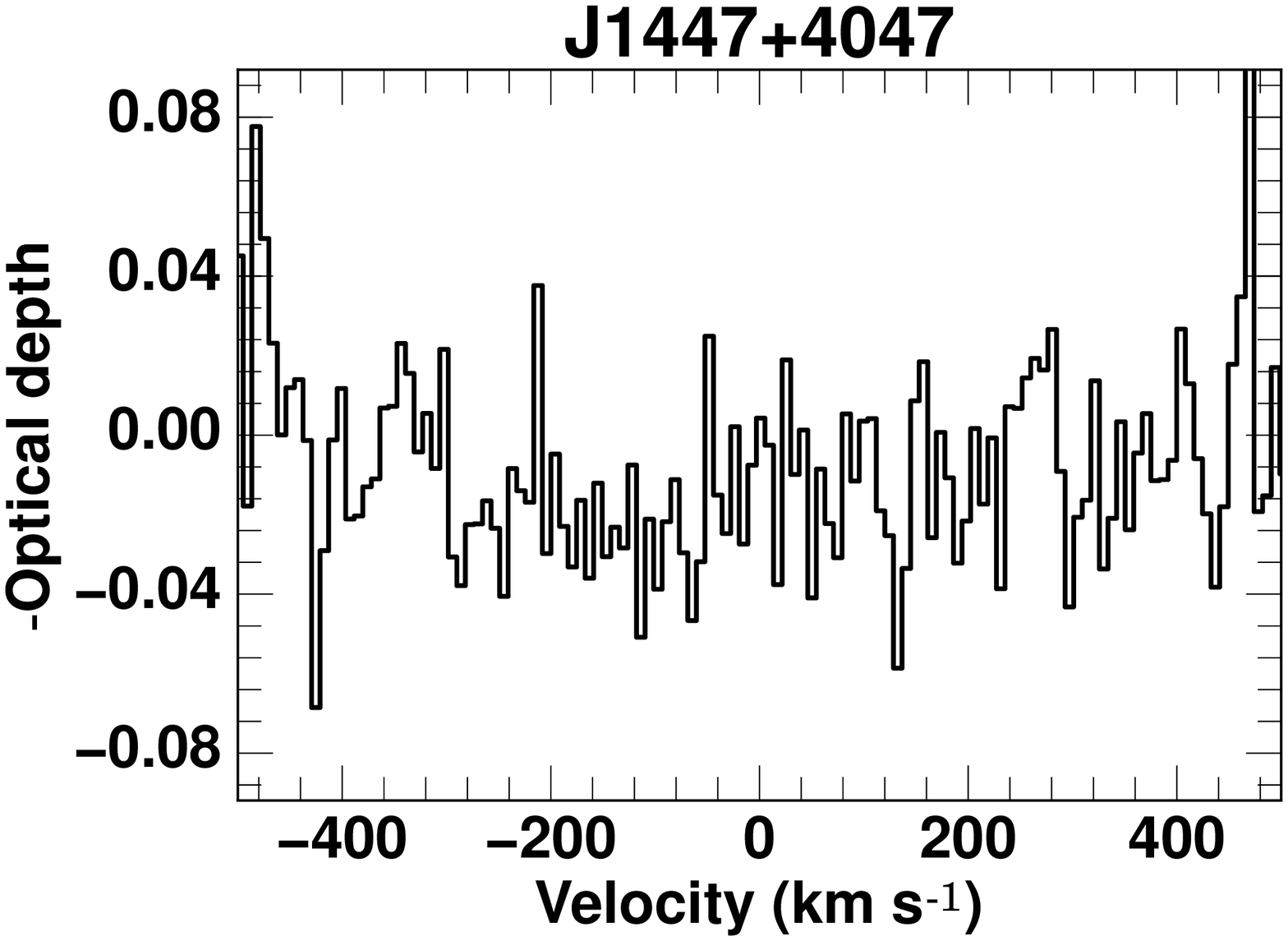}
			\includegraphics[scale=0.3]{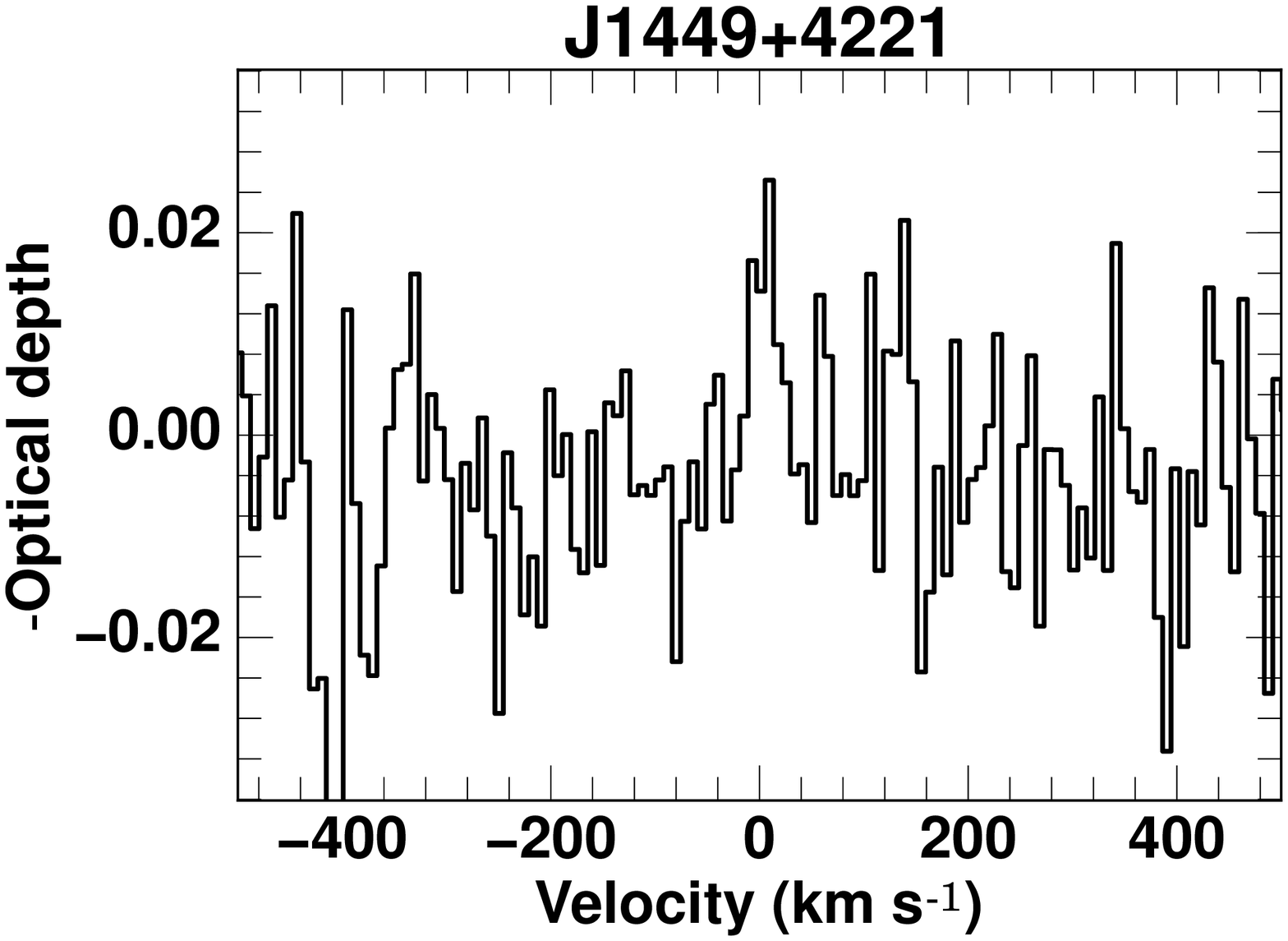}
		}
		
		\hbox{
			\includegraphics[scale=0.3]{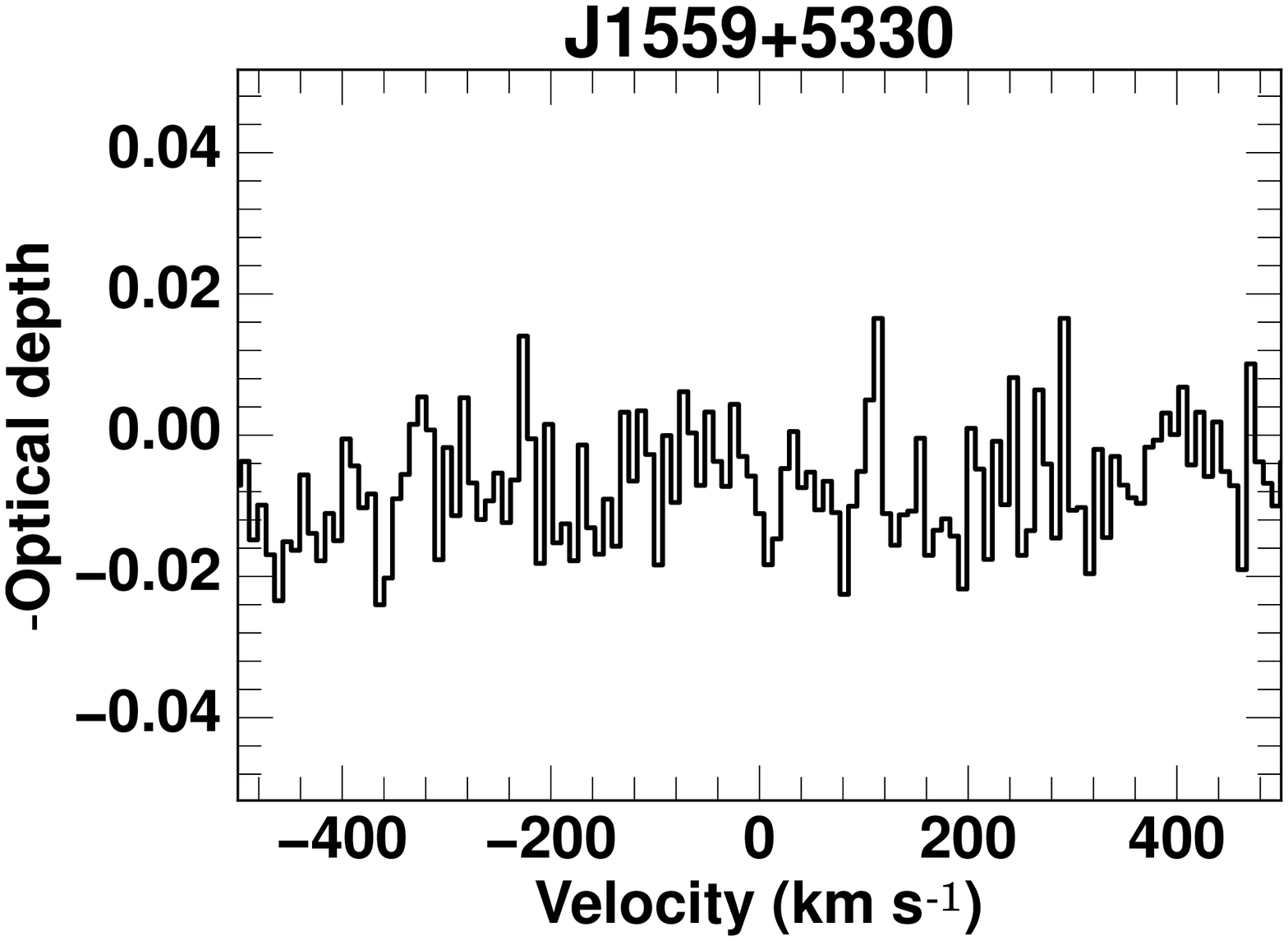}
		}	
		\caption{H{\sc i} absorption profiles towards brightest pixel in sources not  detected with H{\sc i} absorption. X-axes show velocity shift w.r.t. optical systemic velocity and Y-axes show optical depth. Zero represent the optical systemic velocity corresponding to optical redshift.}
		\label{fig3}
	\end{center}
\end{figure*}  

\begin{figure*}
	\begin{center}
		
		\hbox{ 
			\includegraphics[scale=0.3]{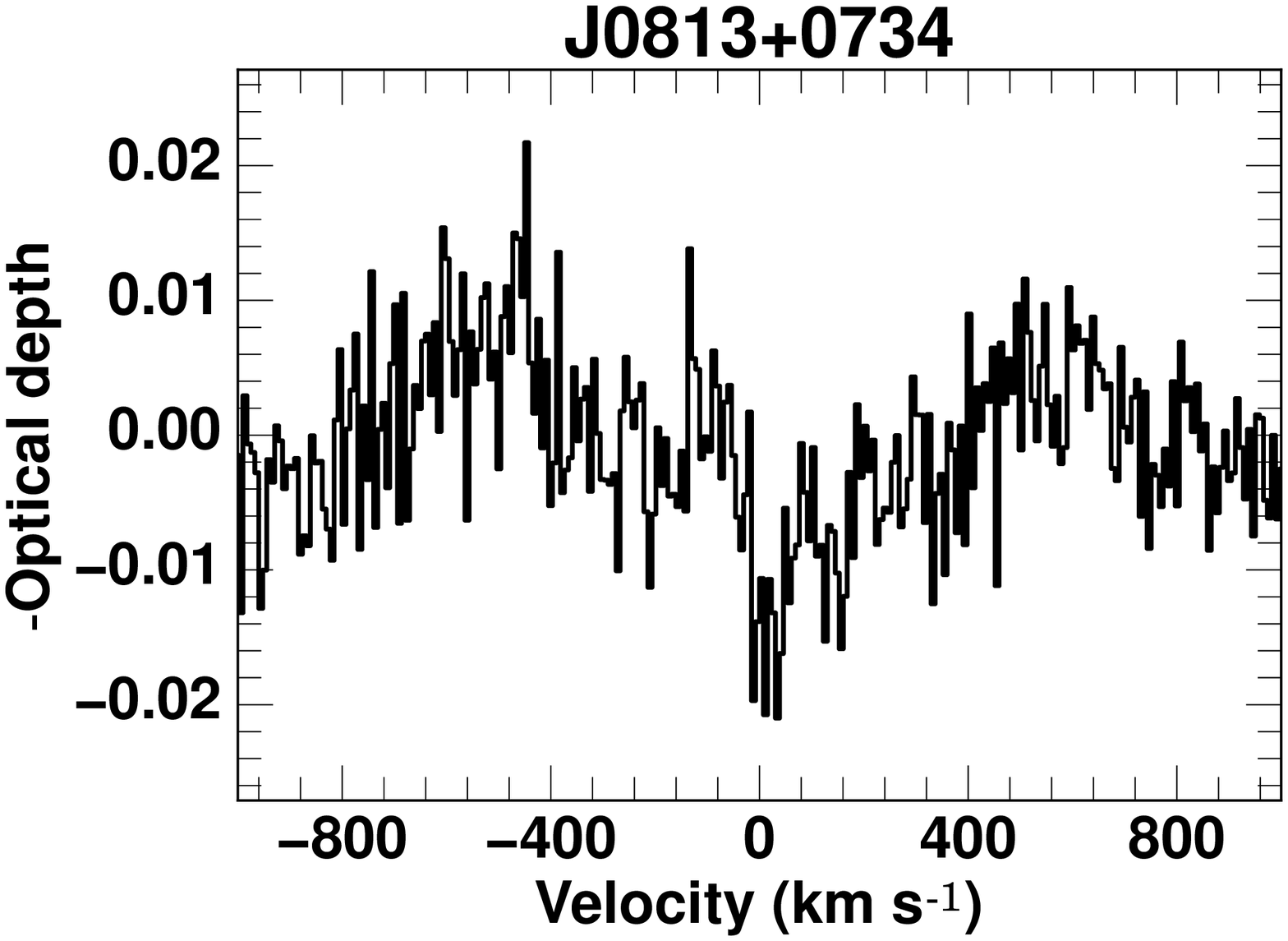}
			\includegraphics[scale=0.3]{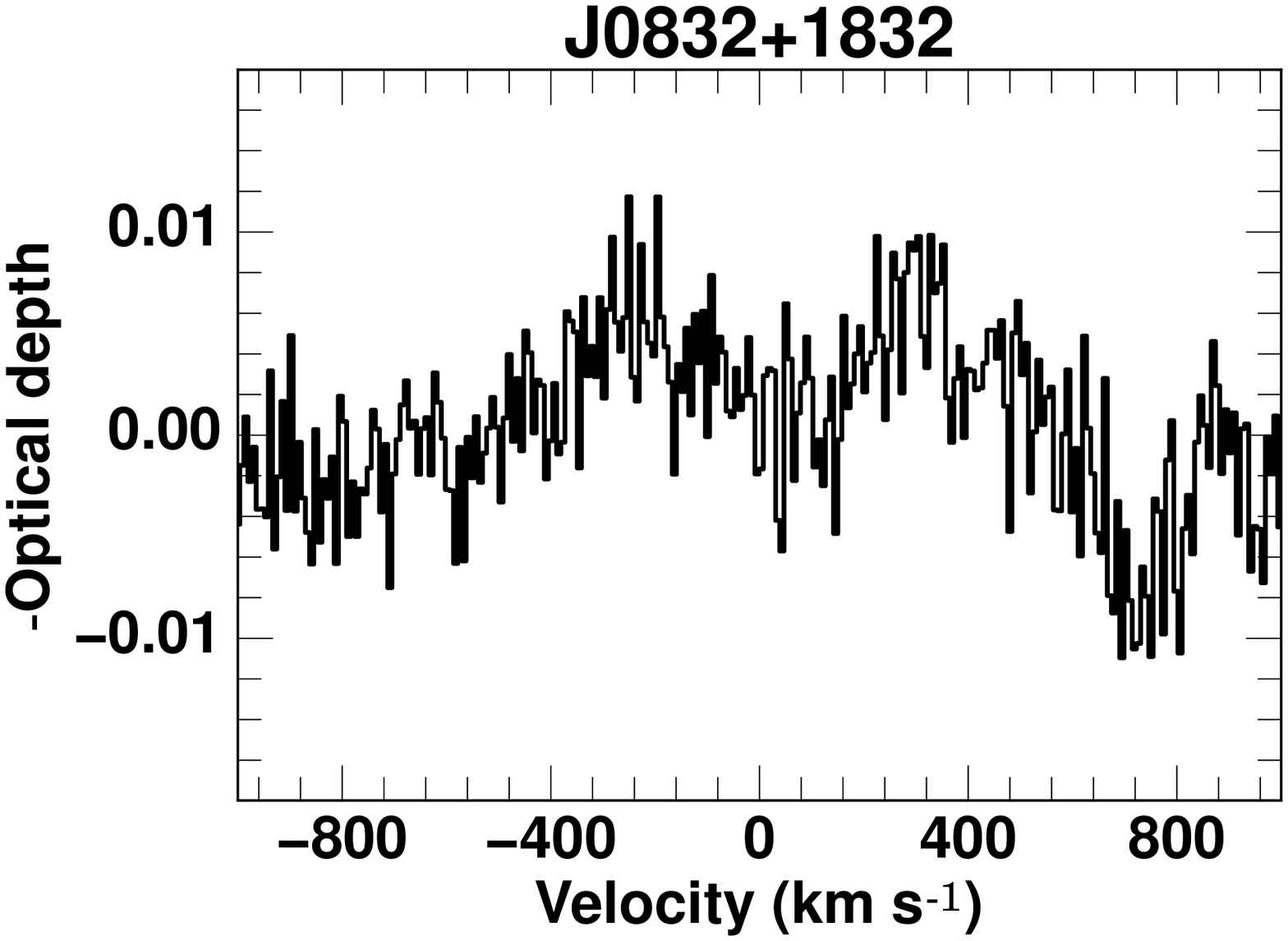}
			\includegraphics[scale=0.3]{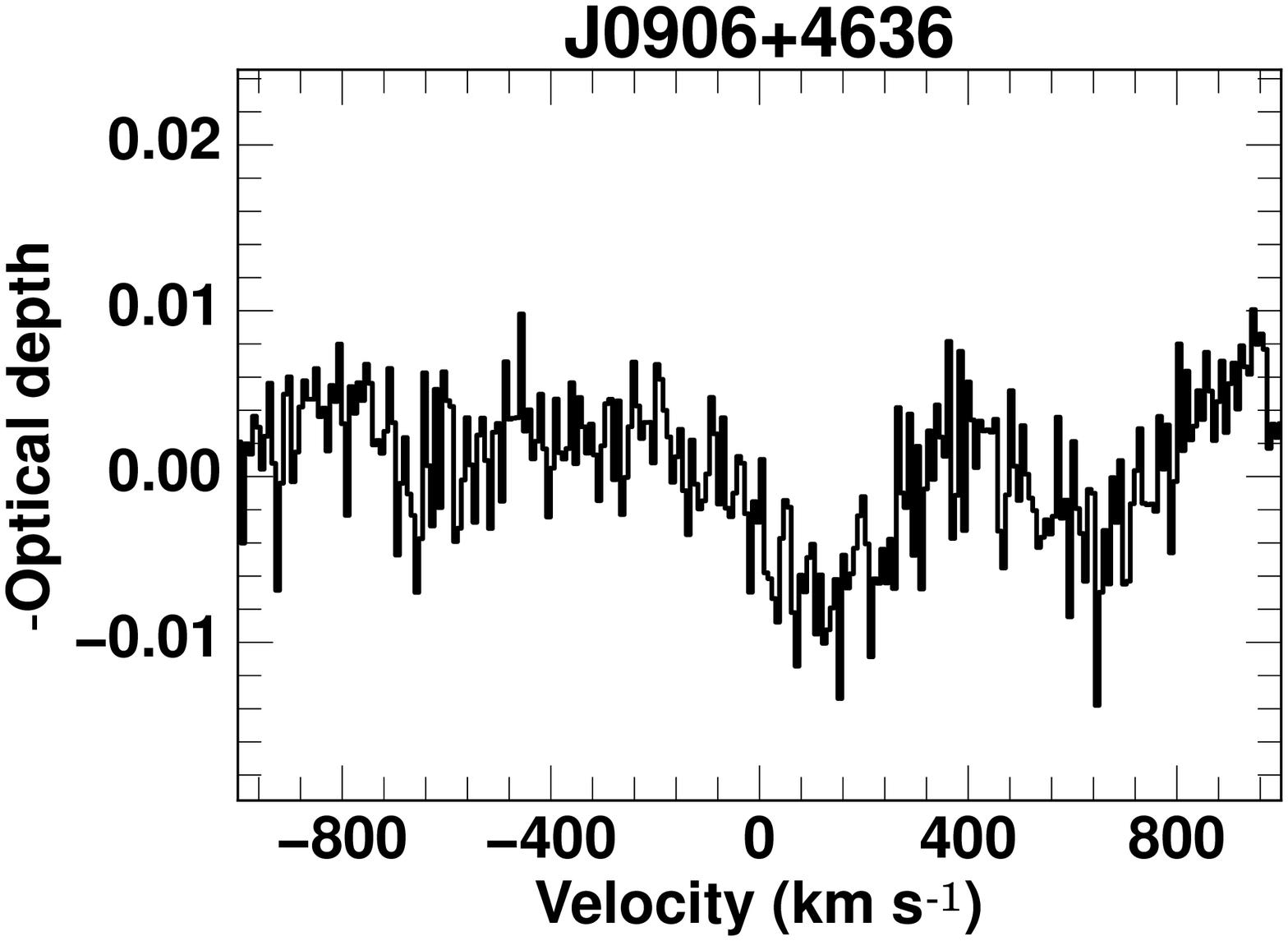}
		}
		
		\hbox{
			\includegraphics[scale=0.3]{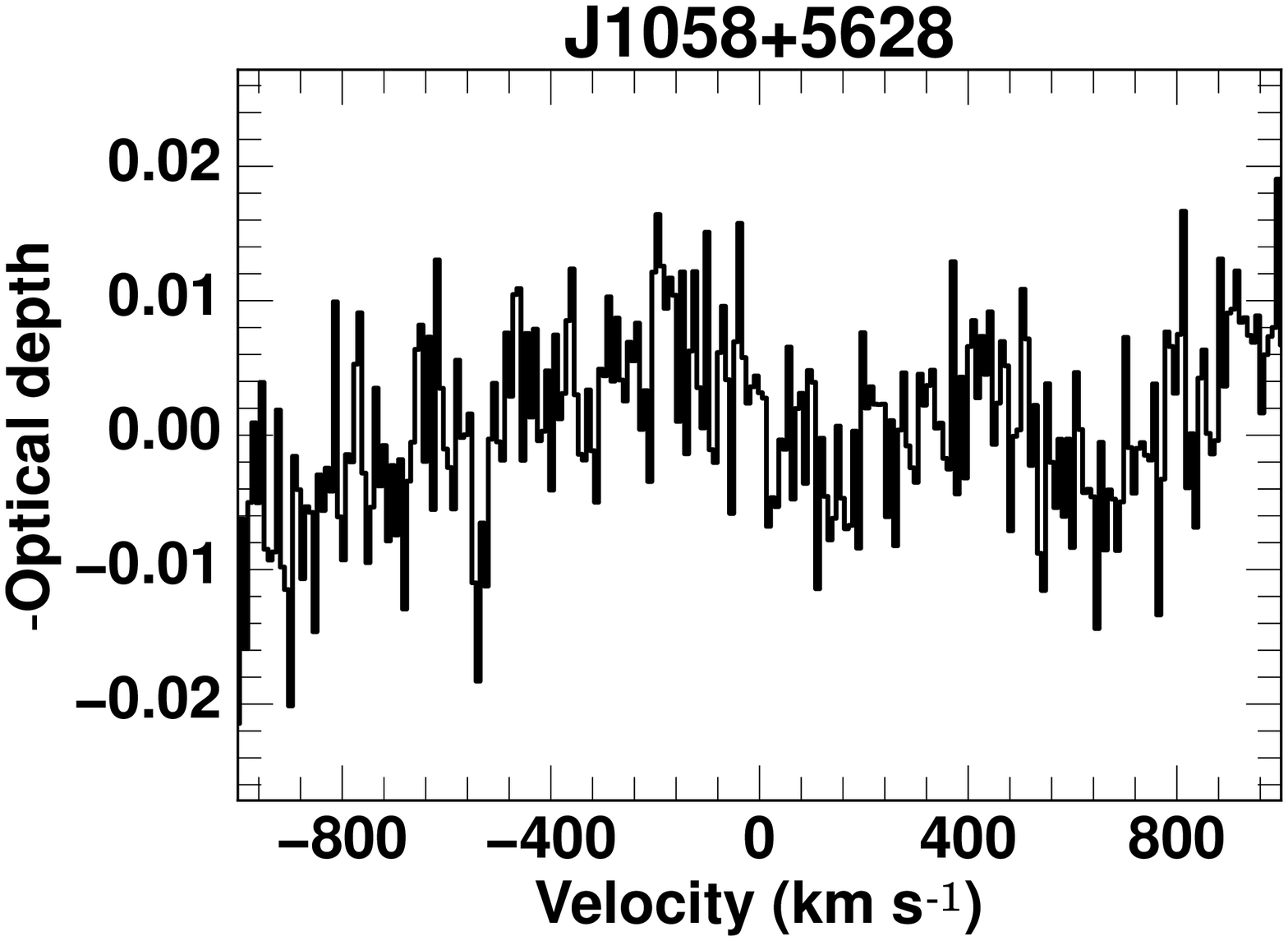}
			\includegraphics[scale=0.3]{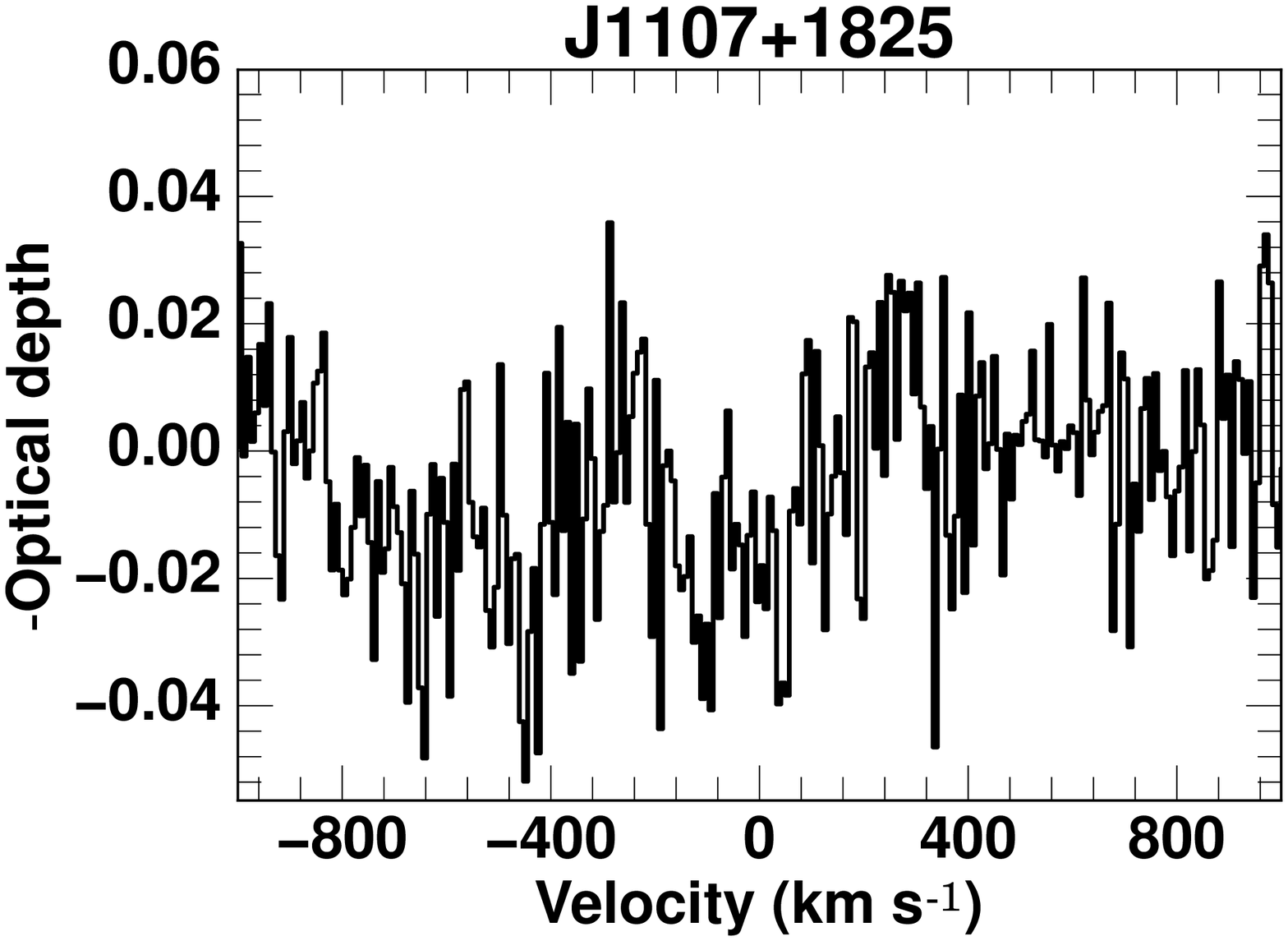}
			\includegraphics[scale=0.3]{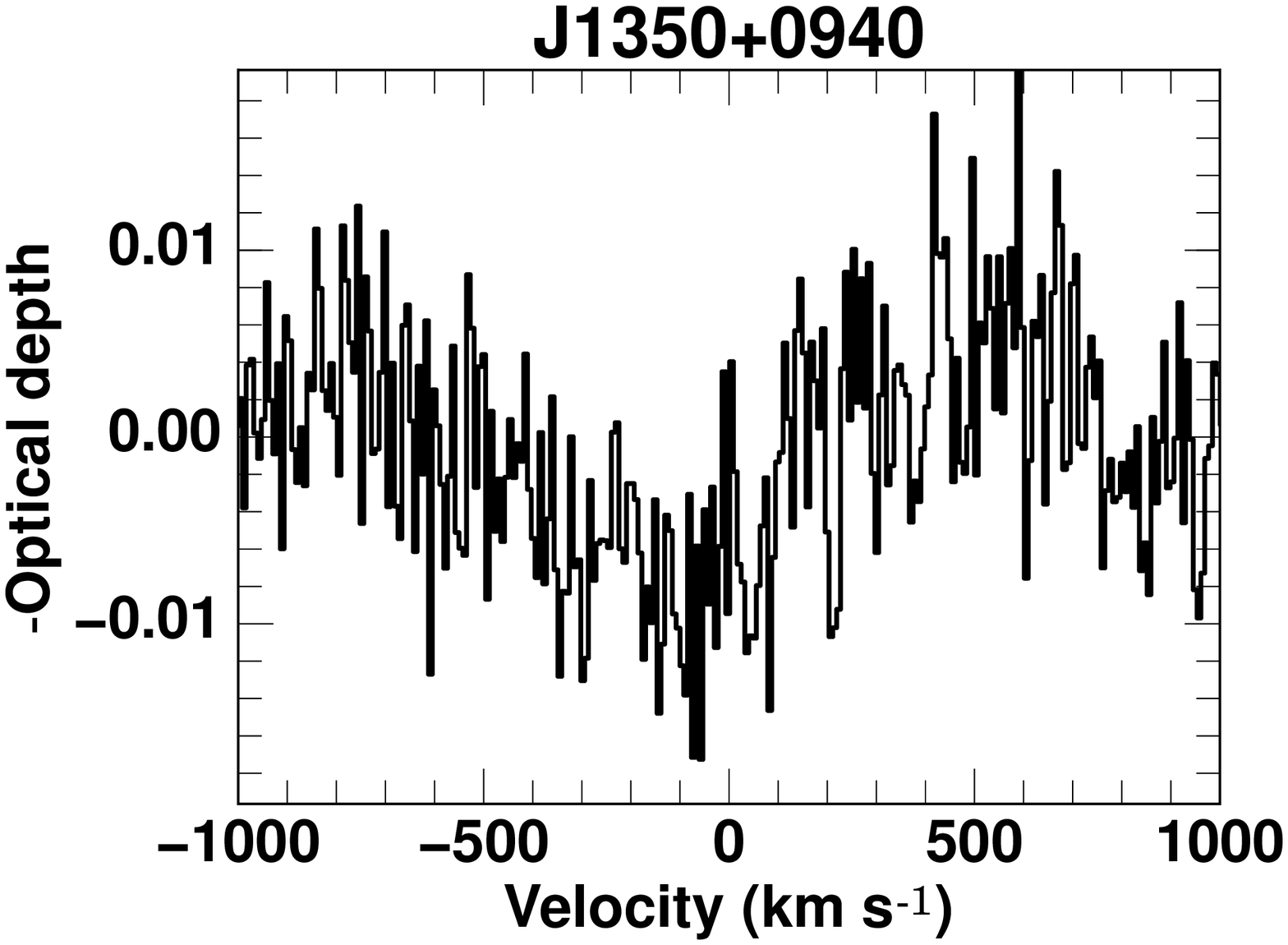}
		}	
		\hbox{
			\includegraphics[scale=0.3]{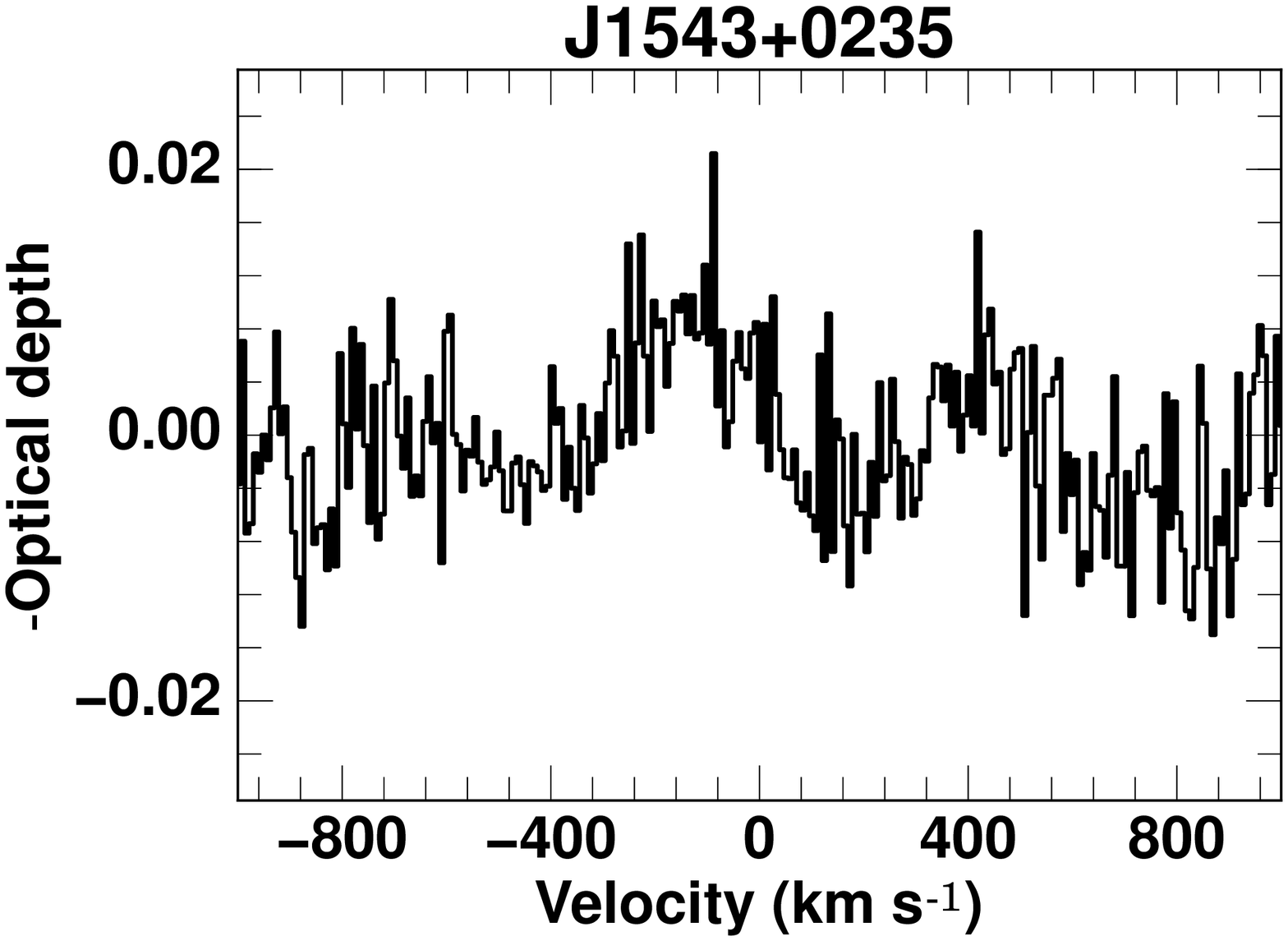}
		}	    
		
		\caption{H{\sc i} absorption profiles towards brightest pixel in sources affected with ripples. X-axes show velocity shift w.r.t. optical systemic velocity and Y-axes show optical depth. Zero represent the optical systemic velocity corresponding to optical redshift.}
		\label{fig3a}
	\end{center}
\end{figure*}  
\begin{figure}
	\centering
	\includegraphics[scale=0.45]{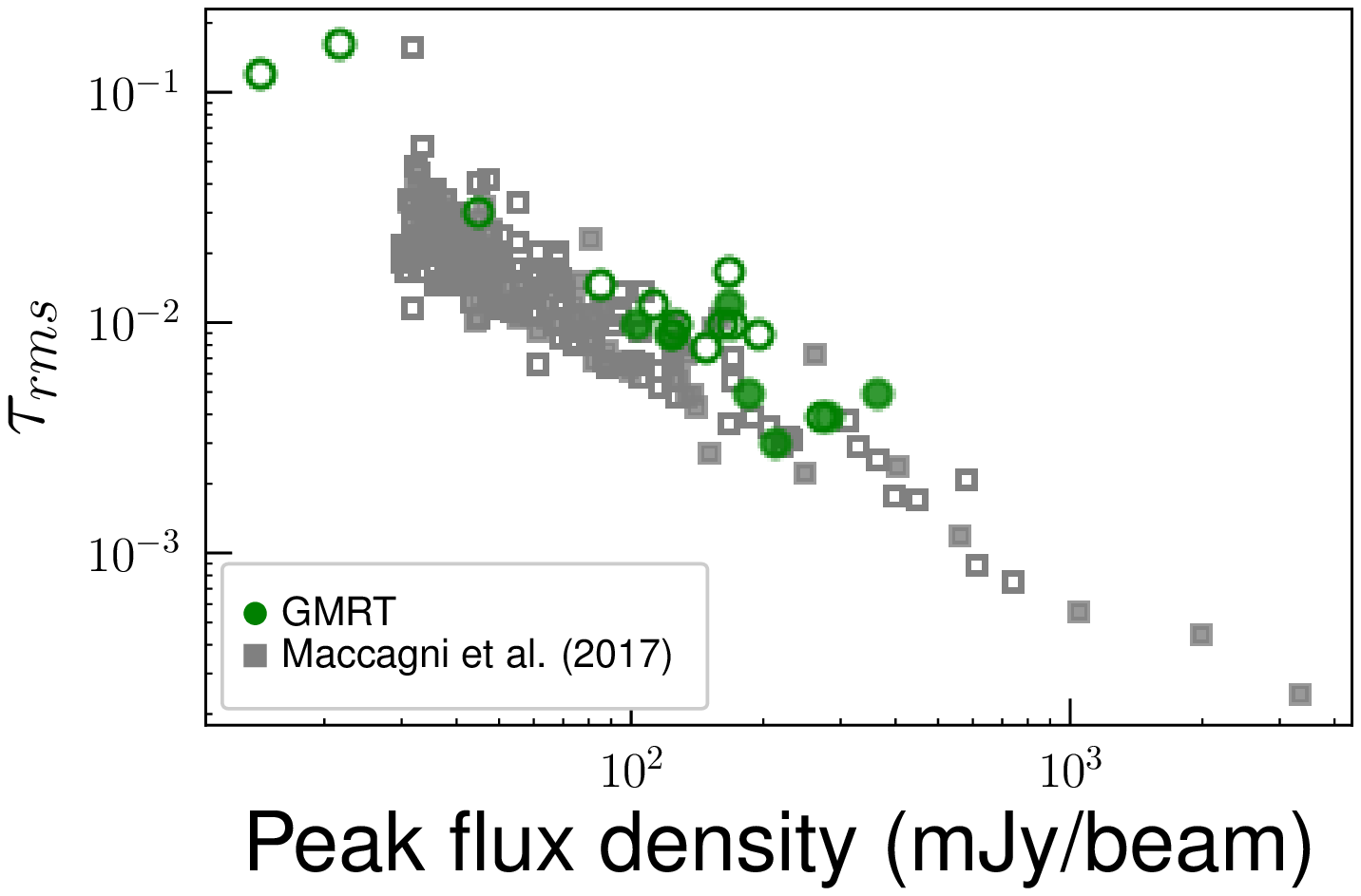}
	\includegraphics[scale=0.45]{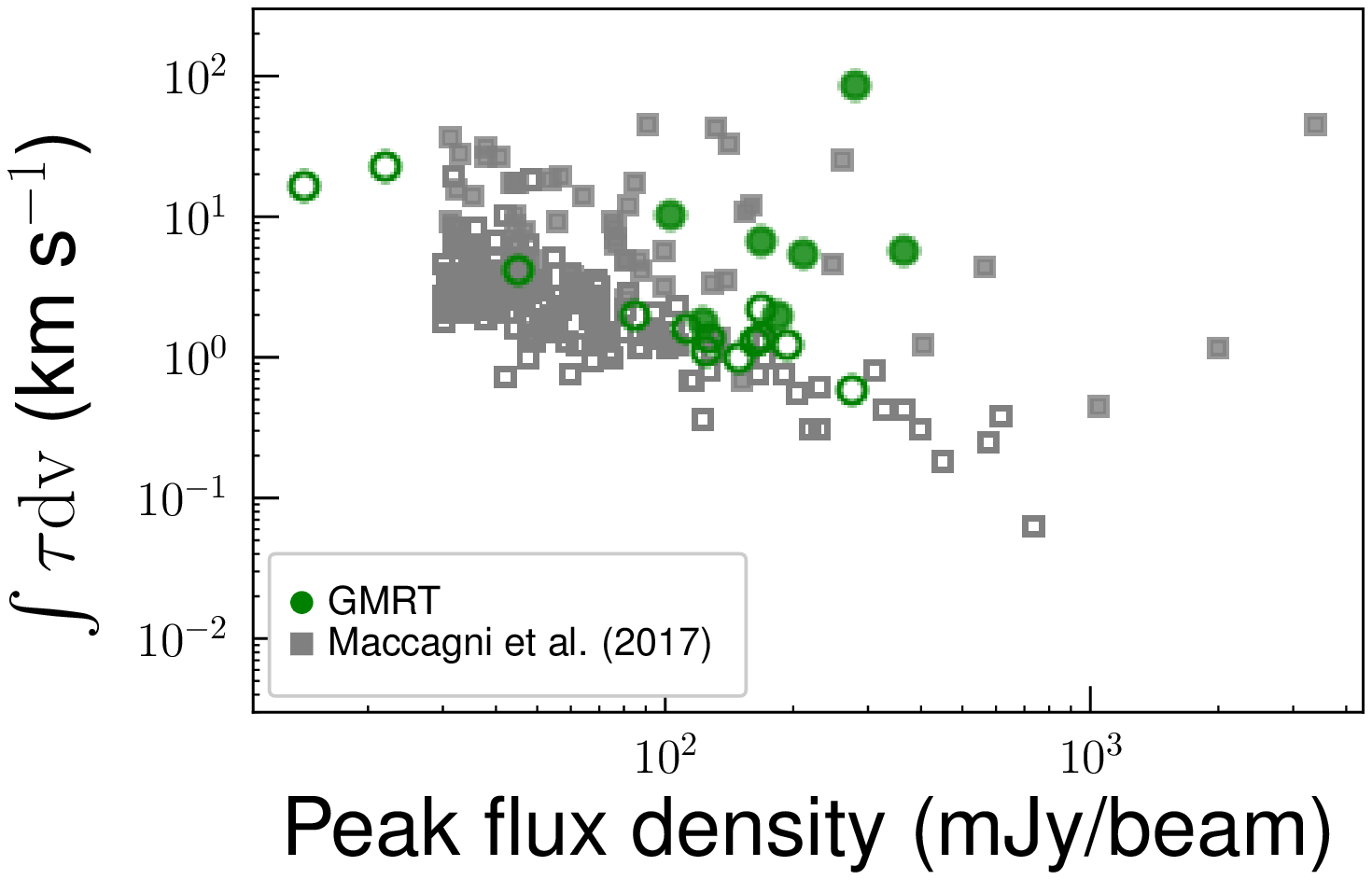}
	
	\caption{\textbf{Top:} Optical depth r.m.s vs peak flux density. 20 sources from the GMRT observations with good spectra are shown with green circles. Grey coloured squares depict 219 sources from \protect\cite{2017A&A...604A..43M}. Detections are shown with filled symbols and non-detection with empty symbols. \textbf{Bottom:} Integrated optical depth vs peak flux density. Empty symbols are 3-$\sigma$ upper limits on integrated optical depth while filled symbols are estimated values from detections.}
	\label{fig4}
\end{figure}
\noindent
   GMRT observations do not find a detection from two independent observations, while \cite{2017A&A...604A..43M} report a detection from a somewhat noisier spectrum compared with most of their sources. Therefore while considering the \cite{2017A&A...604A..43M} and GMRT samples separately we have adopted the results for these two sources as reported by the authors in the respective samples, but have left them out from the combined sample. A further set of observations is required to clarify the discrepancy in the two sources. This does not at all affect the results presented in this paper. Hence, for H{\sc i} analysis, after excluding these two sources and 6 other sources affected by radio frequency interference (RFI) or baseline ripples in our GMRT sample, we have 233 radio sources including 191 LERGs and 27 HERGs for combined sample. The remaining 15 are unclassified as mentioned earlier. This sample henceforth be referred to as the \lq combined sample \rq. In this sample, we have around two and a half times the number of LERGs (80) and  HERGs (11) compared to our previous work \citep{2017MNRAS.465..997C}, which help us to understand the H{\sc i} absorption properties in both types of AGNs better. However, while giving H{\sc i} statistics separately for \cite{2017A&A...604A..43M} and GMRT observed samples, we use 219 sources from their sample, and all 20 sources with good spectra from the GMRT observed sample. 

  \section{Observation and data reduction} 
 H{\sc i} 21cm absorption experiments were done towards these sources with the GMRT during May-December 2016. During these observations, the GMRT full array was used in total intensity mode and base-band bandwidth of 16 MHz divided into 512 channels (except for J2133-0712 where 256 channels were used). This resulted in velocity spectral resolution of $\sim$8 km s$^{-1}$, except for J2133-0712 where it is $\sim$15 km s$^{-1}$. Observational details for these sources are listed in Table~\ref{GMRTobs}. After every 40 minutes of observing time on the target source, phase calibrator was observed for 5 minutes. Flux density or bandpass calibrators were observed for 10-15 minutes after every 3-4 hours of observations. 
 
 Data were reduced mainly using the NRAO AIPS package. Initially, data reduction was done for a single channel. Bad data due to RFI and bad antennas were flagged before any calibration using TVFLG, UVFLG. Phase and gain calibration solutions and bandpass calibration solutions were determined using the task \textsc{CALIB} and the task \textsc{BPASS} respectively. Bad data were iteratively flagged until satisfactory gain solutions were obtained. Then the target source data were split from the multi-source data cube after applying the bandpass and gain solutions. The target source data were examined and bad data flagged before imaging using the task \textsc{IMAGR}. After a few rounds of self-calibration, RFI was removed from the spectral line \textit{uv} datacube using RFI removing software \textsc{AOFLAGGER} \citep{2010ascl.soft10017O}, except for extended sources where it was done manually using \textsc{AIPS} task \textsc{SPFLG}. Next, the final continuum images were made from line free channels. The continuum data were then subtracted from the spectral line \textit{uv} datacube using tasks \textsc{UVSUB} and \textsc{UVLIN}. Task \textsc{CVEL} was used to correct for the motion of the Earth's rotation. Finally, the H{\sc i} absorption spectra were extracted from the H{\sc i} dirty image cube against the pixels corresponding to the peak flux densities towards the central regions of target source images.
  
  \section{Results}
  \label{sec:results}
  In this Section, we report H{\sc i} absorption properties of our sample.  Of the 27 sources, we have obtained reliable spectra towards 20 sources.  Of these 20 sources,  we report 7 detections ($\sim$35 \% detection rate), of which 6 detections are new. The H{\sc i} absorption profiles and Gaussian fits to these are shown in Fig.~\ref{fig2}. Of the 7 detections, 2 H{\sc i} absorption profiles could be fitted with a single Gaussian component, 4 H{\sc i} absorption profiles with two Gaussian components and one, J1534+2330, with 6 Gaussian components. H{\sc i} absorption towards J1534+2330 or Arp 220 has been also reported in earlier studies \citep{1982ApJ...260...75M, 1987ApJ...322...88G, 2001ApJ...560..168M, 2014MNRAS.440..696A}. H{\sc i} absorption towards Arp 220 is due to two counterrotating H{\sc i} disks and bridge of gas connecting them \citep{2001ApJ...560..168M}. We have listed the Gaussian parameters from H{\sc i} absorption profiles and values of column densities in Table~\ref{results}. H{\sc i} profiles towards 13 sources with non-detection are shown in Fig.~\ref{fig3}. Of the 27 sources, 7 have been affected with ripples (Fig.~\ref{fig3a}) and are not included in our analysis. We have also written notes on individual sources in the Appendix.  While Gaussian models of H{\sc i} absorption profiles are used to parameterize the physical conditions and individual components in these absorbers, we also obtain Busy fit parameters  \citep{2014MNRAS.438.1176W}  to describe the average characteristics of the profiles.  The Busy fit parameters are listed in Table~\ref{busyfit}. For single component H{\sc i} absorption profiles, there is no significant difference in velocities for the Gaussian component and centroid of Busy function relative to optical systemic velocity. However, in case of multiple components, centroid velocity of Busy function relative to optical systemic velocity may miss extreme redshifted or blueshifted components. For example, in case of J1534+2330, the velocity at the centroid of Busy function  relative to optical redshift is $\sim -$78 km s$^{-1}$ but the Gaussian component with maximum blueshift has a velocity of $\sim -$326 km s$^{-1}$ relative to optical systemic velocity.  Hence for our analysis, in order to get an idea of extreme blueshifted velocity from Busy function, we  estimated the blueshifted velocity at FW20 relative to optical systemic velocity  as, $- V_{\rm FW20} = V_{\rm centroid} -$ FW20/2, where $V_{\rm centroid}$ and FW20 are the velocity at centroid relative to optical systemic velocity and full width at twenty per cent of maximum in km s$^{-1}$, respectively. We mark both, $V_{\rm centroid}$ and $-V_{\rm FW20}$ with green and blue dashed vertical lines respectively in Fig~\ref{fig2}. 
  
  Optical depths ($\tau$) have been estimated from the line-to-continuum ratio by using the equation
  \begin{equation}
  \tau = -ln(1-\frac{ \rm \Delta I}{f_c \rm I_c})\text{,}
  \end{equation}
  where $\Delta$I is the absorbed flux density, I$_c$ is the continuum flux density and $f_c$ is the fraction of the background source covered by absorbing gas. We have assumed  $f_c =$1 for our calculations. The r.m.s noise on $\tau$ as a function of the peak flux density for our GMRT sample and \cite{2017A&A...604A..43M} sample\footnote{ We used FIRST peak flux densities to estimate the optical depth for \cite{2017A&A...604A..43M} sample.} is shown in Fig~\ref{fig4} top panel. Since most of the sources in our sample have peak flux density $>$ 100 mJy and in their sample $<$ 100 mJy, the median value of $\tau$ r.m.s. achieved from our GMRT observation is 0.009 per channel as compared to 0.015 per channel for the WSRT sample.  However,  for a similar range of peak flux densities in both the samples (S$_{\rm peak} =$ 90-300 mJy/beam) where most (15/20) of our sources are located, our observations are slightly less sensitive (median $\tau_{rms} =$ 0.009) than those of \cite{2017A&A...604A..43M} (median $\tau_{rms}=$ 0.007). It is also to be noted that median velocity spectral resolution for our GMRT sample is $\sim$ 7.9 km s$^{-1}$ as compared to $\sim$16 km s$^{-1}$ in the  \cite{2017A&A...604A..43M} sample.
  
  \subsection{Integrated optical depth and column densities}
  Integrated optical depths are calculated using equation
  \begin{equation}
  \int{\tau dv} = 1.064\times \tau_{peak} \times \rm FWHM \text{,} 
  \end{equation}
  for a Gaussian profile. FWHM and $\tau_{peak}$ are  full width at half maximum and peak optical depth, respectively obtained from the Gaussian fits. The error on integrated optical depth is estimated as
  $\tau_{rms} \times \delta v \times \sqrt{\text{FWZI}/\delta v}$, where $\delta v$, $\tau_{rms}$ and FWZI are  velocity resolution in km s$^{-1}$, rms noise on optical depth and full width at zero intensity, respectively.
  We have assumed FWZI for Gaussian profiles to be 6$\sigma$ i.e. 2.547$\times$ FWHM.  We also estimated \textbf{3$\sigma$} upper limits for the non-detections by assuming a Gaussian profile with FWHM $=$ 100 km s$^{-1}$  and using 
  
  \begin{equation}
  \int{\tau dv} =3 \times \tau_{rms} \times \delta v \times \sqrt{2.547\hspace{1mm} \times \text{FWHM}/\delta v}.
  \end{equation}
  
  For a range of similar peak flux densities (90-300 mJy/beam) in both samples, median integrated optical depth estimated for detections is 7.0 km s$^{-1}$, and for the non-detections the median value for the 3$\sigma$ upper limit is 1.3 km s$^{-1}$ for the GMRT sample. The corresponding values for the \cite{2017A&A...604A..43M} sample are 6.0 km s$^{-1}$ and 1.2 km s$^{-1}$ respectively (Fig~\ref{fig4} bottom). 
  
  We also estimated the H{\sc i} column densities and upper limits  using equations (2) and (3) in the equation \citep{1975ApJ...200..548W}, 
  \begin{equation}
  N (\text{H{\sc i}}) = 1.823 \times 10^{18} T_s \int \tau dv \text{,}        
  \end{equation}
  where $T_{s}$ is the spin temperature. We assume $T_{s} =$ 100 K. The median column density for  H{\sc i} absorption detection from GMRT sample is estimated to be 10.6 $\times$10$^{20}$ cm$^{-2}$. The median 3$\sigma$ upper limit for H{\sc i} absorption non-detections on column density is estimated to be 2.6 $\times$10$^{20}$ cm$^{-2}$. The corresponding values for the \cite{2017A&A...604A..43M} sample are 16.4 $\times$10$^{20}$ cm$^{-2}$ and 3.5 $\times$10$^{20}$ cm$^{-2}$, respectively.
 
   In the combined GMRT and \cite{2017A&A...604A..43M} sample, the median integrated optical depths for H{\sc i} detections are 8.7 km s$^{-1}$ and 5.7 km s$^{-1}$ for LERGs and HERGs respectively.  The corresponding median column densities are 15.9 $\times$10$^{20}$ cm$^{-2}$ and 10.4 $\times$10$^{20}$ cm$^{-2}$ for LERGs and HERGs respectively.
  
  \subsection{Detection rates}
\begin{figure}
        \centering
         \includegraphics[scale=0.35]{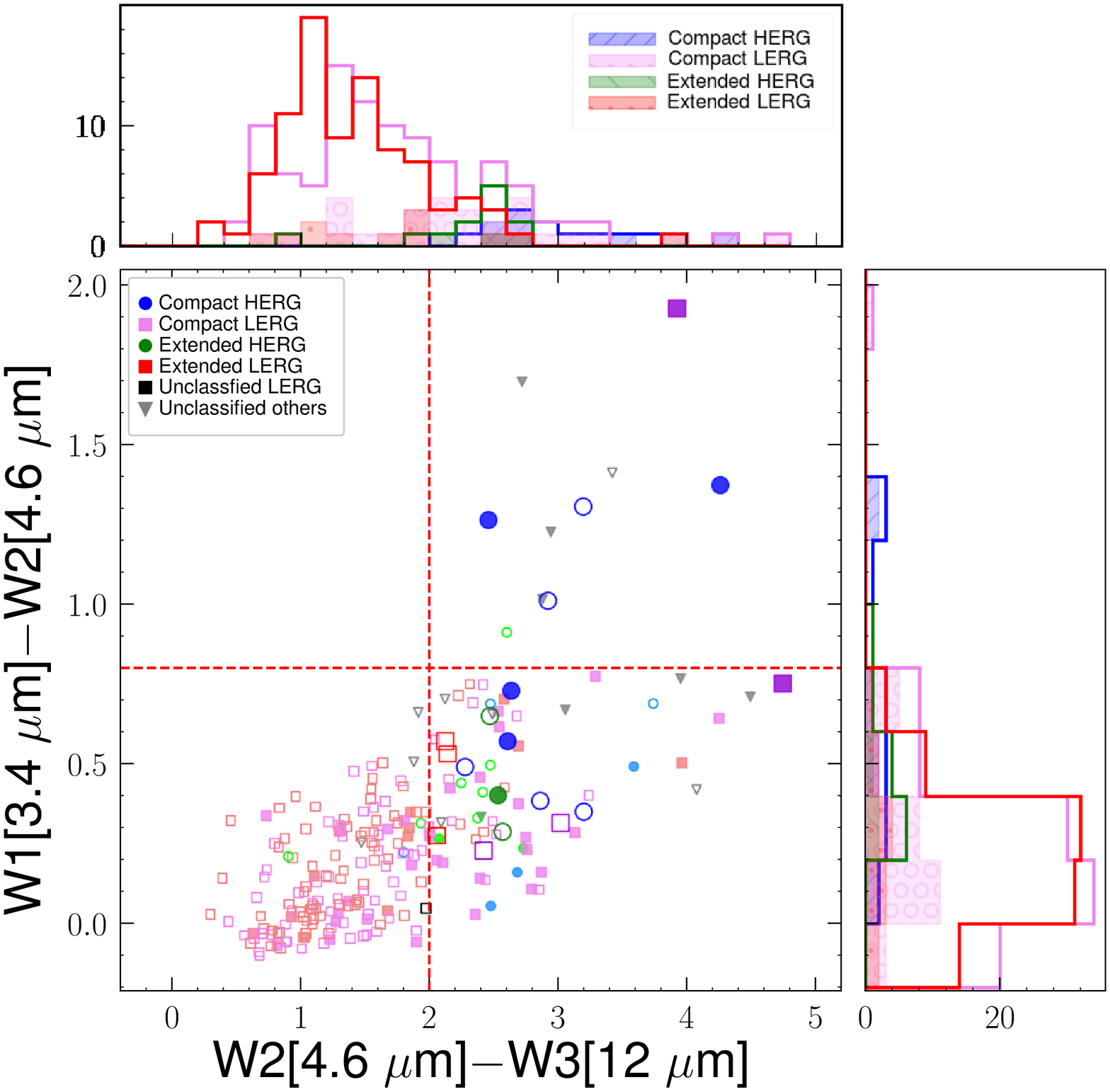}
         \includegraphics[scale=0.35]{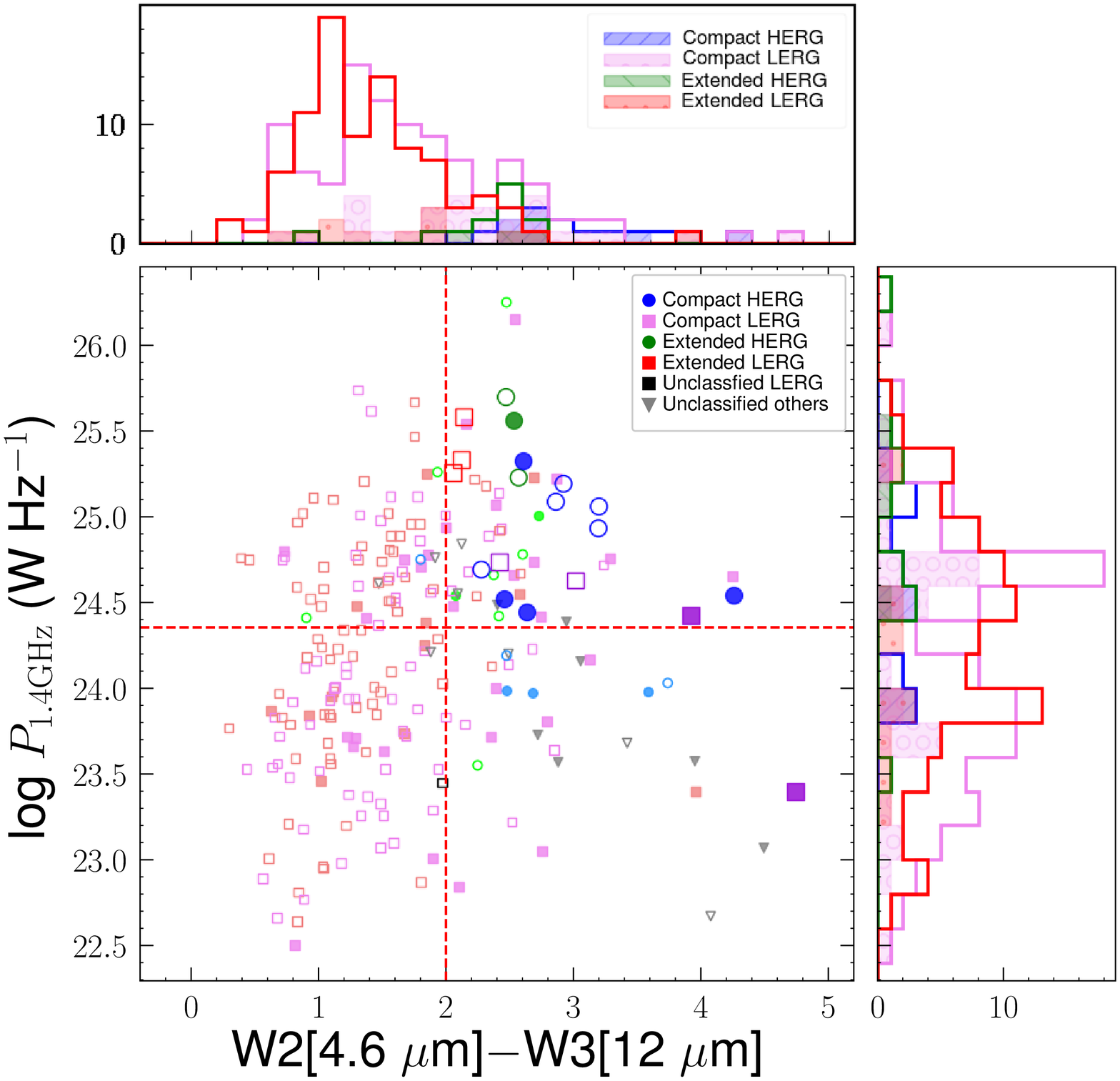}
        \caption{\emph{WISE} W1$-$W2 colour (top) and luminosity at 1.4 GHz (bottom) vs \emph{WISE} W2$-$W3 colour for combined sample of 233 sources. Symbols mean same as in Fig.\ref{fig1}. Filled symbols depict H{\sc i} detections while empty symbols depict H{\sc i} non-detections. Histograms depict the \emph{WISE} W2$-$W3 colour (top of both plots), W1$-$W2 colour (right of top plot) and luminosity at 1.4 GHz (right of botton plot) distribution of LERGs and HERGs.  Histograms in steps (compact HERG: blue, compact LERG: violet, extended HERG: green, extended LERG: red) show distribution of both H{\sc i} detections and non-detections while shaded/hatched histograms (compact HERG:blue shade with right slash hatch, compact LERG: violet shade with circular hatch, extended HERG: green shade with left slash, extended LERG: red shade with dotted hatch) show distribution of only detections. 
        Vertical red coloured dashed lines in both panel mark  W2$-$W3 $=$2. In the upper panel, horizontal red dashed line mark W1$-$W2$=$0.8, for  mid-IR bright AGNs \protect{\citep{2012ApJ...753...30S}}. In the bottom panel, horizontal red coloured line shows the median radio power (at 1.4 GHz) $\sim$ 10$^{24.3}$ W Hz$^{-1}$.}
        \label{fig5}
\end{figure}

  In this subsection, we present the detection rates of radio AGNs according to their classification based on accretion modes, \emph{WISE} colours, and radio properties such as radio power and structures. We have estimated the errors on detection rates using small number Poisson statistics \citep{1986ApJ...303..336G}. We have also given detection rates for different categories of objects in Table~\ref{detectrate}. In Fig.~\ref{fig5}, we have shown the distribution of \emph{WISE} colour W1$-$W2 vs. W2$-$W3 and radio luminosity at 1.4 GHz vs. W2$-$W3 for the combined sample. Detections are shown with filled symbols while non-detections with empty symbols.  
 
 While considering the 219 sources in the sample of \cite{2017A&A...604A..43M}, which are common with that of \cite{2012MNRAS.421.1569B}, those with W2$-$W3$>$2 have a higher detection rate than those with W2$-$W3$<$2, the values being 50.0$^{+10.1}_{-8.7}$ and 13.1$^{+3.6}_{-2.9}$ per cent respectively. The percentages do not change significantly if we exclude the 15 sources, which were unclassified as either LERGs or HERGs by \cite{2012MNRAS.421.1569B}. This is consistent with our earlier work that the radio AGNs with \emph{WISE} colour W2$-$W3 $>$2,  irrespective of their accretion mode and structure, have significantly higher H{\sc i} absorption detection rate than those with W2$-$W3 $<$2 \citep{2017MNRAS.465..997C}.
  
   If we consider only compact sources with W2$-$W3 $>$2 classified as either LERG or HERG in \cite{2017A&A...604A..43M} sample, H{\sc i} absorption detection rate is 21/36 (58.3$^{+15.7}_{-12.6}$ per cent) compared to the detection rate of 5/18 (27.8$^{+18.8}_{-12.0}$ percent) for extended sources. They have classified sources in their sample as extended and compact using ratios of FIRST peak to integrated flux density and major to minor axis in NVSS images following the criterion used by \cite{2015A&A...575A..44G}. In our sample of 20 sources, which were chosen on the basis of their \emph{WISE} colour W2$-$W3$>$2 and have good spectra in our GMRT observations, the detection rates are  42.9$^{+25.6}_{-17.0}$ per cent if we consider only the compact sources with linear projected size less than 20 kpc, and 1/6 (16.7$^{+38.3}_{-13.8}$ percent) if we consider only the extended sources with their linear projected sizes greater than 20 kpc. The higher detection rate for those with W2$-$W3 $>$2 and compact radio structure is consistent with  \cite{2017A&A...604A..43M} sample, within a difference of 1-$\sigma$.
      
       The sample observed with the GMRT, shown with larger symbols in Fig.~\ref{fig5}, all with \emph{WISE} colour W2$-$W3$>$2, has H{\sc i} absorption detection rates of  2/8 (25.0$^{+33.0}_{-16.2}$ per cent) and 5/12 (41.7$^{+28.2}_{-18.0}$ per cent) for LERGs and HERGs respectively (also see Table~\ref{detectrate}). Since detection rates have also dependence on structures, we now consider detection rates for LERGs and HERGs with only compact structures and \emph{WISE} colour W2$-$W3$>$2. The detection rates are 2/5 (40.0$^{+52.8}_{-25.8}$ per cent) and 4/9 (44.4$^{+35.1}_{-21.3}$ per cent) for LERGs and HERGs respectively. This shows for similar radio structure and \emph{WISE} colour W2$-$W3$>$2, detection rates for LERGs and HERGs are similar.
       However for \cite{2017A&A...604A..43M} sample, shown with smaller symbols, detection rates irrespective of their \emph{WISE} colours and structures are 41/189 (21.7$^{+3.9}_{-3.4}$ per cent) for LERGs and 5/15 (33.3$^{+22.5}_{-14.4}$ per cent) for HERGs. For sources with W2$-$W3 $>$2 in their sample, H{\sc i} absorption detection rates are  21/42 (50.0$^{+13.5}_{-10.8}$ per cent) and 5/12 (41.7$^{+28.2}_{-18.0}$ per cent) for LERGs and HERGs respectively. Further limiting their sample to the compact sources and W2$-$W3$>$2, the detection rates are 18/31 (58.1$^{+17.2}_{-13.5}$ per cent) and 3/5 (60.0$^{+58.4}_{-32.7}$ percent) for LERGs and HERGs respectively. This again shows that for similar radio structure and \emph{WISE} colour, there is no difference in detection rates of LERGs and HERGs, albeit within a difference of 1$\sigma$ in detection rate between our sample and the \cite{2017A&A...604A..43M} sample.

  \begin{table*}
  	\centering
  	\caption{Results of Busy Function fit to the H{\sc i} absorption profiles}
  	\begin{tabular}{|l|r|r|r|r|r|r|}
  		\hline
  		
  		\multicolumn{1}{c|}{(1)} &
  		\multicolumn{1}{c|}{(2)} &
  		\multicolumn{1}{c|}{(3)} &
  		\multicolumn{1}{c|}{(4)} &
  		\multicolumn{1}{c|}{(5)} &
  		\multicolumn{1}{c|}{(6)} &
  		\multicolumn{1}{c|}{(7)} \\

  		\multicolumn{1}{c|}{Source} &
  		\multicolumn{1}{c|}{$V_{\rm centroid}$ } &
  		\multicolumn{1}{c|}{$\tau_{\rm peak}$} &
  		\multicolumn{1}{c|}{FWHM} &
  		\multicolumn{1}{c|}{FW20} &
  		\multicolumn{1}{c|}{$-V_{\rm FW20}$} &
  		\multicolumn{1}{c|}{$\int \tau$ dv} \\

  		\multicolumn{1}{c|}{name} &
  		\multicolumn{1}{c|}{km s$^{-1}$ } &
  		\multicolumn{1}{c|}{} &
  		\multicolumn{1}{c|}{km s$^{-1}$} &
  		\multicolumn{1}{c|}{km s$^{-1}$} &
  		\multicolumn{1}{c|}{km s$^{-1}$} &
  		\multicolumn{1}{c|}{km s$^{-1}$} \\
  		\hline

  		J0853+0927 & $-$258.5$\pm$10.0 & 0.049$\pm$0.003 & 201.2$\pm$20.8 & 317.3$\pm$22.1 & $-$417.2$\pm$14.9 &10.5$\pm$0.6\\
  		
  		J0912+5320 & 43.7$\pm$11.8 & 0.039$\pm$0.006 & 51.2$\pm$12.0 & 92.9$\pm$16.6 & $-$2.7$\pm$14.4 &2.4$\pm$0.4\\
  		
  		J1056+1419 & 16.3$\pm$15.0& 0.035$\pm$0.004 & 183.7$\pm$28.0 & 290.2$\pm$34.4 & $-$128.8$\pm$22.8 & 7.0$\pm$0.7\\
  		J1352-0156 & 63.8$\pm$6.4 & 0.049$\pm$0.003 & 49.4$\pm$4.9 & 206.7$\pm$27.7 & $-$39.6$\pm$15.3&5.2$\pm$0.3\\
  		J1534+2330 & $-$77.7$\pm$0.7 & 0.290$\pm$0.001 & 287.7$\pm$1.1 & 443.3$\pm$1.6 & $-$299.3$\pm$1.1 &89.6$\pm$0.3\\
  		J1538+5525 & $-$36.4$\pm$21.6 & 0.014$\pm$0.002 & 118.1$\pm$33.7 & 221.4$\pm$41.03& $-$147.1$\pm$29.8 & 2.0$\pm$0.3\\
  		J2133$-$0712 & 51.2$\pm$0.6 & 0.117$\pm$0.003 & 53.6$\pm$1.6 & 78.2$\pm$1.7 & 12.1$\pm$1.0 &6.5$\pm$0.1\\
  		\hline
  	\end{tabular}
  	\begin{flushleft}
  		Column 1: source name; column 2: velocity at centroid relative to the optical systemic velocity in km s$^{-1}$; column 3:  peak optical depth; column 4: full width at half maximum (FWHM) in km s$^{-1}$; 
  		column 5: full width at twenty percent (FW20) in km s$^{-1}$; column 6:  blueshifted velocity at FW20 relative to the optical systemic velocity in km s$^{-1}$; 
  		column 7:  integrated optical depth in km s$^{-1}$\\
  	\end{flushleft}
  	\label{busyfit}
  \end{table*}
  \begin{table*}
  	\centering
  	\caption{H{\sc i} absorption detection rates for different class of objects}
  	\begin{tabular}{|r|r|l|r|r|r|}
  		\hline
  		\multicolumn{1}{|c|}{} &
  		\multicolumn{1}{|c|}{} &
  		\multicolumn{1}{c|}{LERGs} &
  		\multicolumn{1}{c|}{HERGs} &
  		\multicolumn{1}{c|}{Unclassified} &
  		\multicolumn{1}{c|}{Total} \\
  		\hline
  		\multicolumn{6}{c|}{\cite{2017A&A...604A..43M}  common with \cite{2012MNRAS.421.1569B}} \\   
  		\hline 
  		W2$-$W3 $<$2 & Compact  & 11/69    & 0/1    & 0/2   & 11/72 \\
  		             & Extended & 9/ 77    & 0/2    & 0/1   & 9/80 \\
  		             & Unclassified & 0/1  & -      & -    &  0/1 \\
  		             & Subtotal & 20/147   & 0/3    & 0/3   & 20/153 \\
  	   
  		W2$-$W3 $>$2 & Compact  & 18/31$^{\dagger}$  & 3/5    &  2/7     &  22/40  \\
  		             & Extended &  3/11$^{\ast}$   & 2/7   &  1/1     &  6/17  \\
  		             & Unclassified & -  &  -     &  4/4     & 4/4   \\
  		             & Subtotal & 21/42  & 5/12   & 7/12     & 33/66 \\
  		Total        &          & 41/189 & 5/15   & 7/15     & 53/219 \\
  		\hline
  		\multicolumn{6}{c|}{GMRT observed sample after excluding those with ripples} \\
  		\hline
  		W2$-$W3 $>$2 & Compact  & 2/5  & 4/9  & -  & 6/14  \\
  		             & Extended & 0/3  & 1/3  & -  & 1/6  \\
  		Total        &          & 2/8  & 5/12  & - & 7/20  \\
  		\hline
  	\end{tabular}
  		\begin{flushleft}
  		$\dagger$: 3 sources common with our GMRT observed sample, including J1435+5051 and J0906+4636.\\
  		$\ast$: 2 sources common with our GMRT observed sample.
  	\end{flushleft}

  	\label{detectrate}
  	
  \end{table*}
  We also checked dependence of detection rates on radio luminosities. Higher radio luminosities may increase the population of hydrogen atoms with the higher spin energy level, and hence increasing the spin temperature. For the combined sample, the median radio luminosity at 1.4 GHz is 10$^{24.3}$ W Hz$^{-1}$. We classify those below this value as low radio power sources and those above as intermediate radio power sources. According to their \emph{WISE} W2$-$W3 colour and radio power, we divide 219 sources in the \cite{2017A&A...604A..43M} sample into four categories, (a) low radio power and W2$-$W3$<$2, (b) low radio power and W2$-$W3$>$2, (c) intermediate radio power and W2$-$W3$<$2 and (d) intermediate radio power and W2$-$W3$>$2. The detection percentages for these categories are  13.5$^{+5.1}_{-4.5}$, 55.6$^{+18.4}_{-14.2}$, 12.5$^{+6.2}_{-4.3}$ and 46.2$^{+13.6}_{-10.8}$ showing the dependence on W2$-$W3 colour but no significant dependence on radio luminosity. Since radio structure have also significant effect, we repeat the same exercise by limiting the  radio structures to compact ones. We find the detection percentages for compact radio sources in \cite{2017A&A...604A..43M} sample for above categories are 13.0$^{+7.8}_{-5.2}$, 47.4$^{+21.6}_{-15.5}$, 19.2$^{+13.0}_{-8.3}$, 58.3$^{+20.1}_{-15.4}$ respectively. Most of the sources (19/20) in our GMRT sample with good spectra are located in category of intermediate radio power with W2$-$W3$>$2 and have detection rate of 31.6$^{+18.9}_{-12.5}$ per cent. If limited to compact radio structure, the detection rate increases to 38.5$^{+26.0}_{-16.6}$, within 1-$\sigma$ difference with \cite{2017A&A...604A..43M}. 
 
\subsection{Kinematics}
\begin{figure}
        \centering
       
    \includegraphics[scale=0.35]{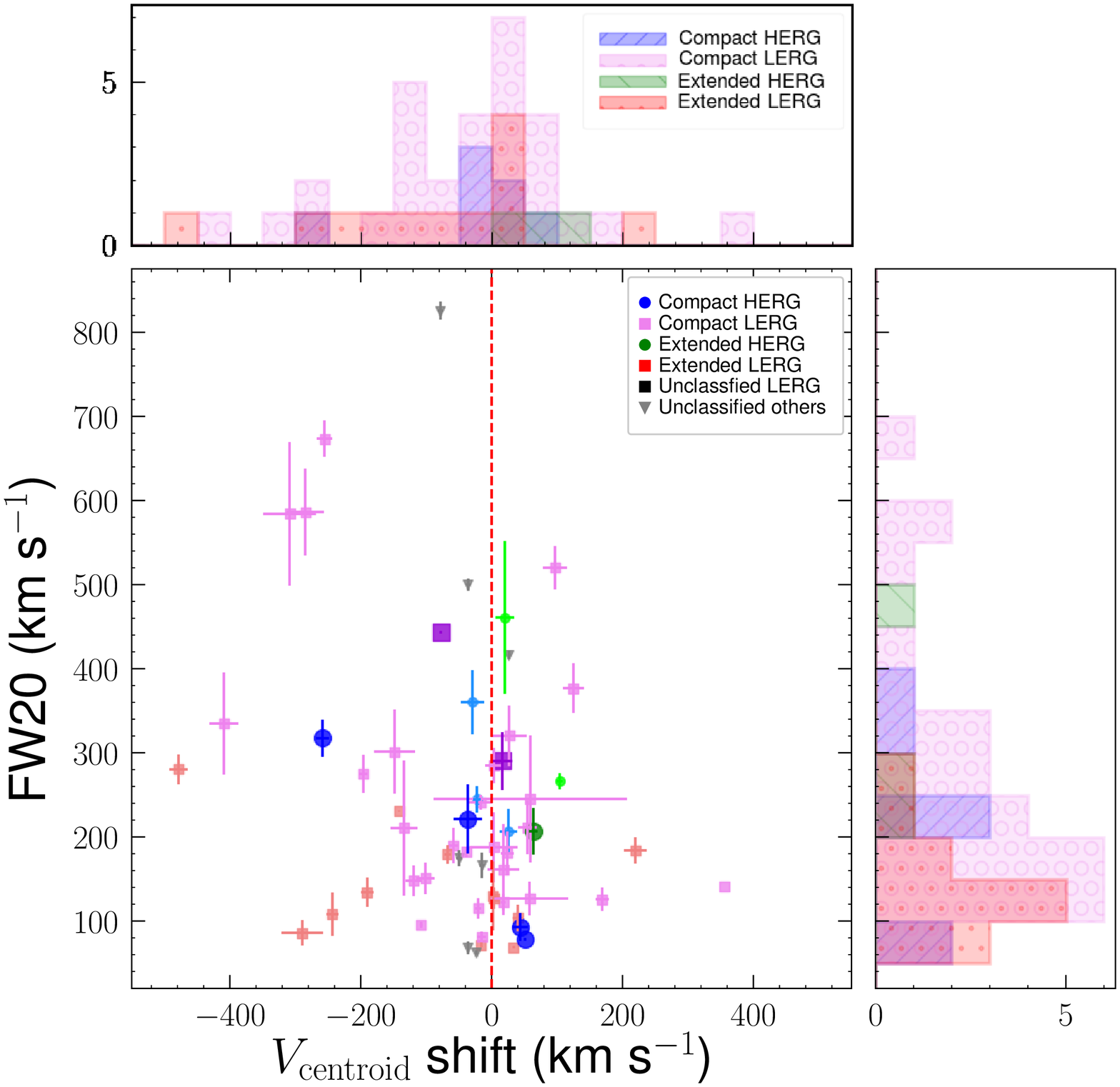}
    \includegraphics[scale=0.35]{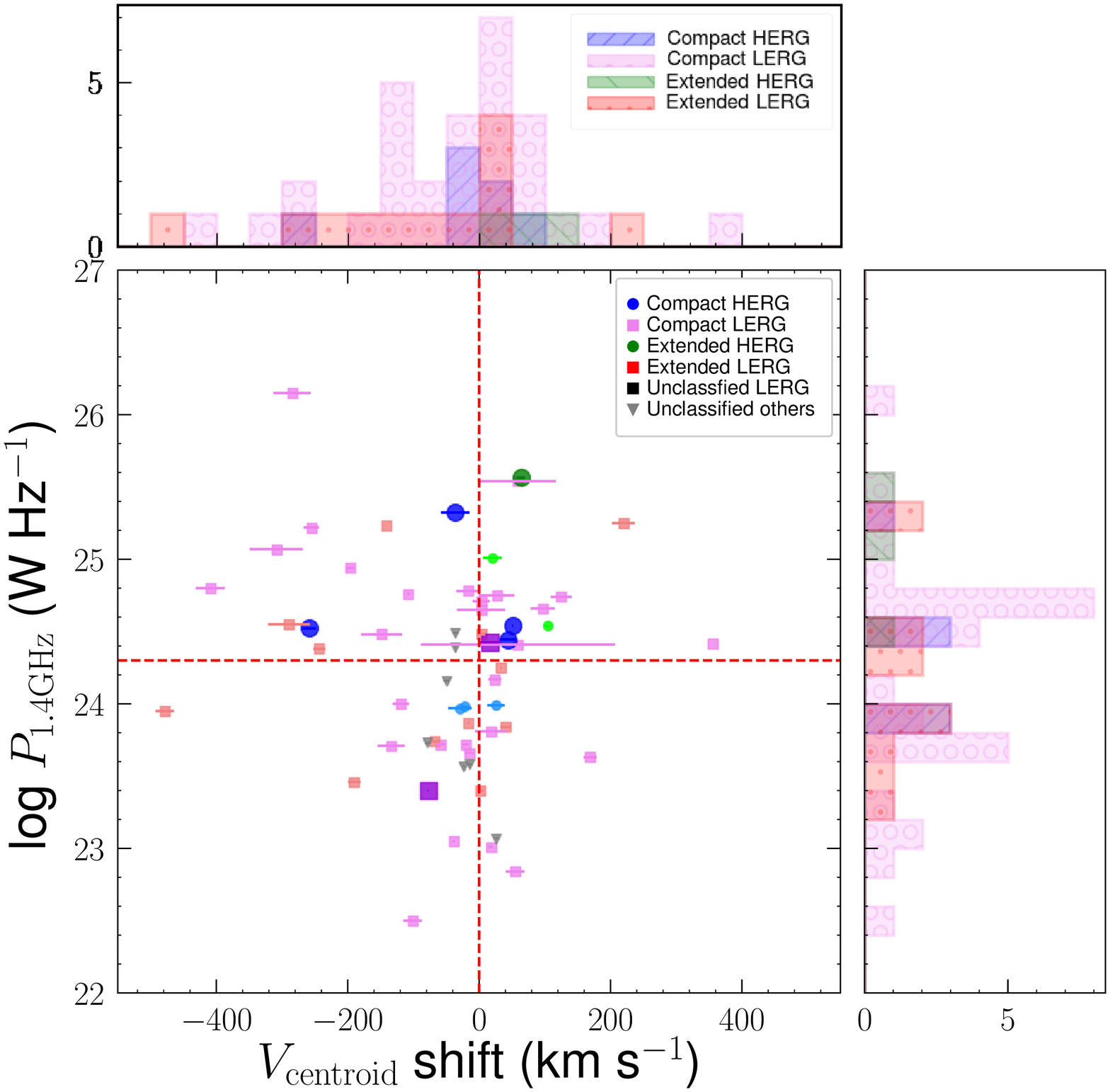}
   \vspace{-0.2 cm}         
        \caption{Full Width at twenty per cent  (top) and luminosity at 1.4 GHz (bottom) vs centroid velocity relative to optical systemic velocity for 59 detections from combined sample. Symbols mean same as in Fig.\ref{fig1}.  Histograms depict the centroid shift (top of both plots), FW20 (right of top plot) and luminosity at 1.4 GHz (right of botton plot) distribution of LERGs and HERGs  from only H{\sc i} detections (compact HERG:blue shade with right slash hatch, compact LERG: violet shade with circular hatch, extended HERG: green shade with left slash, extended LERG: red shade with dotted hatch).  Vertical red coloured dashed lines in both panel mark  $V_{\rm centroid}$ $=$ 0 km s$^{-1}$. In the bottom panel, horizontal red coloured line shows the median radio power (at 1.4 GHz) $\sim$ 10$^{24.3}$ W Hz$^{-1}$.}
        \label{fig6}
\end{figure}

\begin{figure}
        \centering
       
    \includegraphics[scale=0.35]{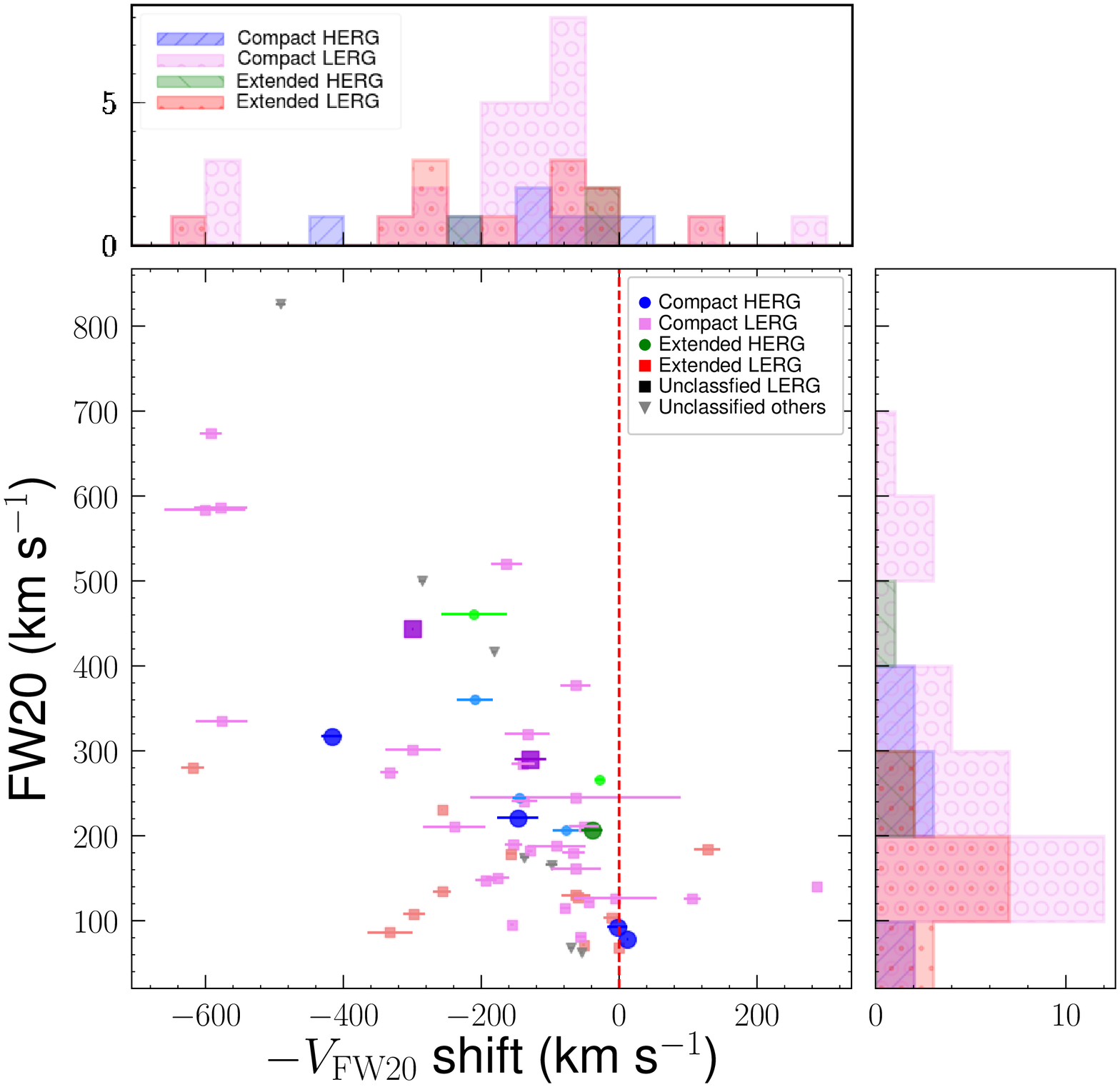}
    \includegraphics[scale=0.35]{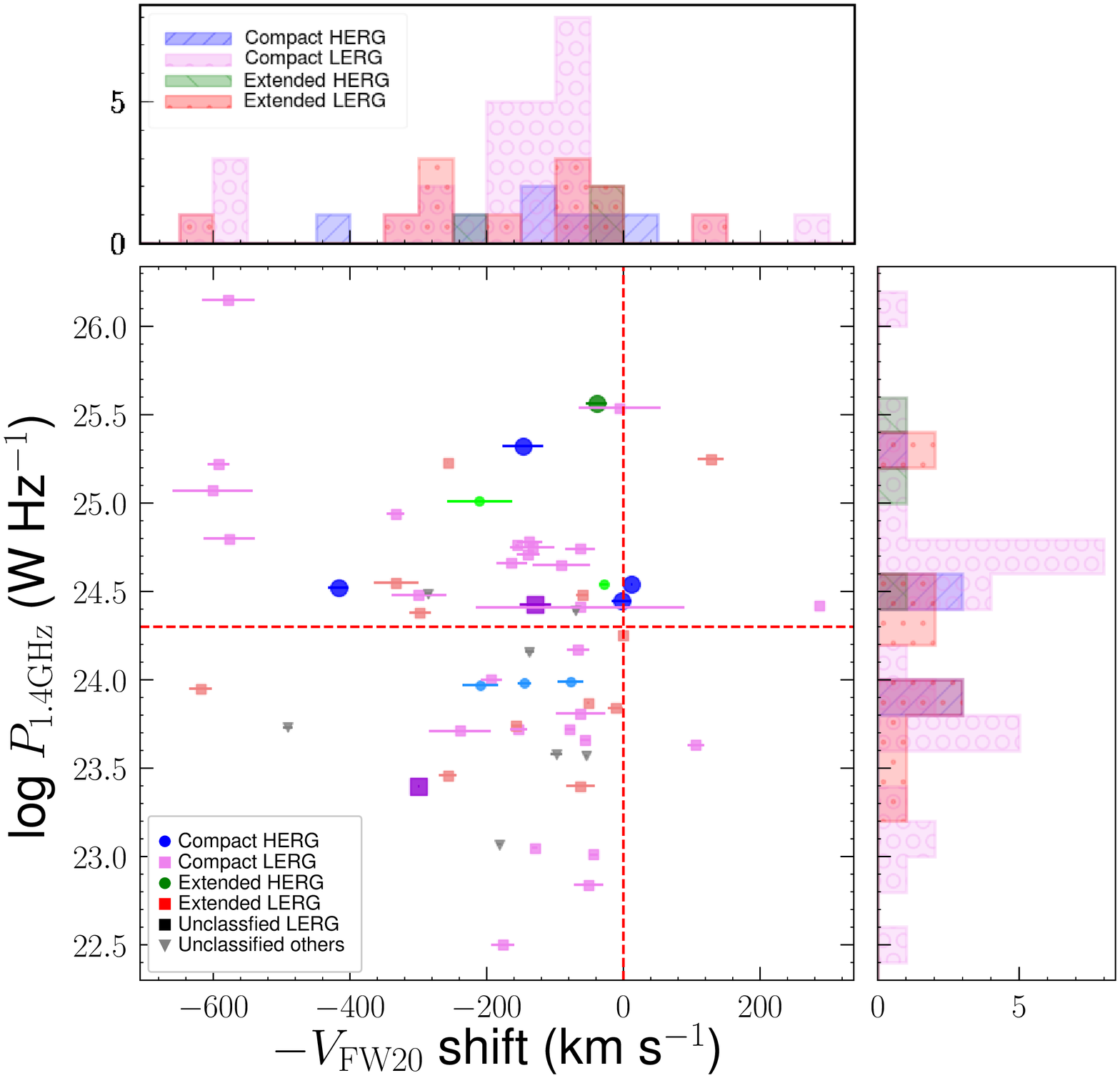}
   \vspace{-0.2 cm}         
        \caption{Full Width at twenty per cent  (top) and luminosity at 1.4 GHz (bottom) vs $-V_{\rm FW20}$ relative to optical systemic velocity for 59 detections from combined sample. Symbols mean same as in Fig.\ref{fig1}. Histograms depict the $-V_{\rm FW20}$ shift (top of both plots), FW20 (right of top plot) and luminosity at 1.4 GHz (right of botton plot) distribution of LERGs and HERGs from only H{\sc i} detections (compact HERG:blue shade with right slash hatch, compact LERG: violet shade with circular hatch, extended HERG: green shade with left slash, extended LERG: red shade with dotted hatch). Vertical red coloured dashed lines in both panel mark  $-V_{\rm FW20}$ $=$ 0 km s$^{-1}$. In the bottom panel, horizontal red coloured line shows the median radio power (at 1.4 GHz) $\sim$ 10$^{24.3}$ W Hz$^{-1}$.}
        \label{fig7}
\end{figure}

     In order to see differences in the H{\sc i} absorption profiles of two types of AGNs, we compare the parameters obtained from fitting the Busy function to the profiles such as FW20 (Full Width at 20 percent of peak), \textbf{$- V_{\rm FW20}$} (extreme blueshifted velocity relative to optical systemic velocity)  and $V_{\rm centroid}$ (centroid shift relative to optical systemic velocity) in Fig.~\ref{fig6} and Fig.~\ref{fig7}. We notice that there is no significant difference in the distribution of FW20 for H{\sc i} profiles from HERGs and LERGs. However, for the HERGs, of the 10 systems only one (10$^{+23.0}_{-8.3}$ per cent) has a shift in the centroid w.r.t. optical systemic velocity more than 200 km s$^{-1}$. For LERGs, H{\sc i} absorption profiles have a wider range ($-$479 km s$^{-1}$ to $+$356 km s$^{-1}$) of centroid shift values with 9/42 (21.4$^{+9.8}_{-7.0}$ per cent) having shifts either towards the blue or red end by more than 200 km s$^{-1}$. Most of these LERGs are intermediate radio power sources. However, for most of the HERGs with intermediate radio power except J0853+0927, this shift is less than 200 km s$^{-1}$.  The centroid shift for most of the low power LERGs and HERGs is less than 200 km s$^{-1}$, except for one LERG J1638+2754. We also notice that minimum of the extreme blueshifted velocities ($-V_{\rm FW20}$) relative to optical systemic velocities for  LERGs is $\sim$ $-$619 km s$^{-1}$ compared to $\sim$ $-$417 km s$^{-1}$ for  HERGs.
\begin{figure}
        \centering
    \includegraphics[scale=0.5]{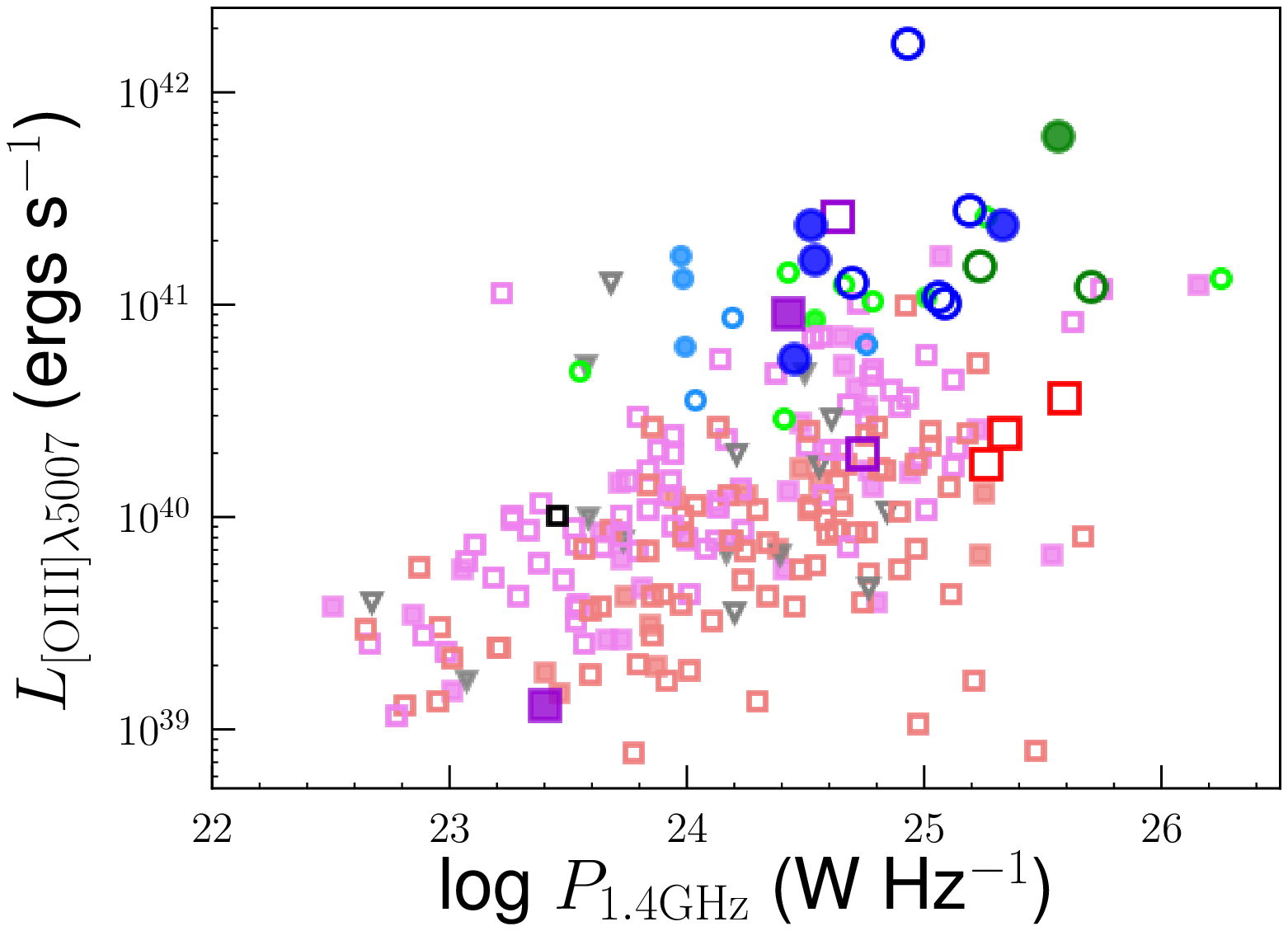}
   \vspace{-0.2 cm}         
        \caption{[O{\sc iii}] $\lambda$5007 spectral line luminosity versus radio luminosity at 1.4 GHz for combined sample of 233 sources. Symbols mean same as in Fig.\ref{fig1}. Filled symbols depict H{\sc i} detections while empty symbols depict H{\sc i} non-detections.}
        \label{fig8}
\end{figure} 
%
\section{Discussion}
    In this Section, we discuss the observed H{\sc i} absorption detection rates and possible effect of AGN feedback on H{\sc i} kinematics. 
    
\subsection{H{\sc i} absorption detection rates}
     As mentioned earlier  in the Section~\ref{sec:results}, we find significant dependence of H{\sc i} absorption detection rates on the radio structures and \emph{WISE} colour, which is consistent with our earlier work \citep{2017MNRAS.465..997C}. It has been discussed in previous papers in the literature that the compact radio sources allow the observer to see through higher density regions and have higher covering factor as compared to the larger radio sources, and hence increasing the detection rates \citep{2003A&A...404..871P,2013MNRAS.431.3408C}. Also, a larger fraction of compact radio AGNs compared to extended ones are known to be residing in the gas rich  host galaxies \citep{2010MNRAS.406..987E}. 
     
     \emph{WISE} late-type galaxies with W2$-$W3 $>$ 2 \citep{2014MNRAS.438..796S} have relatively higher specific star formation rates than those with W2$-$W3 $<$2 \citep{2017MNRAS.465..997C}, and hence the dust is heated due to relatively younger stellar population resulting in redder \emph{WISE} W2$-$W3 colour \citep{2012ApJ...748...80D}. Hence, these are relatively younger systems and are expected to have larger amount of cold gas. Of the 204 sources classified as either LERGs or HERGs in \cite{2017A&A...604A..43M} sample  common with \cite{2012MNRAS.421.1569B}, there are 189 LERGs, of which $\sim$80 per cent have W2$-$W3$<$2, while of the 15 HERGs, $\sim$ 80 per cent have W2$-$W3$>$2. Since the host galaxies of most of the HERGs are\emph{WISE} late type galaxies, this implies sufficient fuel in the form of cold H{\sc i} gas for larger fraction of HERGs compared to LERGs. This effect of \emph{WISE} colour along with radio structure is also reflected in the detection rates of  LERGs (22.0$^{+3.9}_{-3.4}$ per cent)  and  HERGs (37.0$^{+15.8}_{-11.5}$ per cent) for combined GMRT  and \cite{2017A&A...604A..43M} sample. However, as mentioned in the previous Section~\ref{sec:results}, we find similar H{\sc i} detection rate for \emph{WISE} late type LERGs and HERGs with compact radio structure. This implies that the detection of H{\sc i} gas may not necessarily suggest the presence of high excitation mode AGN. It depends on how the cold gas is transported to the vicinity of the central engine which in turn depends on different feeding and feedback mechanisms \citep{2004IAUS..222..235M}. Due to limitations of resolution of our observations and those of \cite{2017A&A...604A..43M}, we are not able to put constraints on precise location of cold H{\sc i} gas absorbers, and hence the mechanism for fuel transport.
     
\subsection{Is cold gas kinematics affected by radio or optical AGN properties?}
      Cold gas kinematics of radio AGNs can tell us about the nature of feedback to their host galaxies. Both, radiative winds from central optical AGN, and radio jets can affect the gas kinematics in host galaxies. In this section, in order to understand the effect of both radiation and jet on gas kinematics, we compare the radio luminosity at 1.4 GHz and [O{\sc iii}] $\lambda$5007 line luminosity (using the line fluxes from MPA-JHU group) for LERGs and HERGs. While radio luminosity at 1.4 GHz can be used as tracer of jet power \citep{2010ApJ...720.1066C}, [O{\sc iii}] $\lambda$5007 line luminosity can be used as a tracer for AGN bolometric luminosity  \citep{2010ApJ...720..786L}. We find that there exists a weak but significant correlation (Kendall's tau$=$0.4, p$=$2.4$\times$10$^{-16}$) between the radio luminosities and the [O{\sc iii}] $\lambda$5007 line luminosities of LERGs (Fig.~\ref{fig8}), which is consistent with earlier studies \citep{1989ApJ...336..702B,1998MNRAS.298.1035T}, implying a common source of both luminosities. A similar correlation is noticed for HERGs but with higher [O{\sc iii}] $\lambda$5007 luminosity than LERGs for a particular value of radio luminosity, which could mean the extra energy in form of wind driven by radiation from accretion disk to cause the turbulence in host galaxy.  However, in our sample of radio AGNs, while for low radio power ($P_{1.4 \rm GHz}$ $<$ 10$^{24.3}$ W Hz$^{-1}$) LERGs and HERGs, there is almost no difference in distribution of centroid shifts in the absorption profiles relative to the systemic velocity, with all but one within $\pm$200 km s$^{-1}$, a larger number of intermediate radio power ($P_{1.4 \rm GHz}$ $\sim$ 10$^{24.3}$-10$^{26.0}$ W Hz$^{-1}$) LERGs show  centroid velocity shifts greater than 200 km s$^{-1}$  compared to HERGs. Also the range of centroid velocity shift for LERGs is wider  (-479 km s$^{-1}$ to $+$356 km s$^{-1}$) than HERGs. Although, we have doubled the number of detections compared to our previous analysis \citep{2017MNRAS.465..997C}, the overall number of sources with H{\sc i} absorption detections for HERGs used for kinematic analysis is still small. Hence there are large statistical uncertainties to draw a firm conclusion about H{\sc i} gas kinematics in HERGs as a population.
     
       We further discuss possible reasons for lesser centroid velocity shifts in our sample of HERGs. We first consider the effect of radiative feedback on the H{\sc i} gas kinematics. In literature, there are examples of HERGs like Mrk 231, where multiphase cold gas outflows are reported to be driven by wide angle strong radiative winds from AGN \citep{2015A&A...583A..99F, 2015ApJ...801L..17A,2016A&A...593A..30M}.
       The radiative winds in HERGs in our sample ($L_{\rm AGN}$(max.) $\sim$ 10$^{46}$ ergs s$^{-1}$; using the bolometric correction factor of $\sim$ 3500 from \citealt{2004ApJ...613..109H} to $L_{\rm O{[III]}}$(max.) $\sim$ 2$\times$10$^{42}$ ergs s$^{-1}$), are capable of driving the galaxy scale H{\sc i} outflows \citep{2011MNRAS.415L...6K,2012ApJ...745L..34Z} of  kinetic luminosity $\sim$10$^{42}$-10$^{43}$ ergs s$^{-1}$ as seen in Mrk 231 \citep{2016A&A...593A..30M}. However, H{\sc i} gas outflows like in Mrk 231 are of very shallow optical depth \citep{2016A&A...593A..30M}, and hence less likely to be detected with the sensitivity of our observations. Also, radio jets are the dominant force which can drive the turbulence in host galaxies of radio AGNs in this range of luminosities \citep{2013MNRAS.433..622M,2019MNRAS.485.2710J, 2019A&A...631A.132M}. However, it depends upon how efficiently the jet couples with different phases of gas in the ISM. Simulations  suggest that the coupling of jet-ISM depends on factors such as jet orientation \citep{2018MNRAS.477.1336C}. Some other simulations suggest young radio sources with weaker inclined jets and clumpy dense interstellar medium  have higher chances of interaction due to longer periods of confinement, and hence causing the turbulence \citep{2018MNRAS.476...80M,2018MNRAS.479.5544M}. Also, differences in radio power of AGNs can give rise to the differences in distribution of velocity shifts \citep{2011MNRAS.418.1787C, 2015A&A...575A..44G, 2017MNRAS.465..997C, 2017A&A...604A..43M}. While for the low and intermediate power LERGs, the difference in distribution of centroid shifts could be attributed to change in radio power along with jet inclination angle relative to H{\sc i} disk, almost no difference in distribution for low and intermediate radio power of HERGs in our sample suggests radio power may have a lesser role to play in these HERGs. It is possible that in most of the HERGs in our sample the jets are well collimated and pierces through the interstellar medium with lesser inclination compared to LERGs or radio source is too compact to interact efficiently with ISM. A better picture can be obtained by parsec scale study of these radio AGNs \citep{2005A&A...441...89G, 2019arXiv190101446M, 2019MNRAS.485.2710J}. Also, the number of HERGs used for kinematic analysis is still small and needs to be increased from larger more sensitive surveys.   
\section{Conclusion}
  We summarise our findings as follows.
  \begin{itemize}
      \item We report 7 H{\sc i} absorption detections towards radio AGNs from the GMRT observations,  of  which 6 are new.
      \item We find that H{\sc i} absorption detection rates have significant dependence on \emph{WISE} W2$-$W3 colour and radio structure.  This further establishes our results from our earlier study with a much smaller sample \citep{2017MNRAS.465..997C}.
      
      \item With help of the larger sample than in our previous study, we are able to put better constraints on H{\sc i} absorption detection rates for LERGs and HERGs. We find that  H{\sc i} absorption detection rate for HERGs (37.0$^{+15.8}_{-11.5}$ per cent) is higher than for LERGs (22.0$^{+3.9}_{-3.4}$ per cent), mainly due to a larger fraction of HERGs being hosted by gas and dust rich galaxies with younger stellar population compared to LERGs.
      
      \item For similar \emph{WISE} W2$-$W3 $>$2 colour and compact radio structure, we don't find a significant difference in detection rate between LERGs and HERGs implying detection of H{\sc i} gas doesn't necessarily mean a high accretion mode AGN. This is also consistent with our previous work. 
      
      \item We don't find any significant dependence of H{\sc i} absorption detections rates on radio luminosities.
      
      \item We also find that for sources of similar intermediate radio power, a larger number of LERGs, compared to HERGs, show H{\sc i} absorption profiles with blue or red shift greater than 200 km s$^{-1}$ for the centroid velocities relative to the optical systemic velocities. Also the range of velocity shift for the LERGs is larger than HERGs. This trend is consistent with that we have found in our earlier paper \citep{2017MNRAS.465..997C}. Though we have double the number of detections of HERGs compared to our earlier study, statistical uncertainities are still large in kinematic analysis. Among the possible reasons for observed distribution of centroid velocity shifts the important ones are radio source size, geometry, power, jet collimation and direction.  While radio power along with jet direction may  play a role for the observed difference in distribution of centroid shifts in low- and intermediate power LERGs, the difference in kinematics of both types of accretion mode AGNs could be due to differences in  their interaction with the interstellar medium. This needs to be studied further from parsec scale study of these objects.
  \end{itemize}
\section*{Acknowledgements}
The authors thank an anonymous referee for many valuable comments and suggestions which have helped improve the paper significantly. We thank the staff of the GMRT who have made these observations possible. GMRT is run by the National Centre for Radio Astrophysics of the Tata Institute of Fundamental Research. This work is supported by National Key R\&D Program of China grant No. 2017YFA0402600. This work of YC was sponsored by the Chinese Academy of Sciences (CAS) Visiting Fellowship for Researchers from Developing Countries, Grant No. 2013FFJB0009”. YC also acknowledges support from National Natural Science Foundation of China (NSFC) Grant No. 11550110181 and No. 11690024. YC also thanks Center for Astronomical Mega-Science, CAS, for FAST distinguished young researcher fellowship.  

 This publication makes use of data products from the \emph{Wide-field Infrared Survey Explorer}, which is a joint project of the University of California, Los Angeles, and the Jet Propulsion Laboratory/California Institute of Technology, funded by the National Aeronautics and Space Administration. This research has made use of NASA's Astrophysics Data System. This research has made use of the NASA/IPAC Extragalactic Database (NED) which is operated by the Jet Propulsion Laboratory, California Institute of Technology, under contract with the National Aeronautics and Space Administration. This research has made use of the VizieR catalogue access tool, CDS, Strasbourg, France. The original description of the VizieR service was published in A\&AS 143, 23.
 This work also makes use of \emph{Sloan Digital Sky Survey} (SDSS)-III . Funding for SDSS-III has been provided by the Alfred P. Sloan Foundation, the Participating Institutions, the National Science Foundation, and the U.S. Department of Energy Office of Science. The SDSS-III web site is \url{http://www.sdss3.org/}. SDSS-III is managed by the Astrophysical Research Consortium for the Participating Institutions of the SDSS-III Collaboration including the University of Arizona, the Brazilian Participation Group, Brookhaven National Laboratory, Carnegie Mellon University, University of Florida, the French Participation Group, the German Participation Group, Harvard University, the Instituto de Astrofisica de Canarias, the Michigan State/Notre Dame/JINA Participation Group, Johns Hopkins University, Lawrence Berkeley National Laboratory, Max Planck Institute for Astrophysics, Max Planck Institute for Extraterrestrial Physics, New Mexico State University, New York University, Ohio State University, Pennsylvania State University, University of Portsmouth, Princeton University, the Spanish Participation Group, University of Tokyo, University of Utah, Vanderbilt University, University of Virginia, University of Washington, and Yale University.
 
 This work has  used different Python packages e.g. numpy, scipy and matplotlib. We thank numerous contributors to these packages. We have also used \textsc{TOPCAT} software \citep{2005ASPC..347...29T} for this work. 

\appendix
\newpage
\section{Notes on individual sources}
\subsection{J0028+0055}
J0028+0055 is a LERG \citep{2012MNRAS.421.1569B} at a redshift $\sim$0.10429, classified as radio loud narrow line Seyfert1 by \cite{2016A&A...591A..98B}. It is a compact steep spectrum radio source. We do not find any evidence of H{\sc i} absorption towards this source.
\subsection{J0813+0734 (4C 07.22)}
J0813+0734 is a compact steep spectrum radio AGN at z$=$0.11239 classified as LERG by \cite{2012MNRAS.421.1569B}. The SDSS optical image shows presence of disk like feature and red coloured nucleus. The H{\sc i} spectrum towards this source is affected by ripples (Fig.~\ref{fig3a}). Hence, we do not consider in our main analysis. However it is possible that there is a H{\sc i} absorption towards this source near optical systemic velocity which needs to be  checked from observations with better sensitivity and bandpass stability.
\subsection{J0816+3804}
This is an extended radio source with angular size $\sim$ 27 \arcsec (linear projected size $\sim$ 80 kpc). Host galaxy is a bright cluster galaxy (BCG). This source has been also searched by \cite{2017A&A...604A..43M} earlier for H{\sc i} absorption, where they did not find any H{\sc i}. Our result is similar to theirs with 3$\sigma$ upper limit on integrated optical depth 22.8 km s$^{-1}$.
\subsection{J0832+1832}
J0832+1832 is a type-2 Seyfert galaxy at a redshift of 0.15411. It is a compact flat spectrum radio source with a possible turn over at around 400 MHz which has been classified as HERG by \cite{2012MNRAS.421.1569B}. Spectrum towards this source has been affected with ripples and hence, we do not consider the H{\sc i} absorption profile towards this source for our analysis. 
\subsection{J0853+0927}
J0853+0927 is a Seyfert type 2 galaxy \citep{2014ApJ...788...45T} classified as HERG by \cite{2012MNRAS.421.1569B} at a redshift of 0.11569. This is compact steep spectrum  radio source with continuum spectra  steep at higher frequency and showing turn over around 350 MHz. We detect blueshifted H{\sc i} absorption line towards this source with two Gaussian components at $-$206 km s$^{-1}$ and $-$351.1 km s$^{-1}$. The integrated optical depth is estimated to be 10.5$\pm$0.8 km s$^{-1}$.
\subsection{J0906+4636}

J0906+4636 is a LERG at a redshift of 0.0847 \cite{2012MNRAS.421.1569B}. In our earlier work we had reported H{\sc i} absorption towards this source \citep{2011MNRAS.418.1787C}. However, with the new GMRT observations we are not able to confirm our earlier result due to baseline ripple and hence do not consider in our analysis. Earlier, \cite{2015A&A...575A..44G} and \cite{2017A&A...604A..43M}, had reported a non-detection towards this radio source. 

\subsection{J0912+5320}
J0912+5320 is a type-2 Seyfert galaxy at a redshift of 0.10173, with double peaked optical emission lines in SDSS spectra, classified as HERG by \cite{2012MNRAS.421.1569B}. Radio image obtained by \cite{2018ApJ...854..169L} using Very Large Baseline Array (VLBA) shows subarcsecond scale structure consisting three components along north-south direction. Total angular size from VLBA image is $\sim$30 mas corresponding to a linear projected size of $\sim$ 56 parsec. We detect H{\sc i} absorption line towards this source at velocity which is red-shifted by $\sim$ 26 km s$^{-1}$ relative to the optical redshift.

\subsection{J1056+1419}
This source has been reported as  a Seyfert-1 type galaxy at redshift 0.08127 by SIMBAD astronomical database \citep{2014ApJ...788...45T}, but classified as LERG  by \cite{2012MNRAS.421.1569B}. In optical SDSS image, this source shows signature of merger. We detect H{\sc i} absorption towards this source with two components, one deeper and broader component near systemic velocity and other shallower, narrower component redshifted by $\sim$ 158.2 km s$^{-1}$. The narrower redshifted component could be due to infalling gas cloud.

\subsection{J1058+5628} 
J1058+5628 is classified as BLLac object by \cite{2008AJ....135.2453P} and LERG by \cite{2012MNRAS.421.1569B}. The redshift of this source in SDSS is 0.14324. This is another source for which the H{\sc i} spectrum has been affected by ripples.

\subsection{J1107+1825} J1107+1825 is a Seyfert type-1 galaxy at redshift 0.17856. It has been classified as HERG by \cite{2012MNRAS.421.1569B}. It has  also a nearby galaxy with unknown redshift in SDSS optical image. H{\sc i} spectrum towards this source is also affected by ripples.

\subsection{J1110+2131} 
J1110+2131 is also a Seyfert type-1 galaxy \citep{2014ApJ...788...45T} and a compact steep spectrum radio source  at a redshift of 0.13461 and classified as HERG by \cite{2012MNRAS.421.1569B}. We do not detect H{\sc i} absorption towards this source as well.

\subsection{J1156+2632}
J1156+2632 is Seyfert type-2 galaxy \citep{2014ApJ...788...45T} at a redshift of 0.15625. It is classified as HERG by \cite{2012MNRAS.421.1569B}. It is a compact flat spectrum radio source. We do not detect H{\sc i} absorption towards this radio source.

\subsection{J1217-0337}
 PKS 1215-033 is a HERG at redshift of 0.18229 \citep{2012MNRAS.421.1569B}. This source has resolved structure with GMRT beam of angular size $\sim$9.68\arcsec, which corresponds to 29.68 kpc in projected linear size. This source has not been detected with  H{\sc i} absorption.

\subsection{J1328+1738}
J1328+1738 is a Seyfert type-1 galaxy \citep{2014ApJ...788...45T} classified as HERG by \cite{2012MNRAS.421.1569B}. It is located at a redshift of 0.18035 and could be classified as compact flat spectrum radio source based on ViZieR radio flux values. We report a H{\sc i} non-detection towards this source. However, it is possible that there is  marginal H{\sc i} absorption towards this source near optical systemic velocity which needs to be checked from observations with better sensitivity. 

\subsection{J1341+5344}
J1341+5344 is a Seyfert type-2 galaxy \citep{2014ApJ...788...45T} at a redshift of 0.14094, classified as HERG by \cite{2012MNRAS.421.1569B}. This is an extended radio source of angular size $\sim$1.52\arcmin which corresponds to a projected linear size of $\sim$ 226 kpc. We do not detect  H{\sc i} absorption towards the brightest pixel in the central region of this source.

\subsection{J1350+0940}
J1350+0940 is bright cluster galaxy (BCG) \citep{2016MNRAS.460.3669Y} at a redshift of 0.13255, classified as Seyfert type-1 galaxy by \citep{2014ApJ...788...45T}. It has been also classified as LERG by \cite{2012MNRAS.421.1569B}. Using radio fluxes obtained from ViZieR, it could classified be as either GPS or CFS radio source. H{\sc i} spectrum towards this source is affected by a broad ripple across the band, and hence not included in our analysis. It is possible that there is a very shallow, wide and blueshifted H{\sc i} absorption profile towards this source. This possibility needs to be checked from more sensitive observation. 
\subsection{J1352-0156}
PKS J1352-0156 is a radio galaxy at a redshift of 0.16694, classified as HERG by \cite{2012MNRAS.421.1569B}. It is an intermediate radio power AGN (L$_{\rm 1.4 GHz}$ $\sim$ 10$^{26}$W/Hz). This source has resolved structure of $\sim$9\arcsec with GMRT beam of $\sim$2\arcsec. This corresponds to a projected linear size of $\sim$26 kpc. The H{\sc i} absorption profile towards this source have two components, one narrow and deep, other wide and shallow, both redshifted w.r.t. the optical emission lines. The wide profile has full width half maximum of $\sim$300 km s$^{-1}$ and is redshifted by $\sim$44 km s$^{-1}$. Narrower profile is redshifted by $\sim$146 km s$^{-1}$ and has FWHM $\sim$39 km s$^{-1}$. The peak optical depths for narrower and broader profiles are 0.041 and 0.013, respectively. Integrated optical depth estimate for this source is 5.8$\pm$0.4 km s$^{-1}$ which corresponds to a column density of 10.6$\pm$0.7 $\times$10$^{20}$ cm$^{-2}$, if we  assume $T_{\rm s}=$100 K and $f_{\rm c}$=1.
\subsection{J1400+5216}
J1400+5216 is a Low Ionization Nuclear Emission-line Region (LINER) galaxy at a redshift of 0.11789, classified as LERG by \cite{2012MNRAS.421.1569B}. This source was also observed by \cite{2017A&A...604A..43M} where they report a non-detection of H{\sc i} absorption. We also do not detect H{\sc i} absorption with a upper limit of 2.3 km s$^{-1}$ on integral optical depth corresponding to 4.1$\times$ 10$^{20}$ cm$^{-2}$ for $T_{\rm s} =$ 100 K and $f_{\rm c}$=1. 

\subsection{J1410+1438}
J1410+1438 is a BLLac object \citep{2014ApJS..215...14D} at a redshift of 0.14419 classified as LERG by \cite{2012MNRAS.421.1569B}. The radio continuum image suggests this is an extended radio source with angular size $\sim$22\arcsec or linear projected size $\sim$55 kpc. We do not detect H{\sc i} absorption towards this source as well.

\subsection{J1435+5051}
J1435+5051 is a radio galaxy at a redshift of 0.09969. It is classified as LERG by \cite{2012MNRAS.421.1569B}. This source was also observed by \cite{2017A&A...604A..43M} where they report a detection of H{\sc i} absorption with a shift on centroid of $-$66.7 km s$^{-1}$ relative to optical systemic velocity. However, from our observation we are not able to confirm it, possibly due to higher noise in our spectra.
 
\subsection{J1447+4047}
J1447+4047 is a radio galaxy at a redshift of 0.19515 which has been classified as LERG by \cite{2012MNRAS.421.1569B}. It is an extended radio source with overall angular size of $\sim$ 14.6\arcsec corresponding to projected linear size of 47.3 kpc. This source has been also searched for H{\sc i} absorption by \cite{2017A&A...604A..43M} with no detection. We also do not detect H{\sc i} absorption towards this source.

\subsection{J1449+4221}
J1449+4047 is a Seyfert type-2, compact flat spectrum radio source at a redshift of 0.17862 classified as LERG by \cite{2012MNRAS.421.1569B}. We do not detect H{\sc i} absorption towards this radio source.

\subsection{J1534+2330}
J1534+2330 or Arp 220, is a dust rich Ultra-luminous infrared galaxy (ULIRG) system of spiral galaxies which has undergone recent merger at redshift of 0.0184. This source has been classified as LERG by \cite{2012MNRAS.421.1569B}. We detected H{\sc i} absorption profile with multiple components towards this source for which the H{\sc i} column density estimate is 163.0$\pm$0.9 $\times$ 10$^{20}$ cm$^{-2}$ for $T_{\rm s}=$100 K and $f_{\rm c}=$1. H{\sc i} absorption towards this source has been detected in earlier works as well \citep{1982ApJ...260...75M, 1987ApJ...322...88G, 2001ApJ...560..168M, 2014MNRAS.440..696A}. \cite{2001ApJ...560..168M} from their parsec scale study reported absorption against two continuum nuclei due to two counter rotating H{\sc i} disks and bridge of gas connecting them. 

\subsection{J1538+5525}
J1538+5525 is type-1 Seyfert galaxy at a redshift of 0.19117, with double peaked narrow optical emission lines detected in SDSS spectra \citep{2012ApJS..201...31G}. It has been classified as HERG by \cite{2012MNRAS.421.1569B}. It is also a compact steep spectrum radio source. We detected H{\sc i} absorption towards this source where the profile can be fitted with two Gaussian components, both blueshifted relative to optical systemic velocity. The narrower and deeper component is blueshifted by $\sim$105 km s$^{-1}$ while broad, shallow component is blueshifted by $\sim$49 km s$^{-1}$.

\subsection{J1543+0235}
J1543+0235, is a Seyfert type-2 galaxy at a redshift of 0.18793, classified as HERG by \cite{2012MNRAS.421.1569B}. This is an extended radio source in GMRT continuum image with resolved structure of $\sim$11.9\arcsec which corresponds to $\sim$37 kpc in linear projected size. H{\sc i} spectrum towards this source is also affected by ripples.

\subsection{J1559+5330}
 J1559+5330 is also a Seyfert type-2 galaxy \citep{2014ApJ...788...45T} at a redshift of 0.17919 which has been classified as HERG by \cite{2012MNRAS.421.1569B}. We do not detect H{\sc i} absorption towards this source.

\subsection{J2133-0712}
J2133-0712 is radio galaxy at a redshift of 0.08654, classified as HERG by \cite{2012MNRAS.421.1569B} with disc shaped morphology in SDSS optical image. We detect H{\sc i} absorption towards this source with single Gaussian component redshifted by $\sim$51 km s$^{-1}$ relative to optical systemic velocity.

\bsp	
\label{lastpage}
\end{document}